\documentclass[review,fleqn,sort&compress]{elsarticle}
\usepackage{amssymb,amsbsy,amsthm,amsmath,amsfonts,amssymb,amscd}
\usepackage{geometry}
\usepackage{babel,blindtext}
\usepackage{lipsum}
\usepackage{bm} 
\usepackage{graphicx}
\graphicspath{{./images/}}
\usepackage{float}
\usepackage{caption}
\usepackage{subcaption}
\usepackage{multirow}
\usepackage{cancel}
\usepackage{lineno}
\usepackage{hyperref}
\usepackage{graphicx}
\usepackage{mathrsfs}
\hypersetup{colorlinks, citecolor=black, filecolor=black, linkcolor=black, urlcolor=black}
\usepackage{booktabs}
\usepackage[ruled,linesnumbered]{algorithm2e}
\usepackage[toc,page,titletoc]{appendix}
\usepackage[capitalise,nameinlink]{cleveref}
\crefname{supp}{Supplement}{Supplements}
\usepackage{algpseudocode}
\usepackage{mathtools,nccmath}
\usepackage{adjustbox}
\usepackage{cleveref}
\usepackage{xparse}
\theoremstyle{definition}

\usepackage{color}
\usepackage{xcolor}
\usepackage{amsmath}
\usepackage{lineno}
\usepackage{soul}
\colorlet{soulred}{red!40}

\definecolor{airforceblue}{rgb}{0.36, 0.54, 0.66}

\begin{document}

\begin{frontmatter}
\title{A new unified arc-length method for damage mechanics problems}

\author{Roshan Philip Saji ${}^{1,2}$}
\author{Panos Pantidis ${}^3$}
\author{Mostafa E. Mobasher ${}^{1,3}$\corref{cor1}}

\cortext[cor1]{Corresponding author. \emph{E-mail address:} \texttt{mostafa.mobasher@nyu.edu} (Mostafa E. Mobasher)}
\address{${}^1$ Mechanical Engineering Department, Tandon School of Engineering, New York University,  6 MetroTech Center, Brooklyn, NY 11201, USA}
\address{${}^2$ Mechanical Engineering Department, New York University Abu Dhabi, Abu Dhabi, P.O. Box 129188, UAE}
\address{${}^3$ Civil and Urban Engineering Department, New York University Abu Dhabi, Abu Dhabi, P.O. Box 129188, UAE}

\begin{abstract}

The numerical solution of continuum damage mechanics (CDM) problems suffers from convergence-related challenges during the material softening stage, and consequently existing iterative solvers are subject to a trade-off between computational expense and solution accuracy. In this work, we present a novel unified arc-length (UAL) method, and we derive the formulation of the analytical tangent matrix and governing system of equations for both local and non-local gradient damage problems. Unlike existing versions of arc-length solvers that monolithically scale the external force vector, the proposed method treats the latter as an independent variable and determines the position of the system on the equilibrium path based on all the nodal variations of the external force vector. This approach renders the proposed solver substantially more efficient and robust than existing solvers used in CDM problems. We demonstrate the considerable advantages of the proposed algorithm through several benchmark 1D problems with sharp snap-backs and 2D examples under various boundary conditions and loading scenarios. The proposed UAL approach exhibits a superior ability of overcoming critical increments along the equilibrium path. Moreover, in the presented examples, the proposed UAL method is 1-2 orders of magnitude faster than force-controlled arc-length and monolithic Newton-Raphson solvers.

\end{abstract}

\end{frontmatter}

\section{Introduction}
\label{Sec:Introduction}

\subsection{Overview}
\label{Sec:IntroOverview}

Damage mechanics studies the behavior of damage in materials and structures by focusing on understanding the mechanisms of crack initiation and propagation that lead to material and structural failure \cite{lemaitre2012course}. Its principles can be applied to model the behavior of a wide variety of natural and man-made materials, such as geomaterials \cite{li2001continuum,mobasher2017non,mobasher2021dual}, metals \cite{bonora1998low,voyiadjis2002modeling}, composites \cite{talreja2006multi,williams2001application}, and biomaterials \cite{zioupos1998recent}, and it can be therefore adopted for various industrial applications \cite{bjorstad2001domain,iankov2003finite}. However, the notorious computational cost of non-linear damage mechanics models necessitates the development of methods with ever-increasing versatility in order to capture complex, non-linear responses with sufficient accuracy, while maintaining industry-acceptable computational effort \cite{oden2006finite,ghrib1995nonlinear,giancane2010fatigue}. In view of the above, a new unified arc-length (UAL) solver is proposed in this work to overcome the high computational cost of damage mechanics problems involving sharp changes in the direction of the equilibrium path while being able to advance beyond the critical points. 

\subsection{{Literature Review}}
\label{Sec:IntroLitReview}

Non-linear mechanics solvers typically face two main challenges: a) stability: whether the solver is capable of bypassing the critical points in the equilibrium path, and b) efficiency: if the computational cost remains reasonably low. These requirements possess an inter-competing nature, since the more advanced and flexible solvers are typically associated with a heavier computational burden and vice versa. With both objectives in mind, we discuss below the strengths and limitations of the prevailing solution techniques for continuum damage mechanics problems, and we identify the literature gap which motivates our work. 

One of the earliest non-linear solvers used was the bisection method \cite{cocks1989inelastic,eiger1984bisection} that iteratively divided an interval until the solution was obtained. Despite being a simple algorithm that ensured convergence in most cases, its slow rate of convergence and the small interval size in non-linear paths limited its wide-scale utilization \cite{demir2008trisection}. 
The fixed point iteration method, built upon the seminal work of L.E. Brouwer in algebraic topology \cite{brouwer1911abbildung}, provides a faster convergence rate than the bisection method \cite{azure2019comparative} and is thus preferable to it \cite{zeid1985fixed}. The drawback of this method however is its sensitivity to the initial guess values of the unknown variables. Several subsequent works \cite{allgower2003introduction,allgower2012numerical,rheinboldt1983locally} have documented the various continuation methods used to solve non-linear equations.

One of the most common non-linear solvers in solid mechanics problems is the Newton-Raphson linearization \cite{bathe1975finite,hughes1978consistent,hartmann2005remark} which has been widely used due to its quadratic rate of convergence and stability compared to earlier approaches \cite{azure2019comparative, ehiwario2014comparative}. Several variations of the Newton-Raphson approach have been proposed over the years, such as the modified Newton-Raphson \cite{jennings1971accelerating,nayak1972note} and the Secant method \cite{wolfe1959secant}, particularly in order to reduce the computational cost. However, in finite element problems with a highly non-linear behavior, the Newton-Raphson and its variations often fail to converge as the solution approaches the critical points of the equilibrium path, due to singularities that arise in the determinant of the system Jacobian matrix \cite{nordmann2019damage}. 

The arc-length methods proposed by Riks \cite{riks1972arclen,riks1979incremental} and Wempner \cite{wempner1971discrete} have been used to overcome the challenges of the Newton-Raphson method in tracing complex equilibrium paths that include snap-back and snap-through behavior \cite{crisfield1981fast}. Significant efforts have been made to optimize the arc-length methods for geometrically non-linear problems over the decades, by:

\begin{itemize}
\item Improving root selection procedures \cite{bathe1983automatic,bellini1987improved,de1999determination}: When using Crisfield's approach for calculating arc-length solution, two root values are obtained. Choosing the 'right' root value is critical to avoid backtracking of the solver.  

\item Controlling the shape of constraint surfaces \cite{park1982family,simo1986finite,skeie1990local} for stability near critical points: The rate of convergence and stability of arc-length solvers are influenced by the choice of the constraint surface, such as cylindrical and spherical.

\item Using algorithmic strategies to improve the performance of the arc-length framework \cite{forde1987improved,de2012nonlinear}: Different implementation schemes have been proposed in the literature to optimize the performance of existing arc-length solvers, especially in the highly non-linear zone of the equilibrium path.
\item Using energy based arc-length methods \cite{may2016new,singh2016fracture,bharali2022robust} that are driven by the internal and/or dissipated energy of the domain to enhance solver performance in highly non-linear problems with sharp changes in the equilibrium path.
\end{itemize}

These approaches, while effective in tracing very complex equilibrium paths, impose a high computational cost \cite{memon2004arc} and require engineering manipulations like those in \cite{bathe1983automatic,bellini1987improved,de1999determination} to overcome the challenges of solver backtracking and non-convergence. Also, many applications of the arc-length method mentioned above are limited to non-linear problems that use a force-controlled or an equivalent force-controlled arc-length approach. The work by Pretti et al. \cite{pretti2022displacement} was the first to introduce a displacement-controlled arc-length method for geometrically non-linear problems with "skeletal" 1D elements, where the constraints are applied on the nodal displacements using strong Dirichlet boundary conditions. However, this work lies in the realm of geometric non-linearities, and an approach that utilizes the strengths of a framework that simultaneously updates both the external force and the displacements for damage mechanics problems has not yet been proposed in the literature. This is the gap addressed by the novel unified arc-length method (UAL) proposed in this work.

Advancing the solution of non-linear problems beyond the bifurcation and inflection points can also be achieved with advanced optimization techniques. Algorithms like the BFGS \cite{byrd1987global,dai2013perfect} and Levenberg-Marquardt \cite{levenberg1944method,marquardt1963algorithm} are particularly useful and widely used, but they have substantial memory requirements and they need a large number of iterations to converge. Therefore, they impose a significant additional cost when used in conjunction with the Newton-Raphson and arc-length methods \cite{kristensen2020phase,bharali2022robust}. Several other techniques have been developed over the years with the goal of reducing the overall cost of non-linear FEM models through algorithmic and mathematical innovations. Such developments include Finite Element Tearing and Interconnecting method (FETI)\cite{farhat1991method,mobasher2016adaptive}, domain decomposition approaches \cite{pebrel2008nonlinear,lloberas2011domain}, multi-grid approaches in space and time \cite{shaidurov2013multigrid,rosam2008adaptive} and staggered solution approaches \cite{miehe2010thermodynamically,hofacker2012continuum}. Overall, while the aforementioned approaches can effectively improve the computational cost of the non-linear damage mechanics models, they do not offer a solution to overcome the critical points in the equilibrium paths which challenge methods like Newton-Raphson and arc-length approaches. 

\subsection{Scope and Outline}
\label{Sec:IntroScopeOutline}

To overcome the aforementioned challenges, we present in this work a new unified arc-length (UAL) method for damage mechanics problems. While this method extends the displacement-controlled arc-length framework proposed by Pretti et. al \cite{pretti2022displacement} from geometric non-linearity to material non-linearity problems with damage, here we coin the term \textit{unified} arc-length method (UAL) to avoid confusion with existing displacement controlled methods that monolithically update only the displacement part \cite{crisfield1983arc,hellweg1998new}. Also, the name UAL better reflects that both all nodal variables including displacement, external force, and any other variable (e.g. non-local strain) contribute to the arc-length equation and to the evolution of the system of equations. The paper presents the derivation of the UAL system of equations for both local \cite{crisfield1982local} and non-local gradient damage laws \cite{peerlings1996gradient}, and appropriate solution schemes are derived and presented in detail for each case. We investigate the performance of our new framework against several benchmark 1D and 2D problems, with the objective of demonstrating its strengths over the Newton-Raphson and traditional force-controlled arc-length (FAL) methods. In all the numerical examples explored in this paper, the UAL method has outperformed the conventional techniques in terms of computational time, number of increments, and robustness, being capable of capturing critical snap-backs in the equilibrium paths at a computational cost which is several orders of magnitude less than the FAL approach.

The structure of this paper is as follows: Section \ref{Continuum Damage Mechanics (CDM)} briefly introduces the basics of Continuum Damage Mechanics and the two damage laws used in this work. Section \ref{Sec:Numerical_Solvers} discusses commonly used non-linear solvers in damage mechanics problems and introduces the concept of the UAL approach in relation to them. Section \ref{Sec:UAL_system_eqns} presents the mathematical derivation of the UAL-based consistent tangent matrices for damage mechanics problems is presented. The different implementation schemes and control parameters used in the implementation of UAL are illustrated in Section \ref{Sec:ImplementationSchemes}. Finally, the numerical results comparing the performance of the UAL method to traditional non-linear solvers are presented in Section \ref{Sec:NumericalExamples}. 


\section{Continuum Damage Mechanics (CDM)}
\label{Continuum Damage Mechanics (CDM)}

Continuum Damage Mechanics (CDM) provides an approach to model fractures in materials through a continuous stiffness degradation zone rather than discrete material discontinuities \cite{murakami2012continuum,kachanov1986introduction}. Damage evolves as a function of material point variables such as the strain or the stress. In the case of isotropic materials, damage is typically represented by a scalar quantity $d$ that ranges from 0 to 1, where 0 represents the intact state and 1 represents complete material failure. The computation of damage as a function of local material point variables leads to the well-known issues of non-unique and mesh-dependent numerical results \cite{pijaudier1987nonlocal}. To overcome the limitations and challenges of the local damage framework, several non-local damage laws have been developed in the literature \cite{peerlings1996gradient,ahmed2021local,pijaudier1987nonlocal}, which are centered around the notion of representing damage as a diffused property over a material area. The size of this diffusion region is typically dictated by an additional variable $l_{c}$, termed as $characteristic \ length$
\cite{pijaudier2004non,jirasek2005non}. In this study, we adopt both the local and the non-local gradient method by Peerlings et al. \cite{peerlings1996gradient}.

\begin{figure}
\centering
\includegraphics[width=1\textwidth,trim = 0cm 2cm 0cm 2cm, clip]{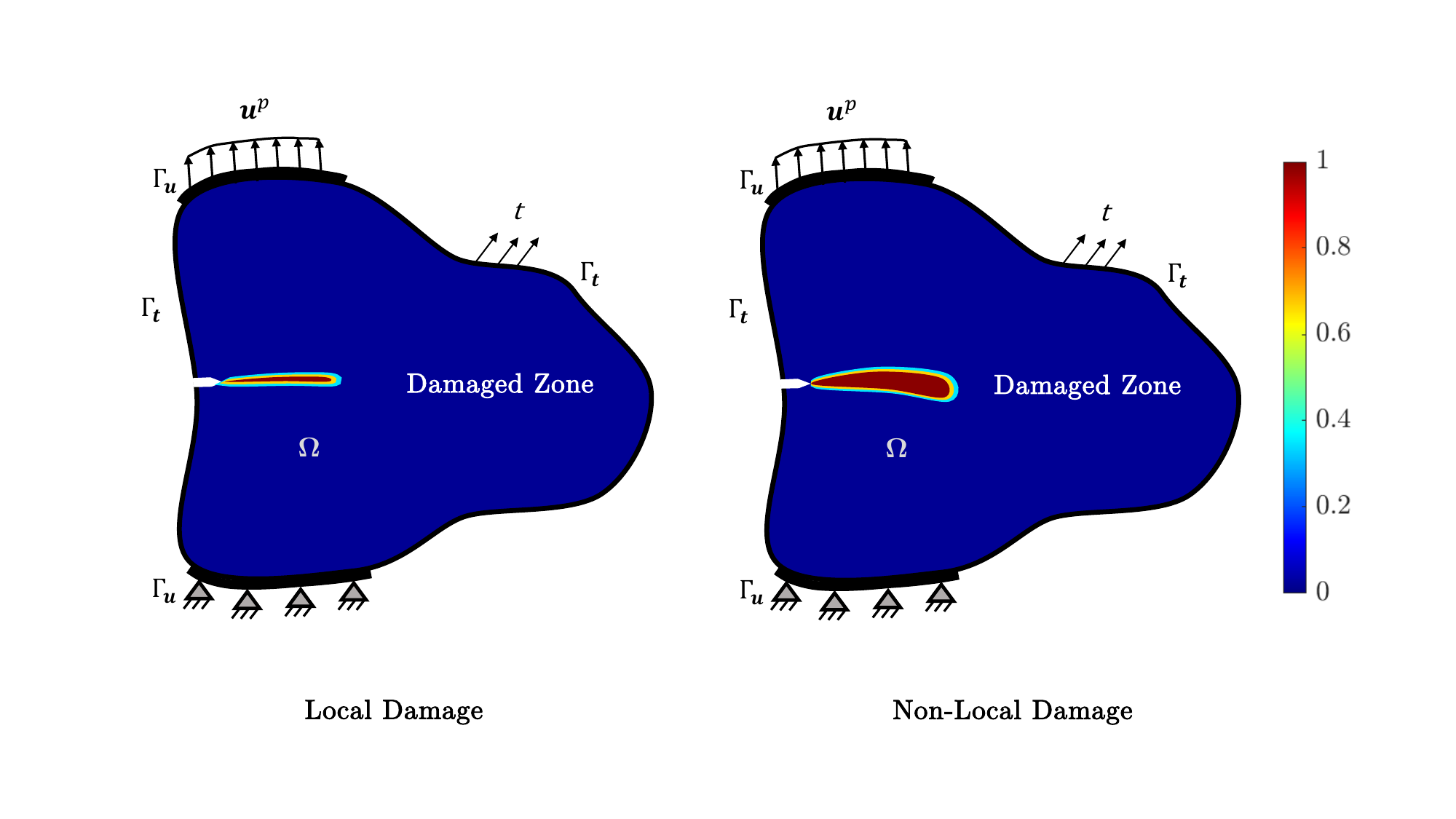}
\caption{Schematic representation of a 2D domain with local (left) and non-local (right) damage idealization.}
\label{CDM Schematic}                                                         
\end{figure}

\subsection{Local Damage}
Consider a domain $\Omega$ as shown in Fig.\ref{CDM Schematic} with the boundary  $\Gamma = \Gamma_u \cup \Gamma_t$, where $\Gamma_u$ is the part of the boundary where prescribed displacements $\boldsymbol{u}^p$ are applied and $\Gamma_t$ is the part of the boundary with prescribed tractions $\boldsymbol{t}$. Using the Einstein indicial notation \cite{chaves2013notes} to represent the various vectors, tensors, and their derivatives, the strong form of the balance of momentum reads as:

\begin{equation}
{\sigma}_{ij,j}=0
\label{CDM Strong Form}
\end{equation}

The stress ${\sigma}_{ij}$ is a function of damage $d$ and can be expressed as: 

\begin{equation}
{\sigma}_{ij} = (1-d) \ C_{ijkl} \epsilon_{kl} 
\label{Stress_func_damage}
\end{equation}

\noindent where, ${\sigma}_{ij}$ is the Cauchy stress tensor, $C_{ijkl}$ is the fourth-order elasticity tensor and $\epsilon_{kl}$ is the local strain tensor. Here, damage $d=d(\epsilon)$ is a function of the local equivalent strain. Assuming small deformations, the strain and elasticity tensor are defined as:

\begin{equation} 
 \epsilon_{ij}= \frac{1}{2} \left[u_{i,j} + u_{j,i} \right]
\label{Strain_tensor_defination}
\end{equation}

\begin{equation}
 C_{ijkl} = \left[k \ {\delta}_{ij} {\delta}_{kl} + \mu \left[ {\delta}_{ik} {\delta}_{jl} + {\delta}_{il} {\delta}_{jk} - \frac{2}{3} {\delta}_{ij} {\delta}_{kl} \right] \right] ; 
\label{Elasticity_tensor_defination}
\end{equation}

\noindent where $k$ is the bulk modulus, $\mu$ is the shear modulus and $\delta_{ij}$ is the Kronecker delta function. To convert the strong form equation to a discretized weak form, we introduce an arbitrary weight function $\boldsymbol{w}^u$ and a shape function ${\boldsymbol{N}}^u$ and its derivative ${\boldsymbol{B}}^u$:

\begin{equation}
\boldsymbol{u} = \boldsymbol{N}^u u^e  \qquad ;  \qquad \boldsymbol{w}^u =  \boldsymbol{N}^u w^{u,e} \qquad ; \qquad \nabla \boldsymbol{w}^u = \boldsymbol{B}^u w^{u,e}
\label{Disp_shape_functions}
\end{equation}

\noindent where the superscript 'e' refers to nodal values. Applying the shape functions to Eqn.\eqref{CDM Strong Form} yields the discretized weak form of the PDE:

\begin{equation}
\boldsymbol{r}^u = \underbrace{\int_\Omega [\boldsymbol{B}^u]^T \boldsymbol{\sigma} d\Omega}_{\boldsymbol{f}^{int}} - \underbrace{\int_\Gamma [\boldsymbol{N}^u]^T t d\Gamma}_{\boldsymbol{f}^{ext}} 
\label{r_u_Local_weak}
\end{equation}

The solution of the non-linear system is obtained when the residual $\boldsymbol{r}^u$ in Eqn. \eqref{r_u_Local_weak} is minimized. We note that the terms $\boldsymbol{f}^{int}$ and $\boldsymbol{f}^{ext}$ refer to the internal and external force vectors respectively. 

\subsection{Non-Local Gradient Damage}

The non-local gradient damage framework proposed by Peerlings et al. \cite{peerlings1996gradient} introduces the following strain-diffusion PDE in the system of governing equations:

\begin{equation}
\epsilon = {\bar{\epsilon}} - c {{\bar {\epsilon}},}_{ii} 
\label{Nonlocal_strain_GvnEq}
\end{equation}

\noindent where $\bar{{\epsilon}}$ is the non-local strain, ${{\bar \epsilon},}_{ii}$ is the second-order spatial partial derivative of the non-local strain, $c$ is a function of the characteristic length $l_{c}$ ($c = {l_{c}}^{2}/2$) and $\epsilon$ is the equivalent local strain at a material point. Here, damage $d=d(\bar {\epsilon})$ is a function of the non-local equivalent strain. Discussion of the natural boundary condition on Equation \eqref{Nonlocal_strain_GvnEq} can be found in \cite{peerlings1996gradient}. The nodal discretization of the non-local strain requires the introduction of weight functions, shape functions, and their derivatives as follows:

\begin{equation}
{\boldsymbol{\bar \epsilon}} = \boldsymbol{N}^{\bar \epsilon} {{\bar \epsilon}}^{e} \qquad ; \qquad \boldsymbol{w}^{\bar \epsilon} = \boldsymbol{N}^{\bar \epsilon} w^{{\bar \epsilon},e} \qquad ; \qquad \nabla \boldsymbol{w}^{\bar \epsilon} = \boldsymbol{B}^{\bar \epsilon} w^{{\bar \epsilon},e}
\label{Nonlocal_shape_functions}
\end{equation} 

Applying Eqns. \eqref{Disp_shape_functions} and \eqref{Nonlocal_shape_functions} to the governing Eqns. \eqref{CDM Strong Form} and \eqref{Nonlocal_strain_GvnEq} yields the following discretized weak form of the governing PDEs: \eqref{r_u_Nonlocal_weak} - \eqref{r_e_Nonlocal_weak}. 

\begin{equation}
\boldsymbol{r}^u = \underbrace{\int_\Omega [\boldsymbol{B}^u]^T \boldsymbol{\sigma} d\Omega}_{\boldsymbol{f}^{int}} - \underbrace{\int_\Gamma [\boldsymbol{N}^u]^T t d\Gamma}_{\boldsymbol{f}^{ext}} 
\label{r_u_Nonlocal_weak}
\end{equation}

\begin{equation}
\boldsymbol{r}^{\bar \epsilon} = \underbrace{{\int_\Omega [\boldsymbol{N}^{\bar \epsilon}]^T \boldsymbol{\bar{\epsilon}} \ d\Omega + \int_\Omega [\boldsymbol{B}^{\bar \epsilon}]^T \ c \ \nabla \boldsymbol{\bar{\epsilon}} \ d\Omega} -{\int_\Omega [\boldsymbol{N}^{\bar \epsilon}]^T \boldsymbol{\epsilon} \ d\Omega}}_{\boldsymbol{f}^{int, \bar\epsilon}} 
\label{r_e_Nonlocal_weak}
\end{equation}

The solution of the non-linear system is obtained by minimizing the residuals $\boldsymbol{r}^u$ and $\boldsymbol{r}^{\bar \epsilon}$. The term $\boldsymbol{f}^{int, \bar \epsilon}$ refers to the internal force vector due to the non-local strain.

\subsection{Clausius-Duhem Inequality}

Damage evolution continues in the positive direction by enforcing the Clausius-Duhem inequality, which is an interpretation of the first and second laws of thermodynamics and ensures that damage growth is irreversible \cite{murakami2012continuum,kachanov1986introduction}. This inequality can be implemented in CDM problems by enforcing the following condition, where $\dot d$ refers to the damage growth rate:

\begin{equation}
\dot d \geq 0 
\label{Kuhn-Tucker_conditions}
\end{equation}


\section{Non-Linear solvers}
\label{Sec:Numerical_Solvers}

This section discusses the fundamentals of three non-linear solvers: the Newton-Rapshon (NR) method, force-controlled arc-length (FAL) method, as well as the newly proposed unified arc-length (UAL) method. A detailed discussion of the advantages and limitations of each method is presented along with their mathematical formulation. The overarching system of equations presented in Eqns. \eqref{General_Nonlinear_solver_Eqns} appertains to all cases and it will be used in the presentation of each solver, where $\boldsymbol{J}$ is the Jacobian matrix of the system defined as $\boldsymbol{J}={\partial\boldsymbol{r}}/{\partial\boldsymbol{x}}$:

\begin{subequations}

\begin{equation}
\{\boldsymbol{x}\}=\begin{cases}
    \{\boldsymbol{x}^{ld}\} = [\boldsymbol{u}], & \text{in the local damage case}.\\
    \{\boldsymbol{x}^{nld}\} = [\boldsymbol{u} \quad \boldsymbol{\bar \epsilon}]^T, & \text{in the non-local gradient damage case}.
  \end{cases}
  \label{General_x}
\end{equation}

\begin{equation}
\{\boldsymbol{r}\}=\begin{cases}
    \{\boldsymbol{r}^{ld}\}, & \text{in the local damage case}.\\
    \{\boldsymbol{r}^{nld}\}, & \text{in the non-local gradient damage case}.
  \end{cases}
  \label{General_r}
\end{equation}

\begin{equation}
\boldsymbol{r}= \boldsymbol{f}^{int}-\boldsymbol{f}^{ext}
\label{General_residual_vector}
\end{equation}

\begin{equation}
\prescript{}{}{\boldsymbol{r}}= - \prescript{}{}{\boldsymbol{J}} \prescript{}{}{\delta \boldsymbol{x}} 
\label{General_corrector_equation}
\end{equation}

\label{General_Nonlinear_solver_Eqns}
\end{subequations}

Eqns. \eqref{General_x} and \eqref{General_r} represent the independent variables and residual vectors respectively, for both the local and non-local gradient damage cases. Eqn. \eqref{General_residual_vector} is a generalized representation of the non-linear weak-form Eqns. \eqref{r_u_Local_weak}, \eqref{r_u_Nonlocal_weak} and \eqref{r_e_Nonlocal_weak}. Finally, Eqn. \eqref{General_corrector_equation} represents the non-linear system of equations that governs the numerical solution and relates the residual vector to the Jacobian matrix and the nodal degrees of freedom of the system.


\subsection{Newton-Raphson (NR)}
\begin{figure}[H]
\centering
\includegraphics[width=0.9\textwidth,trim = 0cm 2.5cm 0cm 1.5cm, clip]{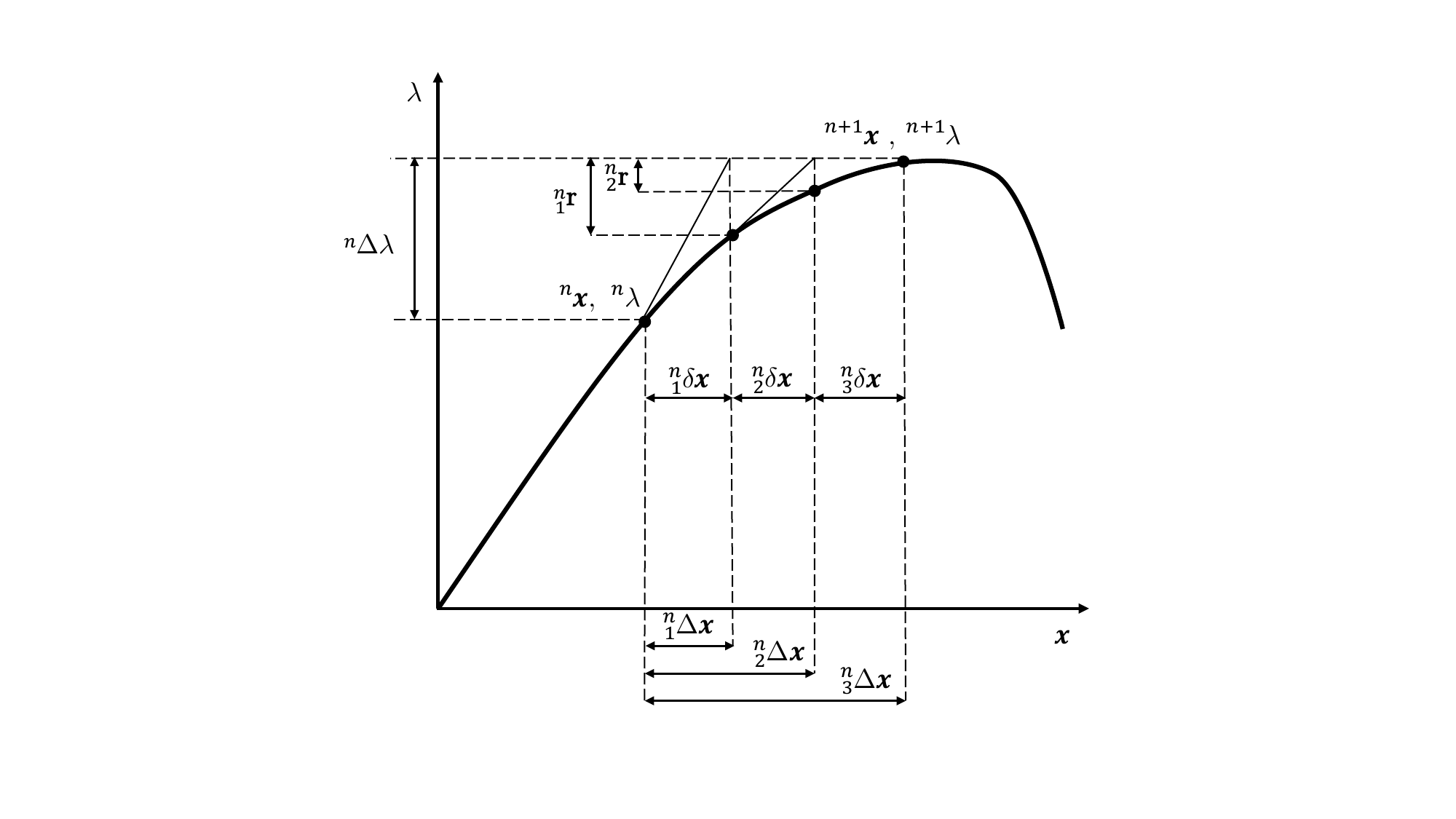}
\caption{Schematic of the Newton-Raphson scheme where $\boldsymbol{x}$ represents the independent variables and $\lambda$ represents the loadfactor. The left superscript is the increment number and the left subscript represents the iteration number. In every increment the system is advanced by $\Delta \lambda$ and an iterative process is required to achieve convergence.}
\label{fig:NR_Schematic}
\end{figure}

Consider the external force vector $\boldsymbol{f}^{ext}=\lambda \boldsymbol{q}$, where $\boldsymbol{q}$ is the total load vector and $\lambda$ is a scalar loadfactor value that ranges between 0 and 1 and represents the applied load level. Fig. \ref{fig:NR_Schematic} shows a schematic representation of the NR solver where an incremental load is applied to the system.

As observed in Fig. \ref{fig:NR_Schematic} the loadfactor $\lambda$ is updated at the beginning of each increment to push the system forward until the total load is applied. For the NR case, the governing system of equations - upon substitution of the Jacobian matrix definition in Eqn. \eqref{General_corrector_equation}  - is given as:

\begin{equation}
    \prescript{}{}{\left[\boldsymbol{r}\right]} = - \prescript{}{}{\left[\frac{\partial \boldsymbol{r}}{\partial \boldsymbol{x}}\right]} \prescript{}{}{\left[\delta \boldsymbol{x}\right]} 
    \label{NR_residual}
\end{equation}

\noindent where for every load incrementation $\prescript{n}{}{\Delta \lambda}$, the non-linear system of equations in Eqn.\eqref{NR_residual} is solved iteratively for $\boldsymbol{x}$  until convergence \cite{bathe2006finite,hughes2012finite} (see Fig.  \ref{fig:NR_Schematic}). Here, we underline that the loadfactor $\lambda$ is a user-specified variable and it is not influenced by the independent variables $\boldsymbol{x}$. The loadfactor can be either a fixed scalar quantity or subject to adaptive load incrementation schemes, in order to account for sharp changes in the equilibrium path and to overcome the associated convergence issues in highly non-linear problems. \cite{padovan1980self, jennings1971accelerating, nayak1972note, wolfe1959secant, diez2003note}. Despite the availability of various adaptive loading methods, the Newton-Raphson method generally faces challenges as the solution approaches critical points in the force-displacement curve. For example, when the system approaches the peak value of the externally applied force, or the direction of the equilibrium path changes in a snap-back or snap-through manner, singularities in the Jacobian matrix arise and NR often fails to converge \cite{nordmann2019damage}. These well-documented challenges of NR have paved the way for more advanced numerical techniques, such as the force-controlled arc-length approach which is discussed next.


\subsection{Force-controlled arc-length (FAL)}
\label{Sec:Force Controlled arc-length}

Arc-length methods were developed to overcome the limitations of the NR solver near critical points, and in this subsection we discuss the fundamental approach of force-controlled arc-length (FAL) method. FAL and other line search methods are commonly used to solve non-linear problems with sharp changes in the direction of the equilibrium path \cite{nocedal2006line,riks1979incremental}. The goal of these methods is to avoid the possibility of singularity of the Jacobian matrix that arises when solving Eqn \eqref{NR_residual}, and their core idea is to solve simultaneously for the independent variable $\Delta \boldsymbol{x}$ and the load level incrementation $\Delta \lambda$. To achieve this goal, the solution at each increment is sought for within a distance $\Delta l$ from the last converged step, where - unlike NR - both the loadfactor $\Delta \lambda$ and the nodal degrees of freedom $\Delta \boldsymbol{x}$ are treated as unknown quantities. Fig. \ref{fig:FAL_Schematic} depicts a schematic representation of the FAL method: 
 
\begin{figure}
    \centering
    \begin{subfigure}{8.2cm}
    \centering     
    \includegraphics[width=1\textwidth,trim = 4cm 2cm 7cm 1cm, clip]{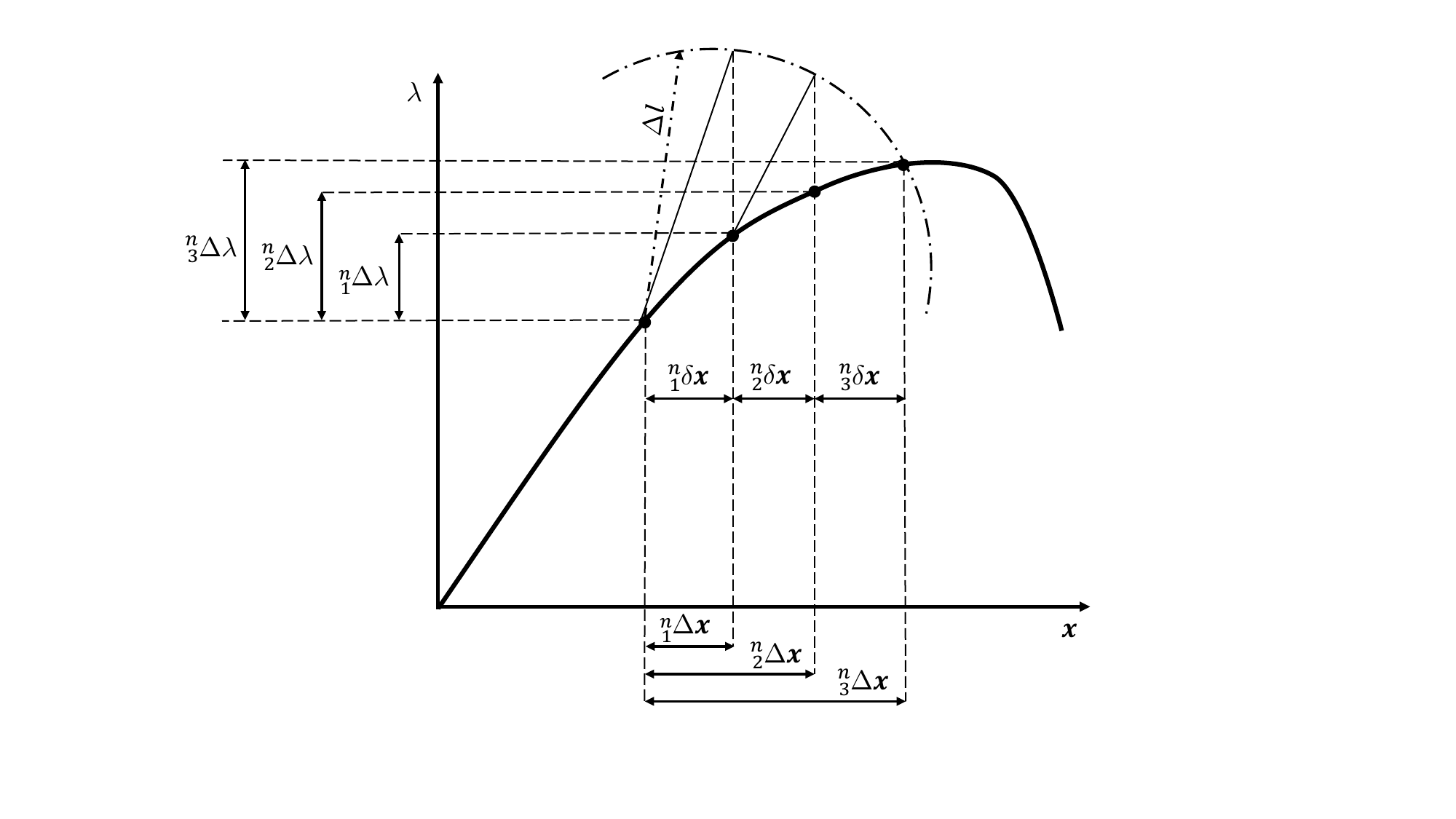}
    \caption{}
    \label{fig:FAL_Schematic}
    \end{subfigure}
    \hfill
    \centering
    \begin{subfigure}{8.2cm}
    \centering     
    \includegraphics[width=1\textwidth,trim = 5cm 1.5cm 6cm 1cm, clip]{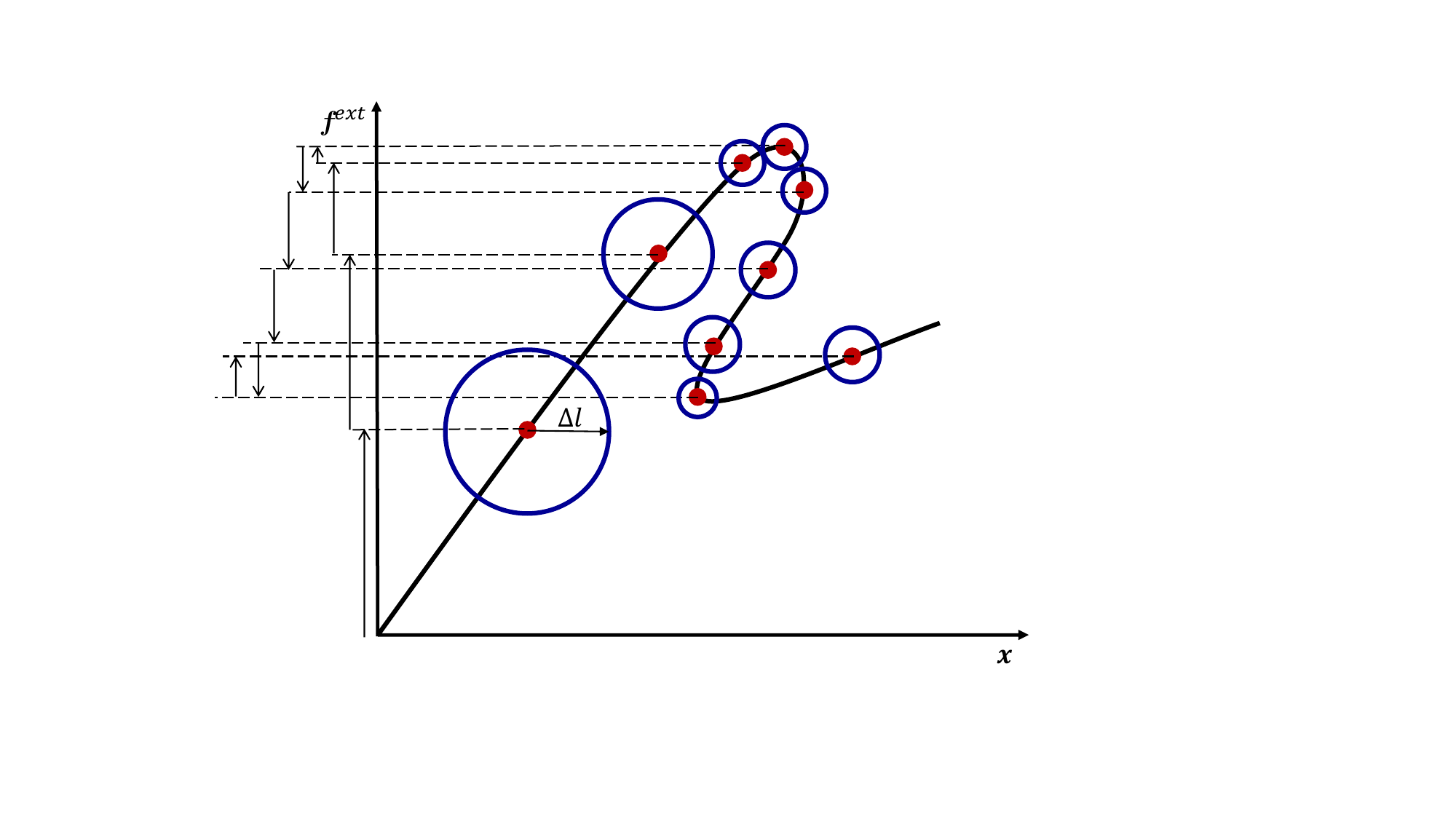}
    \caption{}
    \label{fig:FAL_ArcLength_representation}
    \end{subfigure}
    \caption{(a) Schematic representation of the force-controlled arc-length method where $\boldsymbol{x}$ represents the independent variables and $\lambda$ represents the loadfactor. Unlike the Newton-Raphson method, both $\boldsymbol{x}$ and $\lambda$ are updated simultaneously in each iteration. $\Delta l$ refers to the arc-length used to define the search radius within which $\Delta \boldsymbol{x}$ and $\Delta \lambda$ are calculated; (b) Force-controlled arc-length method with representative arc-length contours at different points on the equilibrium path; In FAL, $\boldsymbol{f}^{ext}$ = $\lambda$ $\boldsymbol{q}$ and $\lambda$ is an unknown.}
    \label{fig:FAL_scheme}
\end{figure}

Since $\lambda$ is unknown, an additional equation is required to complete the definition of the FAL model. This is called the arc-length equation and it is presented below:

\begin{equation}
\prescript{}{}{\ {g}} (\Delta \boldsymbol{x}, \Delta \lambda) = [ \Delta \boldsymbol{x}+ \delta \boldsymbol{x}]^T [\Delta \boldsymbol{x}+  \delta \boldsymbol{x}] +{\beta}^2 (\Delta \lambda + \delta \lambda)^2 \boldsymbol{q}^T . \boldsymbol{q} - \Delta l^2 
\label{FAL_ArcLength_Equation}
\end{equation} 

\noindent where we underline that the initial value of $\Delta l$ is a known, user-defined quantity. Note that Eqn.  \eqref{FAL_ArcLength_Equation} represents the general force-controlled arc-length equation only. The terminology chosen here is based on that used previous works \cite{pretti2022displacement,findeisen2017characteristics} to describe classical arc-length methods. The performance of the FAL solver depends on the locus traced by the arc-length $\Delta l$ and its intersection with the equilibrium path. The location of the intersection points depends on the shape of the constraint surface which is dictated by the value of $\beta$.

Fig. \ref{fig:FAL_Schematic} shows the simultaneous update of $\boldsymbol{x}$ and $\lambda$ at each iteration within a loadstep defined by the radius of the arc-length $\Delta l$. Following Eqns. \eqref{General_residual_vector} and \eqref{FAL_ArcLength_Equation}, the governing system of equations solved by FAL is given as: 

\begin{gather}
\displaystyle 
\begin{bmatrix} \displaystyle \prescript{}{}{\boldsymbol{r}} \\ \\ \prescript{}{}{g} \\ 
\end{bmatrix} = - \begin{bmatrix} \displaystyle  \prescript{}{}{{\frac{\partial \boldsymbol{r}}{\partial x^e}}} & \prescript{}{}{{\dfrac{\partial \boldsymbol{r}}{\partial \lambda}}} \\  & \\ \prescript{}{}{{\dfrac{\partial g}{\partial x^e}}} & \prescript{}{}{{\dfrac{\partial g}{\partial \lambda}}}
\end{bmatrix}
\begin{bmatrix} \displaystyle \prescript{}{}{\delta \boldsymbol{x}} \\ \\ \prescript{}{}{\delta \lambda}
\end{bmatrix} 
\end{gather}

Several implementation schemes and variations of the FAL solver have been developed over the years. The derivation of the FAL solver for local and non-local gradient damage laws, along with the relevant implementation schemes and algorithms which are used in this work are presented in detail in the supplementary material to this manuscript. Even though the FAL method can effectively capture highly non-linear paths it can also face stability issues, for which several algorithmic remedies have been explored \cite{bathe1983automatic,bellini1987improved,yang1990solution,de1999determination,park1982family,simo1986finite,skeie1990local,forde1987improved,de2012nonlinear}. For damage mechanics problems in particular, existing FAL solvers are too demanding from a computational standpoint and often require an excessive number of increments until they converge, especially when sharp snap-backs are involved \cite{crisfield1983arc,carrera1994study,hellweg1998new}. In view of the above challenges, there is a true need to develop an alternative arc-length-based approach that can accurately trace the highly non-linear paths in CDM problems while reducing substantially the computational cost compared to the FAL approach. This provides fruitful ground to our newly proposed unified arc-length method, which is presented in the next subsection.


\subsection{Unified arc-length (UAL)}
\label{Sec:Unified arc-length}

The work by Pretti et al. \cite{pretti2022displacement} introduced a displacement-controlled arc-length framework in the context of geometrically non-linear problems with 1D skeletal elements. By drawing inspiration from this study we expand this framework to continuum damage mechanics problems, but we annotate this framework as the unified arc-length method to emphasize its ability to simultaneously update the displacement and the external force , and other nodal variables such as the non-local strain. The general formulation of the UAL method is presented in this subsection.

\begin{figure}
\centering
\includegraphics[width=0.75\textwidth,trim = 0cm 0.5cm 1cm 2.25cm, clip]{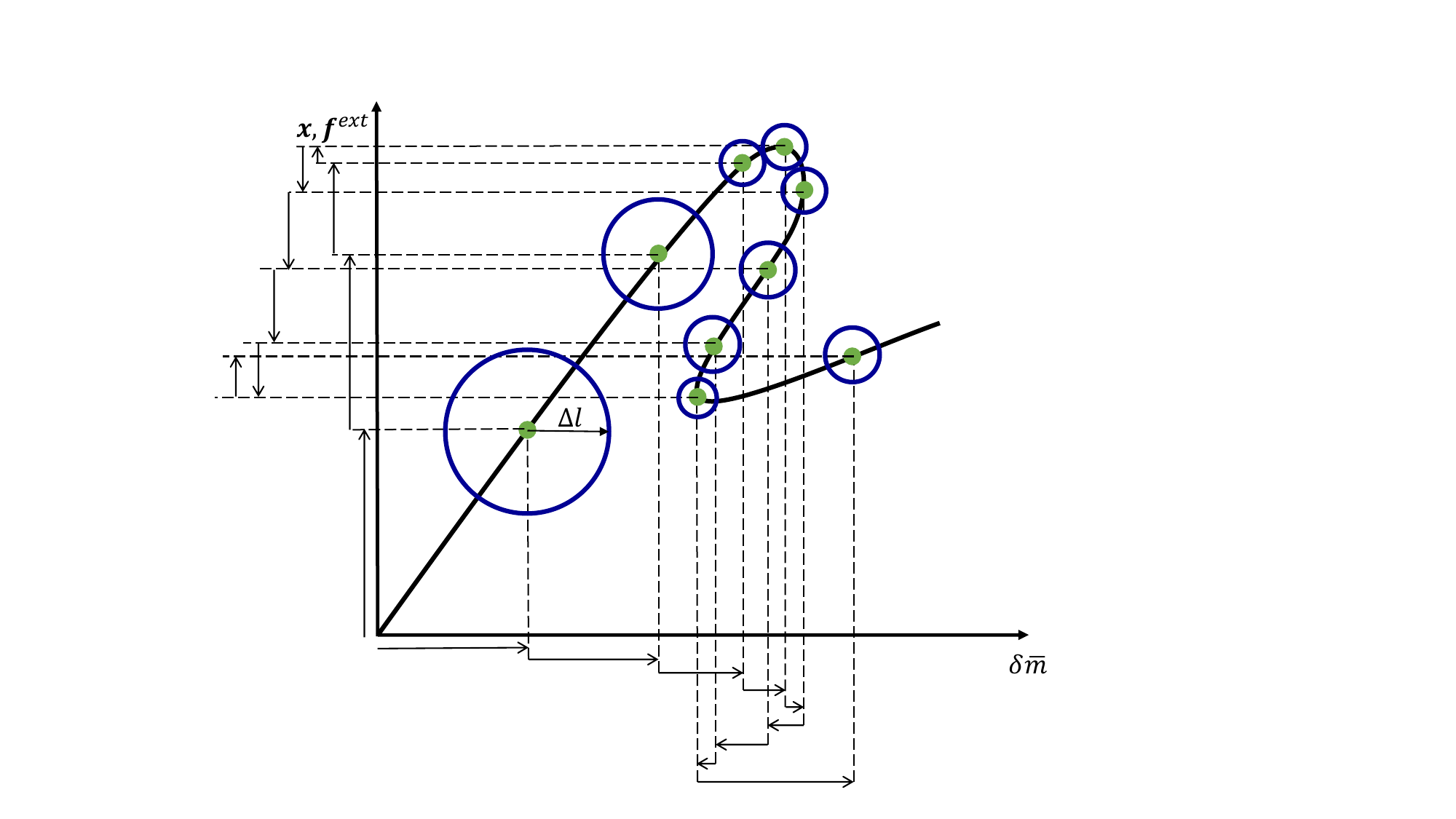}
\caption{Schematic representation of the unified arc-length method, where $\delta \bar m$ is the fraction of the total applied displacement load $\boldsymbol{u}^p$ acting on the system at a given increment.}
\label{fig:UAL_schematic}
\end{figure}

Fig \ref{fig:UAL_schematic} graphically illustrates the equilibrium path of a representative problem solved using UAL. The UAL method targets to minimize the same residual Eqns.\eqref{General_residual_vector} as the NR and FAL method. However, the key difference here is that the entire $\boldsymbol{f}^{ext}$ vector is treated as an independent variable, and different variations are allowed for its entries. To better illustrate the conceptual differences, we begin by partitioning the general residual Eqn. \eqref{General_residual_vector} based on the free and prescribed nodes, denoted by the right superscripts $f$ and $p$ respectively. Note that as the displacement is applied only on the prescribed nodes, $\boldsymbol{f}^{ext}$ is zero at the free nodes. Thus, Eqn. \eqref{General_residual_vector} is expressed in the context of the UAL method as: 

\begin{equation}
{\boldsymbol{r}}={[\boldsymbol{r}^p \ \ \boldsymbol{r}^f]}{^T} \qquad ; \qquad
\prescript{}{}{\boldsymbol{r}^p} = \prescript{}{}{\boldsymbol{f}^{int,p}} - \prescript{}{}{\boldsymbol{f}^{ext,p}} \ ,at \ \Gamma_u \qquad ; \qquad \prescript{}{}{\boldsymbol{r}^f} = \prescript{}{}{\boldsymbol{f}^{int,f}} \  \qquad 
\label{UAL_General_res_eq_partitioned}
\end{equation}

\noindent where, $\prescript{}{}{\boldsymbol{r}}$ is a function of the independent variable vector $\boldsymbol{x}$ which is partitioned as $\boldsymbol{x}={[\boldsymbol{x}^p \ \ \boldsymbol{x}^f]}{^T}$. The displacement vector $\boldsymbol{u}$ is also partitioned as $\boldsymbol{u}={[\boldsymbol{u}^p \ \ \boldsymbol{u}^f]}{^T}$. Consequently, Eqn. \eqref{General_corrector_equation} in the context of the UAL approach becomes: 

\begin{equation}
\prescript{}{}{\boldsymbol{r}^p} = - \prescript{}{}{\boldsymbol{J}^p} \prescript{}{}{\delta \boldsymbol{x}^p} + \prescript{}{}{\delta \boldsymbol{f}^{ext,p}} \qquad ; \qquad  \prescript{}{}{\boldsymbol{r}^f} = - \prescript{}{}{\boldsymbol{J}^f} \prescript{}{}{\delta \boldsymbol{x}^f} 
\label{UAL_residual_equation}
\end{equation}

Note that Eqn. \eqref{UAL_residual_equation} represents $m^p+m^f$ equations and $2 m^p+m^f$ unknowns. To reduce the number of unknowns, the displacement of the prescribed nodes is defined in terms of a known scalar variable $\delta \bar m$. This reduces the number of unknowns to $m^p+m^f+1$. To complete the definition of the UAL model while incorporating $\boldsymbol{f}^{ext}$ as an independent variable, the UAL arc-length equation is introduced as:  

\begin{equation}
\prescript{}{}{{g}} ({\Delta \boldsymbol{x}}, {\Delta \boldsymbol{f}^{ext}}) = \prescript{}{}{[{\Delta \boldsymbol{x}}]^T} \ \prescript{}{}{[{\Delta \boldsymbol{x}}]} + {\beta}^{2} \prescript{}{}{[{\Delta \boldsymbol{f}^{ext}}]^T} \ \prescript{}{}{[{\Delta \boldsymbol{f}^{ext}}]} - \Delta l^2 
\label{UAL_ArcLength_Equation}
\end{equation} 

Eqn. \eqref{UAL_ArcLength_Equation} represents the general UAL arc-length constraint equation and $g$ refers to the residual of this expression. Here it is important to note that the residual $g$ is a scalar variable, which is independent of the dimensions of the independent variables and loading. This number is a sum of scalar terms, since $\Delta l^2$ is a scalar value and the product of a vector with its transpose ($A^{T} \cdot A$) also yields a scalar number. Therefore $g$ is independent of the size of the problem. Also, we note that the term $\boldsymbol{f}^{ext}$ exists only on the nodes with prescribed displacements.

In UAL, the system is moved forward by contributions from both the $\boldsymbol{f}^{ext}$ as well as the other independent variables $\boldsymbol{x}$ as shown in Fig. \ref{fig:UAL_schematic} and in Eqn. \ref{UAL_residual_equation}. By treating the entire $\boldsymbol{f}^{ext}$ vector as an independent variable, the UAL method can account for the variation of all the nodal values of $\boldsymbol{f}^{ext}$. This characteristic is expected to improve its computational performance in damage mechanics problems over the FAL method where a loadfactor $\lambda$ scales the applied load $\boldsymbol{q}$ uniformly across all the nodes to obtain the $\boldsymbol{f}^{ext}$ vector. Following Eqns. \eqref{UAL_residual_equation} and \eqref{UAL_ArcLength_Equation}, the below-mentioned general system of equations is solved by UAL at the end of each iteration; the solution schemes and the implementation of the UAL solver are discussed in the next sections. 

\begin{gather}
\displaystyle 
\begin{bmatrix} \displaystyle \prescript{}{}{\boldsymbol{r}^p} \\ \\ \prescript{}{}{\boldsymbol{r}^f} \\ \\ \prescript{}{}{g} \\ 
\end{bmatrix} = - \begin{bmatrix} \displaystyle  
\prescript{}{}{{\frac{\partial \boldsymbol{r}^p}{\partial x^{p,e}}}} & \displaystyle
\prescript{}{}{{\frac{\partial \boldsymbol{r}^p}{\partial x^{f,e}}}} & \displaystyle  
\prescript{}{}{{\dfrac{\partial \boldsymbol{r}^p}{\partial f^{ext,e}}}} 
\\  & \\ \displaystyle  
\prescript{}{}{{\frac{\partial \boldsymbol{r}^f}{\partial x^{p,e}}}} & \displaystyle  
\prescript{}{}{{\frac{\partial \boldsymbol{r}^f}{\partial x^{f,e}}}} & \displaystyle  
\prescript{}{}{{\dfrac{\partial \boldsymbol{r}^f}{\partial f^{ext,e}}}} 
\\  & \\ 
\prescript{}{}{{\dfrac{\partial g}{\partial x^{p,e}}}} & 
\prescript{}{}{{\dfrac{\partial g}{\partial x^{f,e}}}} & 
\prescript{}{}{{\dfrac{\partial g}{\partial f^{ext,e}}}}
\end{bmatrix}
\begin{bmatrix} \displaystyle 
\prescript{}{}{\delta \boldsymbol{x}^p} \\ \\
\prescript{}{}{\delta \boldsymbol{x}^f} \\ \\ 
\prescript{}{}{\delta \boldsymbol{f}^{ext}}
\end{bmatrix}
\label{UAL_general_tangent_J}
\end{gather}

Finally, we note here that the Jacobian matrix of UAL is non-symmetric, as evidently shown in Eqn. \ref{UAL_general_tangent_J}. Therefore, it is expected to bear an additional expense when used within an iterative solver. However, based on Pretti's work \cite{pretti2022displacement} and the in-built setup of UAL, this theoretical disadvantage is expected to be counteracted by the ability of the method to maintain larger arc-length values throughout the analysis. In fact, this will be demonstrated in detail in Section \ref{Sec:Computational_performance}, where the overall computational cost of UAL will be substantially smaller due to the smaller total number of increments and iterations.

\section{UAL system of equations for damage mechanics problems}
\label{Sec:UAL_system_eqns}

In this section, the mathematical derivation for the UAL system of equations for CDM problems is presented for local and non-local gradient damage. Throughout the paper, the left subscript $i$ and the left superscript $n$ represent the iteration and increment number respectively, and the left superscript $n$ represents the increment number. Note that in the section below, the prescribed displacements are calculated using the equation $\delta \boldsymbol{u}^p = \delta \bar m \ \boldsymbol{u}^p$, where $\boldsymbol{u}^p$ refers to the total applied displacement load, $\delta \boldsymbol{u}^p$ refers to the correction to the prescribed displacement values at a given iteration and $\delta \bar m$ refers to a scalar load fraction of $\boldsymbol{u}^p$ that is applied within an increment. 


\subsection{Local damage}

Based on the discretized weak form Eqn. \eqref{r_u_Local_weak}, partitioning of the residual system of equations in Eqn. \eqref{UAL_General_res_eq_partitioned} and the UAL arc-length Eqn. \eqref{UAL_ArcLength_Equation}, the following system of residual equations for the local damage law is obtained:

\begin{subequations}
\begin{equation}
\prescript{n}{i}{\boldsymbol{r}^p} = \prescript{n}{i}{\left[{\underbrace{\int_\Omega [\boldsymbol{B}^u]^T \boldsymbol{\sigma} d\Omega}_{\boldsymbol{f}^{int,p}} - \underbrace{\int_\Gamma [\boldsymbol{N}^u]^T t d\Gamma}_{\boldsymbol{f}^{ext,p}}}\right]}
\label{Local_Damage_Res1}
\end{equation}

\begin{equation}
\prescript{n}{i}{\boldsymbol{r}^f} = \prescript{n}{i}{\left[{\underbrace{\int_\Omega [\boldsymbol{B}^u]^T \boldsymbol{\sigma} d\Omega}_{\boldsymbol{f}^{int,f}}}\right]}
\label{Local_Damage_Res2}
\end{equation}

\begin{equation}
\prescript{n}{i}{\ {g}} = \prescript{n}{i}{[\Delta \boldsymbol{x}]^T} \ \prescript{n}{i}{[\Delta \boldsymbol{x}]} + {\beta}^{2} \prescript{n}{i}{[{\Delta \boldsymbol{f}^{ext}}]^T} \ \prescript{n}{i}{[{\Delta \boldsymbol{f}^{ext}}]} - \Delta l^2 
\label{Local_Damage_Res3}
\end{equation} 
\label{Local_damage_Res_Eqns}
\end{subequations}

Here we clarify that Eqn.\eqref{Local_Damage_Res3} represents the special case of the general UAL arc-length equation \eqref{UAL_ArcLength_Equation} that is specific to the local damage model, and $\boldsymbol{x}$ therein refers to $\boldsymbol{u}^p$ and $\boldsymbol{u}^f$. The linearized form of Eqns. \eqref{Local_damage_Res_Eqns} expressed as $\boldsymbol{J}\delta \boldsymbol{x}=-\boldsymbol{r}$ is presented below, and the detailed derivation of each entry in the consistent Jacobian matrix $\boldsymbol{J}$ is provided in \ref{Appendix: UAL Jacobian (Local Damage)}. 

\begin{equation}
\resizebox{0.95\hsize}{!}{$
 \begin{bmatrix} {\dfrac{\partial {\boldsymbol{r}^p}}{\partial f^{ext,e}}} & {\dfrac{\partial {\boldsymbol{r}^p}}{\partial u^{f,e}}} & {\dfrac{\partial {\boldsymbol{r}^p}}{\partial \bar m}} \\  & \\ {\dfrac{\partial {\boldsymbol{r}^f}}{\partial f^{ext,e}}} & {\dfrac{\partial {\boldsymbol{r}^f}}{\partial u^{f,e}}} & {\dfrac{\partial {\boldsymbol{r}^f}}{\partial \bar m}} \\  & \\ {\dfrac{\partial g}{\partial f^{ext,e}}} & {\dfrac{\partial g}{\partial u^{f,e}}} & {\dfrac{\partial g}{\partial \bar m}} \end{bmatrix}
\begin{bmatrix} {\delta \boldsymbol{f}^{ext}} \\ \\ {\delta \boldsymbol{u}^f} \\ \\ \delta \bar {m} \end{bmatrix}
 =\underbrace{ \begin{bmatrix} \boldsymbol{I} & \prescript{n}{i}{\left[\boldsymbol{J}^{pf}\right]} & \prescript{n}{i}{\left[{{\boldsymbol{J}^{pp}} \ \boldsymbol{u}^p}\right]} \\  & \\ \boldsymbol{0} & \prescript{n}{i}{\left[\boldsymbol{J}^{ff}\right]} & \prescript{n}{i}{\left[{\boldsymbol{J}^{fp}} \ \boldsymbol{u}^p\right]} \\  & \\  2 {\beta}^{2} {\left[\prescript{n}{i}{\Delta \boldsymbol{f}^{ext}}\right]}^T & 2 {\left[\prescript{n}{i}{\Delta \boldsymbol{u}^f}\right]}^T &  2 {\left[\prescript{n}{i}{\Delta \boldsymbol{u}^p}\right]}^T \ \boldsymbol{u}^p \end{bmatrix}}_{\boldsymbol{J}}
\underbrace{\begin{bmatrix} \prescript{n}{i+1}{\delta \boldsymbol{f}^{ext}} \\ \\ \prescript{n}{i+1}{\delta \boldsymbol{u}^f} \\ \\ \prescript{n}{i+1}{\delta \bar {m}} \end{bmatrix}}_{\boldsymbol{\delta x}}
 =
 - \underbrace{\begin{bmatrix} \prescript{n}{i}{\boldsymbol{r}^p} \\ \\ \prescript{n}{i}{\boldsymbol{r}^f} \\ \\ \prescript{n}{i}{g} \end{bmatrix}}_{\boldsymbol{r}}
 \label{UAL_Local_Jacobian}$}
 \end{equation}


\subsection{Non-local gradient damage}
Based on the discretized weak form Eqn. \eqref{r_u_Nonlocal_weak}, discretized weak form of the non-local strain governing Eqn. \eqref{r_e_Nonlocal_weak}, partitioning of the residual system of equations in Eqn. \eqref{UAL_General_res_eq_partitioned} and the UAL arc-length Eqn. \eqref{UAL_ArcLength_Equation}, the following system of residual equations for the non-local gradient damage law is obtained:

\begin{subequations}
\begin{equation}
\prescript{n}{i}{\boldsymbol{r}^p} = \prescript{n}{i}{\left[{\underbrace{\int_\Omega [\boldsymbol{B}^u]^T \boldsymbol{\sigma} d\Omega}_{\boldsymbol{f}^{int,p}} - \underbrace{\int_\Gamma [\boldsymbol{N}^u]^T t d\Gamma}_{\boldsymbol{f}^{ext,p}}}\right]}
\label{NonLocal_Damage_Res1}
\end{equation}

\begin{equation}
\prescript{n}{i}{\boldsymbol{r}^f} = \prescript{n}{i}{\left[{\underbrace{\int_\Omega [\boldsymbol{B}^u]^T \boldsymbol{\sigma} d\Omega}_{\boldsymbol{f}^{int,f}}}\right]}
\label{NonLocal_Damage_Res2}
\end{equation}

\begin{equation}
\prescript{n}{i}{\boldsymbol{r}^{\bar\epsilon}} = \underbrace{{\prescript{n}{i}{\left[{{\int_\Omega [\boldsymbol{N}^\epsilon]^T \boldsymbol{\bar{\epsilon}} \ d\Omega + \int_\Omega [\boldsymbol{B}^\epsilon]^T \ c \ \nabla \boldsymbol{\bar{\epsilon}} \ d\Omega} -{ \int_\Omega [\boldsymbol{N}^\epsilon]^T \boldsymbol{\epsilon} \ d\Omega }}\right]}}}_{\boldsymbol{f}^{int, \bar\epsilon}}
\label{NonLocal_Damage_Res3}
\end{equation}

\begin{equation}
\prescript{n}{i}{\ g} = \prescript{n}{i}{[\Delta \boldsymbol{x}]^T} \ \prescript{n}{i}{[\Delta \boldsymbol{x}]} + {\beta}^{2} \prescript{n}{i}{[\Delta \boldsymbol{f}^{ext}]^T} \ \prescript{n}{i}{[\Delta \boldsymbol{f}^{ext}]} - \Delta l^2 
\label{NonLocal_Damage_Res4}
\end{equation} 
\label{NonLocal_damage_Res_Eqns}
\end{subequations}

In Eqns. \eqref{NonLocal_damage_Res_Eqns}, $\boldsymbol{r}^{\bar\epsilon}$ refers to the non-local strain residual. Eqn.\eqref{NonLocal_Damage_Res4} represents the UAL arc-length equation specific to the non-local gradient damage, and $\boldsymbol{x}$ therein refers to $\boldsymbol{u}^p$, $\boldsymbol{u}^f$ and $\boldsymbol{\bar \epsilon}$. Though Eqns. \eqref{NonLocal_Damage_Res1} and \eqref{NonLocal_Damage_Res2} look similar to Eqns. \eqref{Local_Damage_Res1} and \eqref{Local_Damage_Res2}, it is important to draw a distinction between them. In Eqns. \eqref{Local_Damage_Res1} and \eqref{Local_Damage_Res2}, the variables damage $d$ and stress $\boldsymbol{\sigma}$ are calculated as a function of local strain values while those in Eqns. \eqref{NonLocal_Damage_Res1} and \eqref{NonLocal_Damage_Res2} are calculated as a function of the non-local strain. The linearized form of Eqns. \eqref{NonLocal_damage_Res_Eqns} expressed as $\boldsymbol{J}\delta \boldsymbol{x}=-\boldsymbol{r}$ is presented below and the detailed derivation for all entries of the consistent Jacobian matrix $\boldsymbol{J}$ is provided in \ref{Appendix: UAL Jacobian (Non-Local Damage)}.

\begin{equation}
\resizebox{0.95\hsize}{!}{$
 \begin{bmatrix} {\dfrac{\partial {\boldsymbol{r}^p}}{\partial f^{ext,e}}} & {\dfrac{\partial {\boldsymbol{r}^p}}{\partial u^{f,e}}} & {\dfrac{\partial {\boldsymbol{r}^p}}{\partial \bar m}} & {\dfrac{\partial {\boldsymbol{r}^p}}{\partial {\bar \epsilon}^e}} \\  & \\ {\dfrac{\partial {\boldsymbol{r}^f}}{\partial f^{ext,e}}} & {\dfrac{\partial {\boldsymbol{r}^f}}{\partial u^{f,e}}} & {\dfrac{\partial {\boldsymbol{r}^f}}{\partial \bar m}} & {\dfrac{\partial {\boldsymbol{r}^f}}{\partial {\bar \epsilon}^e}}\\  & \\ {\dfrac{\partial g}{\partial f^{ext,e}}} & {\dfrac{\partial g}{\partial u^{f,e}}} & {\dfrac{\partial g}{\partial \bar m}} & {\dfrac{\partial g}{\partial {\bar \epsilon}^e}} \\& \\ {\dfrac{\partial {\boldsymbol{r}^{\bar\epsilon}}}{\partial f^{ext,e}}} & {\dfrac{\partial {\boldsymbol{r}^{\bar\epsilon}}}{\partial u^{f,e}}} & {\dfrac{\partial {\boldsymbol{r}^{\bar\epsilon}}}{\partial \bar m}} & {\dfrac{\partial {\boldsymbol{r}^{\bar\epsilon}}}{\partial {\bar \epsilon}^e}}
 \end{bmatrix}
\begin{bmatrix} {\delta \boldsymbol{f}^{ext}} \\ \\ {\delta \boldsymbol{u}^{f}} \\ \\ \delta \bar {m} \\ \\ \delta \boldsymbol{\bar \epsilon} \end{bmatrix}
 =
 \underbrace{\begin{bmatrix}
 \boldsymbol{I} & {\prescript{n}{i}{\boldsymbol{J}^{pf}}} & {\prescript{n}{i}{\boldsymbol{J}^{pp}} \ \boldsymbol{u}^p} & {\prescript{n}{i}{\left[{\boldsymbol{J}^{u \bar\epsilon}}\right]}^p} 
 \\  & \\ 
 \boldsymbol{0} & {\prescript{n}{i}{\boldsymbol{J}^{ff}}} & {\prescript{n}{i}{\left[{\boldsymbol{J}^{fp}}\right]} \ \boldsymbol{u}^p} & {\prescript{n}{i}{\left[{\boldsymbol{J}^{u \bar\epsilon}}\right]}^f}
 \\  & \\ 
 {2 {\beta}^{2} {\left[\prescript{n}{i}{\Delta \boldsymbol{f}^{ext}}\right]}^T} & {2 {\left[\prescript{n}{i}{\Delta \boldsymbol{u}^f}\right]}^T} & {2 {\left[\prescript{n}{i}{\Delta \boldsymbol{u}^p}\right]}^T \boldsymbol{u}^p} & {2 {\prescript{n}{i}{\Delta \boldsymbol{\bar\epsilon}}}^T} 
 \\& \\
 \boldsymbol{0} & {\prescript{n}{i}{\left[{\boldsymbol{J}^{\bar \epsilon u}}\right]}^f} & {\prescript{n}{i}{\left[{\boldsymbol{J}^{\bar \epsilon u}}\right]}^p \boldsymbol{u}^p} & {\prescript{n}{i}{\boldsymbol{J}^{{\bar \epsilon} {\bar \epsilon}}}}
 \end{bmatrix}}_{\boldsymbol{J}}
\underbrace{\begin{bmatrix} \prescript{n+1}{i}{\ \delta \boldsymbol{f}^{ext}} \\ \\ \prescript{n+1}{i}{\ \delta \boldsymbol{u}^f} \\ \\ \prescript{n+1}{i}{\ \delta \bar {m}} \\ \\ \prescript{n+1}{i}{\delta \boldsymbol{\ \bar \epsilon}} \end{bmatrix}}_{\delta \boldsymbol{x}}
=
 - \underbrace{\begin{bmatrix} \prescript{n}{i}{\boldsymbol{r}^p} \\ \\  \prescript{n}{i}{\boldsymbol{r}^f} \\ \\ \prescript{n}{i}{g} \\ \\ \boldsymbol{r}^{\bar\epsilon} \end{bmatrix}}_{\boldsymbol{r}}
 $}
 \label{UAL_NonLocal_Jacobian}
 \end{equation}
 
Eqns. \eqref{UAL_Local_Jacobian} and \eqref{UAL_NonLocal_Jacobian} represent the UAL system of equations that need to be solved to trace the equilibrium path of the problem. The different implementation schemes, solution algorithms, and control parameters used to this end are presented in the next section. 


\section{Implementation Schemes}
\label{Sec:ImplementationSchemes}

In this section, we present the implementation schemes used to solve Eqns. \eqref{UAL_Local_Jacobian} and \eqref{UAL_NonLocal_Jacobian}, the algorithm for implementing the novel UAL method for damage mechanics, and the control parameters used in aiding the stability of the UAL solver. A monolithic inversion of the final system of Eqns. \eqref{UAL_Local_Jacobian} and \eqref{UAL_NonLocal_Jacobian} is challenging due to the large variations between the Jacobian matrix components, especially along the diagonals. This large variation is due to the difference in the magnitude of the terms originating from the stiffness matrix compared to those related to the displacement and non-local strain. To this end, two solution schemes are presented in this paper. They are termed Partitioned Consistent (PC) and Partitioned Non Consistent (PNC), and they follow the work of Pretti et al. \cite{pretti2022displacement} and the earlier works of Crisfield \cite{crisfield1983arc,crisfield1979faster,crisfield1981fast} and Riks \cite{riks1972arclen,riks1979incremental}. Note that the proposed methods are implemented in MATLAB and the matrix inversion operations in the derivations are performed by using mldivide \cite{mldivide} to solve the system of equations.

In subsection  \ref{PC_scheme} and \ref{PNC_scheme} below, the predictor and corrector values of each independent variable for the PC and PNC schemes are presented. The predictor and corrector values at each iteration are related by the following expression for both schemes:

\begin{equation}
\prescript{n}{i}{\Delta (.)} = \prescript{n}{i-1}{\Delta (.)} + \prescript{n}{i}{\delta (.)} 
\label{Imp_Sch_Update_Delta}
\end{equation}

\noindent where $\Delta$ is the predictor value, $\delta$ is the corrector value and $(.) = [\boldsymbol{f}^{ext} \ \boldsymbol{x}]^T$. This iterative update of the predictor values continues until convergence is reached within an increment. If convergence is not achieved within the maximum allowed number of iterations, the arc-length is updated and the predictors are reset to their last converged values. Below, the PC and PNC schemes are discussed.


\subsection{Partitioned Consistent scheme (PC)}
\label{PC_scheme}


\subsubsection{Local Damage}
\label{PC_Local}
PC adapted to local damage is the first scheme used in this work to implement the UAL method, and it is used to solve the system of Eqns. \eqref{Local_damage_Res_Eqns} as described below: 

$\underline{\textbf{Predictor Values:}}$ At the start of the analysis ($n=1$), $\Delta \bar m$ is set to a scalar positive user-defined value $0<\alpha<1$, and the residuals $\boldsymbol{r}^p$, $\boldsymbol{r}^f$ and $g$ are initialized to zero in Eqn. \eqref{UAL_Local_Jacobian}. From the first two rows of Eqn. \eqref{UAL_Local_Jacobian} the following expressions for the predictor values of the independent variables are obtained: 

\begin{equation}
    \prescript{1}{1}{\Delta \bar m} = \alpha    
    \qquad ; \qquad    
    \prescript{1}{1}{{\boldsymbol{J}^{ff}}} \prescript{1}{1}{\Delta \boldsymbol{u}^f}
    = -  \prescript{1}{1}{\boldsymbol{J}^{fp}} \ \boldsymbol{u}^p \ \alpha    
    \qquad ; \qquad    
    \prescript{1}{1}{\Delta \boldsymbol{f}^{ext}} 
    = - \left[ \prescript{1}{1}{\boldsymbol{J}^{pp}} \ u^p + \prescript{1}{1}{\boldsymbol{J}^{pf}}  \  \prescript{1}{1}{{\Delta \boldsymbol{u}^{f,B}}}   \right] \ \alpha       
\label{PC_Delta_n1}
\end{equation}

where, $\prescript{1}{1}{{\boldsymbol{J}^{ff}}} \ \prescript{1}{1}{{\Delta \boldsymbol{u}^{f,B}}} = - \prescript{1}{1}{\boldsymbol{J}^{fp}} \ {\boldsymbol{u}^p}$
For all other increments ($n>1$), the last converged predictor value $\prescript{n-1}{}{\Delta (.)}$ is used such that $\prescript{n}{1}{\Delta} (.) = \prescript{n-1}{}{\Delta} (.)$. The predictor values are updated at each iteration using Eqn. \eqref{Imp_Sch_Update_Delta}.

$\underline{\textbf{Corrector Values:}}$ The corrector values are calculated from Eqn. \eqref{UAL_Local_Jacobian} at each iteration until convergence. It follows from the second row of Eqn.\eqref{UAL_Local_Jacobian} that $\delta \boldsymbol{u}^f$ can be expressed as:
\begin{subequations}
\begin{equation}
\prescript{n}{i+1}{\delta \boldsymbol{u}^f} = \prescript{n}{i}{{\delta \boldsymbol{u}^{f,A}}}+ \prescript{n}{i}{{\delta \boldsymbol{u}^{f,B}}} \ \prescript{n}{i+1}{\delta \bar {m}}
\label{PC_del_u^f}
\end{equation} 

\noindent where, $\prescript{n}{i}{{\boldsymbol{J}^{ff}}} \ \prescript{n}{i}{{\delta \boldsymbol{u}^{f,A}}} = -  \prescript{n}{i}{\boldsymbol{r}^f}$ and $ \prescript{n}{i}{{\boldsymbol{J}^{ff}}} \ \prescript{n}{i}{{\delta \boldsymbol{u}^{f,B}}} = - \prescript{n}{i}{\boldsymbol{J}^{fp}} \ {\boldsymbol{u}^p}$. This expression of $\delta \boldsymbol{u}^f$ is used in conjunction with the first row of  Eqn.\eqref{UAL_Local_Jacobian} to express $\delta \boldsymbol{f}^{ext}$ as: 
\begin{equation}
\prescript{n}{i+1}{\delta \boldsymbol{f}^{ext}} = \prescript{n}{i}{\delta \boldsymbol{f}^{ext,A}} + \prescript{n}{i}{\delta \boldsymbol{f}^{ext,B}} \prescript{n}{i+1}{\delta \bar m}
\label{PC_del_f_ext}
\end{equation}  

\noindent where,  $\prescript{n}{i}{\delta \boldsymbol{f}^{ext,A}} =  - \left[ \prescript{n}{i}{\boldsymbol{r}^p} +  \prescript{n}{i}{\boldsymbol{J}^{pf}} \prescript{n}{i}{{\delta \boldsymbol{u}^{f,A}}}\right]$ and $\prescript{n}{i}{\delta \boldsymbol{f}^{ext,B}} = - \left[ \prescript{n}{i}{\boldsymbol{J}^{pp}} \ \boldsymbol{u}^p + \prescript{n}{i}{\boldsymbol{J}^{pf}}  \  \prescript{n}{i}{{\delta \boldsymbol{u}^{f,B}}}   \right]$. Finally, the $\delta \bar m$ value is computed using Eqn.\eqref{PC_del_u^f}, Eqn.\eqref{PC_del_f_ext} along with the last row of Eqn. \eqref{UAL_Local_Jacobian} as: 

\begin{equation}
\displaystyle
\prescript{n}{i+1}{\delta \bar m} = \frac{{\prescript{n}{i}{\ g}} + \prescript{n}{i}{\left({\frac{\partial g}{\partial \boldsymbol{u}^f}}\right)} \prescript{n}{i}{{\delta \boldsymbol{u}^{f,A}}} + \prescript{n}{i}{\left({\frac{\partial g}{\partial \boldsymbol{f}^{ext}}}\right)} {\prescript{n}{i}{\delta \boldsymbol{f}^{ext,A}}}}
{ \prescript{n}{i}{\left({\frac{\partial g}{\partial \boldsymbol{u}^p}}\right)} \boldsymbol{u}^p + \prescript{n}{i}{\left({\frac{\partial g}{\partial \boldsymbol{u}^f}}\right)} \prescript{n}{i}{{\delta \boldsymbol{u}^{f,B}}} + \prescript{n}{i}{\left({\frac{\partial g}{\partial \boldsymbol{f}^{ext}}}\right)}  \prescript{n}{i}{\delta \boldsymbol{f}^{ext,B}} }
\label{PC_del_bar_m}
\end{equation}
\label{PC_Local_correctors}
\end{subequations}

 This iterative process is continued until $\boldsymbol{u}^p$ is applied to the system completely. The implementation algorithm for this scheme is presented in Section \ref{Implementation_Algorithm}.


\subsubsection{Non-local gradient damage}
\label{PC_Nonlocal}
PC adapted to the non-local gradient damage law is the second scheme used in this work to implement the UAL method. It is used to solve the system of Eqns.\eqref{NonLocal_damage_Res_Eqns} as described below: 

$\underline{\textbf{Predictor Values:}}$ At the start of the analysis ($n=1$), $\Delta \bar m$ is set to a scalar positive user defined value $\alpha$, and the residuals $\boldsymbol{r}^p$, $\boldsymbol{r}^f$, $\boldsymbol{r}^{\bar \epsilon}$ and $g$ are set to zero in Eqn. \eqref{UAL_NonLocal_Jacobian}. Following the first, second and fourth row of Eqn. \eqref{UAL_NonLocal_Jacobian} the following expressions for the predictor values of the independent variables are obtained: 

\begin{equation}
\prescript{1}{1}{\Delta \boldsymbol{\bar \epsilon}} = \boldsymbol{B} \ \alpha \qquad;\qquad \prescript{1}{1}{\Delta \boldsymbol{u}^f} = \boldsymbol{D} \ \alpha \qquad;\qquad \prescript{1}{1}{\Delta \boldsymbol{f}^{ext}} = \boldsymbol{F} \ \alpha
\label{PC_Delta_n2}
\end{equation} 

\noindent For all other increments ($n>1$), the last converged predictor value $\prescript{n-1}{}{\Delta (.)}$ is used such that $\prescript{n}{1}{\Delta} (.) = \prescript{n-1}{}{\Delta} (.)$. The predictor values are updated at each iteration using Eqn. \eqref{Imp_Sch_Update_Delta}.

$\underline{\textbf{Corrector Values:}}$ Now, the expressions for the corrector values of the independent variables for the PC non-local gradient damage scheme are presented. It follows from the first, second and fourth row of Eqn.\eqref{UAL_NonLocal_Jacobian} that $\delta \bar {\boldsymbol{\epsilon}}$, $\delta \boldsymbol{u}^f$ and $\delta \boldsymbol{f}^{ext}$ can be expressed as:

\begin{subequations}
\begin{equation}
\prescript{n}{i+1}{\delta \boldsymbol{\bar \epsilon}} = \boldsymbol{A} \ + \boldsymbol{B} \ \delta \bar m \qquad;\qquad \prescript{n}{i+1}{\delta \boldsymbol{u}^f} = \boldsymbol{C} \ + \boldsymbol{D} \ \delta \bar m \qquad;\qquad \prescript{n}{i+1}{\delta \boldsymbol{f}^{ext}} = \boldsymbol{E} \ + \boldsymbol{F} \ \delta \bar m
\label{PC_NL_del_bar_epsilon}
\end{equation} 

The values of $\boldsymbol{A}$ to $\boldsymbol{F}$ are calculated at the start of each iteration. This is followed by calculating the $\delta \bar m$ value from the third row of Eqn. \eqref{UAL_NonLocal_Jacobian} as:
\begin{equation}
\boldsymbol{G} \ \prescript{n}{i+1}{\delta \bar m} = \boldsymbol{H}
\label{PC_NL_delta_m_bar}
\end{equation} 
\label{PC_NL_correctors}
\end{subequations}

The definition of $\boldsymbol{A}$ to $\boldsymbol{H}$ are given in  \ref{Appendix:Coeff_PC_gradient}. The value of $\delta \bar m$ obtained from Eqn. \eqref{PC_NL_delta_m_bar} is used to calculate the remaining corrector values using Eqns. \eqref{PC_NL_del_bar_epsilon}. This iterative process is continued until $
\boldsymbol{u}^p$ is applied to the system completely. The implementation algorithm for this scheme is presented in Section \ref{Implementation_Algorithm}.


\subsection{Partitioned Non-Consistent scheme}
\label{PNC_scheme}
 The PNC scheme is based on the earlier work of Crisfield \cite{crisfield1981fast} for problems involving sharp snap-backs. This algorithm has been utilized in several other studies in the literature where it has showcased its effectiveness \cite{schweizerhof1986consistent, watson1983quadratic, krishnamoorthy1996post} and it is adopted here as an alternative option to the PC scheme, in case the latter faces challenges in capturing the post-peak response of the domain.
 

\subsubsection{Local Damage}
\label{PNC_Local_Damage}

$\underline{\textbf{Predictor Values:}}$ In PNC, the predictor values at the first iteration ($n = 1$) are calculated using Eqns. \eqref{PC_Delta_n1} and for $n > 1$ using the last converged values of $\Delta (.)$ such that $\prescript{n}{1}{\Delta} (.) = \prescript{n-1}{}{\Delta} (.)$. The predictor values are updated at each iteration using Eqn. \eqref{Imp_Sch_Update_Delta}. 

$\underline{\textbf{Corrector Values:}}$ In each increment, for all iterations $i > 1$, the  corrector values are obtained by substituting the expression Eqn. \eqref{Imp_Sch_Update_Delta} in Eqn. \eqref{UAL_ArcLength_Equation}. Following this, the quadratic equation presented below is obtained: 

\begin{equation}
a \ \prescript{n}{i+1}{{\delta \bar m}^2} + b \ \prescript{n}{i+1}{{\delta \bar m}} + c = 0
\label{PNC_Quadratic_Eq}
\end{equation} 

The definition of $a$, $b$ and $c$ are given in \ref{Appendix:Coeff_PNC_local}. Once roots of the quadratic Eqn.\eqref{PNC_Quadratic_Eq} are found, the value of $\delta \bar m$ is chosen such that it moves the system forward to the next point on the equilibrium path based on the cosine rule described in \cite{crisfield1983arc}. This iterative process is continued until $\boldsymbol{u}^p$ is applied to the system completely. The algorithm in Section \ref{Implementation_Algorithm} is used to implement this scheme.
 
The PNC scheme for non-local gradient damage is not presented here as the PC scheme effectively captured the equilibrium path for all the UAL problems modeled using the non-local gradient damage law as presented in Section \ref{Sec:NumericalExamples}. Nonetheless, PNC scheme for non-local gradient damage can be derived by following the same approach presented above.

\subsection{Control Parameters}
\label{Control_parameters}
The performance of the UAL solver is influenced by several control parameters, the main objective of which are to aid the UAL solver in overcoming the critical points in highly non-linear problems. Below we present these control parameters, and a numerical investigation of their impact is reported in Section \ref{NumEx:SNT}. 


\subsubsection{Strain-Tolerance (ST)}
The Clausius-Duhem inequality is enforced to ensure positive damage growth, following the condition mentioned in Eqn. \eqref{Kuhn-Tucker_conditions}. In some cases, the difference between the equivalent strain at the current increment $\prescript{n}{i}{\varepsilon^{*}_{eq}}$ and that of the last converged increment $\prescript{n-1}{}{\varepsilon^{*}_{eq}}$ is sufficiently small that the introduction of the Clausius-Duhem condition will lead to unnecessary discontinuities in the flow of the algorithm and interrupt convergence. This difference is mathematically expressed as $\epsilon_{diff} = \prescript{n}{i}{\varepsilon^{*}_{eq}} - \prescript{n-1}{}{\varepsilon^{*}_{eq}}$. Under these conditions, it is not practical to enforce the Clausius-Duhem inequality directly. Instead, a strain tolerance (ST) is defined such that:

\begin{equation}
\{\prescript{n}{i}{d}\}=\begin{cases}
    \prescript{n}{i}{d} , & \text{when $\epsilon_{diff}> ST$}.\\
    \prescript{n-1}{}{d} , & \text{when $\epsilon_{diff}< ST$}. \qquad \qquad
  \end{cases}
 \left\{\prescript{n}{i}{\frac{\partial{d}}{\partial{\varepsilon^{*}_{eq}}}}\right\}=\begin{cases}
    \prescript{n}{i}{\frac{\partial{d}}{\partial{\varepsilon^{*}_{eq}}}} , & \text{when $\epsilon_{diff}> ST$}.\\
    0 , & \text{when $\epsilon_{diff}< ST$}.
  \end{cases}
  \label{ST_cases}
\end{equation}

Eqn. \eqref{ST_cases} describes the manner in which the Clausius-Duhem inequality is enforced in the presented UAL framework. When the $\epsilon_{diff} > ST$, the damage and its derivatives are assigned the corresponding values calculated at the current iteration. However, when  $\epsilon_{diff} < ST$, damage is reset to its last converged value from the previous increment, and its derivative is set to zero. This approach prevents the solver from getting stuck at the critical points where the step sizes continue to decrease until the termination condition (i.e. smallest allowable arc-length value) is reached without convergence. The relevance of ST in the UAL model is presented in the Algorithm \ref{alg:two}. 


\subsubsection{Arc-length Limits ($\Delta l_{min}$,$\Delta l_{max}$)}

\begin{figure}
\centering
\includegraphics[width=0.7\textwidth,trim = 2.5cm 0.5cm 2cm 0cm, clip]{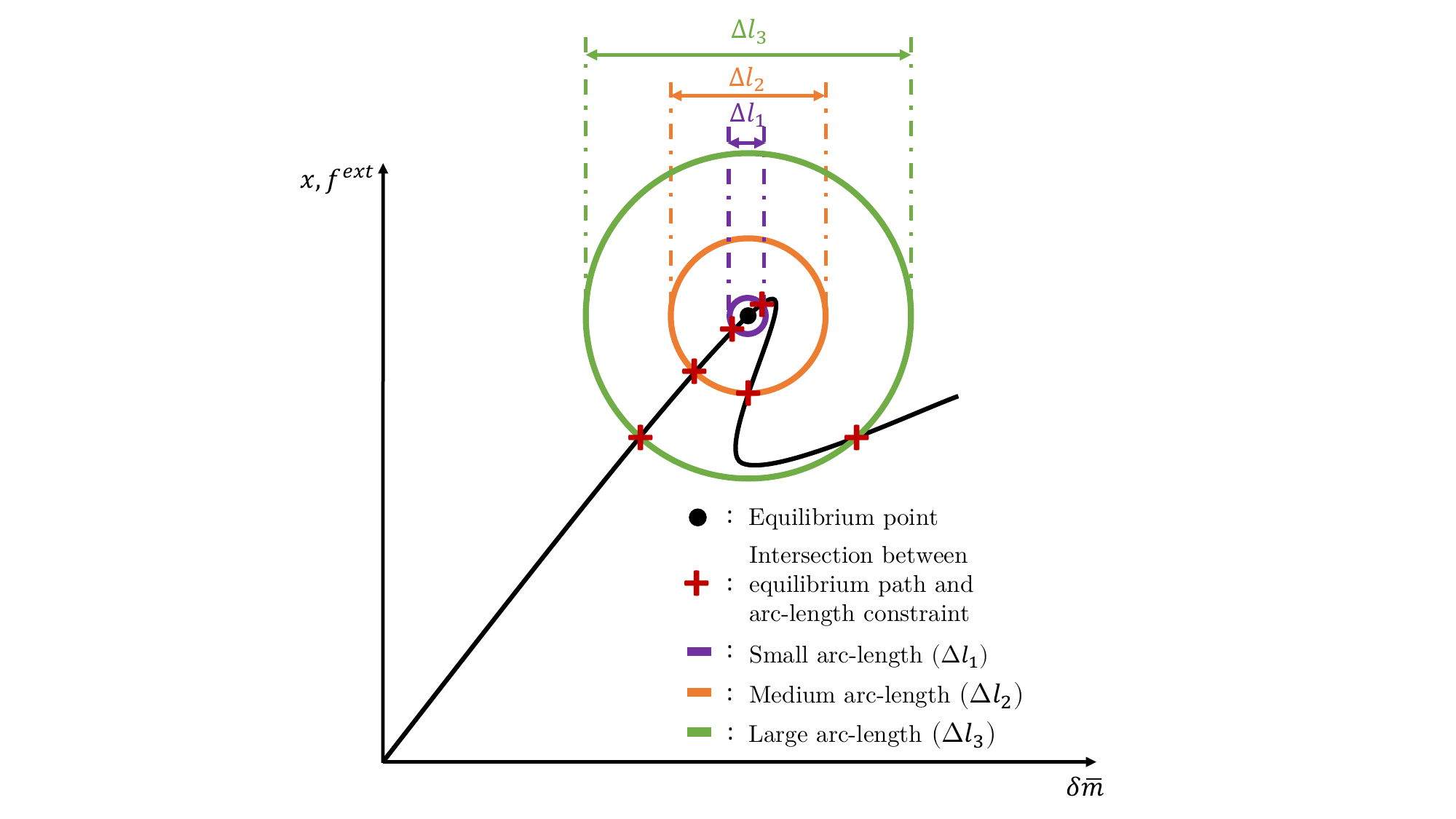}
\caption{Schematic representation of the influence of different arc-length values on the UAL performance.}
\label{fig:AL_Schematic}
\end{figure}

The arc-length value at a given increment impacts the equilibrium path traced. Fig. \ref{fig:AL_Schematic} is a schematic representation of the influence of three different arc-length values namely $\Delta l_1, \Delta l_2$ and $\Delta l_3$, in order of increasing magnitude respectively. Choosing $\Delta l_3$ could lead to the system bypassing one or more critical points on the equilibrium path, while $\Delta l_1$ could lead to issues like backtracking in problems with very sharp snap-backs. As the numerical problems discussed in this paper are path-dependent, missing even one critical point can lead to divergence or tracing an incorrect equilibrium curve \cite{zienkiewicz2005finite}. Thus, a lower limit ($\Delta l_{min}$) and upper limit ($\Delta l_{max}$) of arc-length are defined to aid in the stability of the solver. The value of $\Delta l_{min}$ is set to zero and the initial value of arc-length ($\Delta l_0$) is set to $\Delta l_{max}$ for all the numerical examples presented in this work. Inspired by the built-in adaptive time stepping utilized in \cite{taylor2014feap} and adapted in CDM models such as \cite{mobasher2017non} and \cite{mobasher2021dual}, the arc-length $\Delta l$ is updated as:

\begin{equation}
\{\prescript{n+1}{}{\Delta l}\}=\begin{cases}
    min(\Delta l_{max}, 10^{log_{10} (\prescript{n}{}{\Delta l}) + 0.2}) , & \text{when} \ i<i_{min}\\
    max(\Delta l_{min}, 10^{log_{10} (\prescript{n}{}{\Delta l}) - 0.2}) , & \text{when} \ i>i_{max}
  \end{cases}
  \qquad ; 
  \label{AL_limit_cases}
\end{equation}

Here, $i_{min}$ and $i_{max}$ represent the lower and upper limit of the $ideal$ number of iterations for an increment. When the number of iterations is outside this range, the arc-length is updated. Another term $i_{total}$ is introduced and it represents the maximum allowable number of iterations within an increment. In the numerical examples presented in this work, $i_{min}$=5, $i_{max}$=12 and $i_{total}$=30. Finally, the term $tol$ in Algorithm \ref{alg:one} refers to the convergence tolerance chosen for the analysis.

\subsubsection{Maximum Damage ($d_{max}$)}

The maximum allowable value of damage within an element is another control parameter that affects the response of the UAL solver. The influence of this parameter has also been investigated in previous works by Mobasher et al. \cite{mobasher2017non,mobasher2021dual} and Londono et al. \cite{londono2016prony}, to aid in numerical stability. Damage within an element is typically not allowed to be exactly 1, as complete loss of the element stiffness leads to singularity issues. Usually an upper bound for $d_{max}$ is established, which is close to but not equal to unity, and therefore the impact of the residual stiffness in the damaged element on the UAL solver remains to be investigated. Similar to the ST control variable, $d_{max}$ is also introduced in the Algorithm \ref{alg:two} below and its influence is studied in Section \ref{NumEx:SNT}.

\newpage
\subsection{Implementation Algorithm}
\label{Implementation_Algorithm}
Both the PC and PNC schemes mentioned above are implemented using the following algorithm:

\begin{algorithm}[H]
\caption{Implementation algorithm for PC and PNC schemes}\label{alg:one}
\linespread{0.5}\selectfont
Initialize Inputs \; 
Initialize control parameters ST, $\Delta l_{min}$, $\Delta l_{max}$ and $d_{max}$ \;
Initialize n = 1 \;
\While{$\delta \bar m < 1$}{
Set $i = 0$ \;
Set ${\lVert \boldsymbol{r} \rVert}_2$ = 2 x $tol$ \;
Calculate $\beta$ \;
\While{${\lVert \boldsymbol{r} \rVert}_2$ $>$ $tol$ and $i<i_{max}$}
{Update $i = i+1$ \;
Calculating Jacobian matrix and damage (Algorithm \ref{alg:two}) \; 
\eIf {i==1}{
\eIf {n==1}{
Calculate $\prescript{1}{1}{\Delta (.)}$ predictor values, \ Eqn. \eqref{PC_Delta_n1} / \eqref{PC_Delta_n2} ;
}{Recover last converged $\prescript{n}{1}{\Delta(.)}=\prescript{n-1}{}{\Delta(.)}$ and $\prescript{n}{1}{(.)}=\prescript{n-1}{}{(.)}$}
}{Calculate $\prescript{n}{i}{\delta(.)}$, Eqns. \eqref{PC_Local_correctors}/ \eqref{PC_NL_correctors} and Eqn. \eqref{PNC_correctors}\;
Update $\prescript{n}{i}{\Delta(.)}$, Eqn. \eqref{Imp_Sch_Update_Delta}}
Update variables $\prescript{n}{i}{(.)}$ = $\prescript{n-1}{}{(.)}$ + $\prescript{n}{i}{\Delta(.)}$ \;
Calculating Jacobian matrix and damage (Algorithm \ref{alg:two})\; 
Calculate ${\lVert \boldsymbol{r} \rVert}_2$ \;
}
\eIf{${\lVert \boldsymbol{r} \rVert}_2$ $<$ tol and $i<i_{total}$}
{Update $n = n+1$ \;Save converged $\prescript{n}{i}{\Delta(.)}$ and $\prescript{n}{i}{(.)}$ \; Update arc-length, Eqn. \eqref{AL_limit_cases}}{Update arc-length, Eqn. \eqref{AL_limit_cases}}
}
\end{algorithm}

\begin{algorithm}[H]
\caption{Calculating Jacobian matrix and damage}\label{alg:two}
\linespread{1}\selectfont
\For{each finite element}
{
\For{each material point}{
Calculate ${\varepsilon^{*}_{eq}}$ \;
Calculate $\epsilon_{diff}$ \;
\eIf {$\epsilon_{diff}>$ST}{$\prescript{n}{i}{d}=min(\prescript{n}{i}{d},d_{max})$ \\
$\prescript{n}{i}{\frac{\partial{d}}{\partial{\varepsilon^{*}_{eq}}}} = \prescript{n}{i}{\frac{\partial{d}}{\partial{\boldsymbol{\varepsilon}^{*}_{eq}}}}$}{$\prescript{n}{i}{d}=\prescript{n-1}{}{d}$ \\
$\prescript{n}{i}{\frac{\partial{d}}{\partial{{\varepsilon}^{*}_{eq}}}} = 0$} 
Calculate $\boldsymbol{\sigma}$ \;
}
Calculate Jacobian matrix $\boldsymbol{J}$ \;
Calculate residual vector $\boldsymbol{r}$ \;
}
Assemble Jacobian matrix $\boldsymbol{J}$\; 
Assemble residual vector $\boldsymbol{r}$\;
Partition Jacobian matrix $\boldsymbol{J}$\;
\end{algorithm}



\section{Numerical Examples}
\label{Sec:NumericalExamples}

In this section, the performance of the new UAL approach is compared against the NR and FAL methods through a series of benchmark numerical examples. These include a 1D bar under tension and several 2D geometries under various loading conditions: Single Notch Tension (SNT), Symmetric Single Notch Tension (SSNT), Two Notch Tension (TNT), and Single Notch Shear (SNS). In all problems we report the geometry, material properties, control parameters, implementation schemes, convergence criteria, force-displacement curves and damage contours. In all cases the Mazars damage model is implemented, which is detailed in \ref{Appendix: Mazar_Damage_model}. A cylindrilcal arc-length constraint, i.e. $\beta=0$, is used in all analyses. All the models deploy either linear 2-noded elements in 1D or bilinear 4-noded elements in 2D. All problems follow a quasi-static displacement-driven loading setup, which is commonly used in the literature \cite{singh2016fracture,chen2021phase,bharali2022robust}.


\subsection{1D bar under tension}
\label{1D_bar_problem}

\begin{figure}[H]
    \centering
    \begin{minipage}{15cm}
    \centering     
    \includegraphics[width=1\textwidth,trim = 3.5cm 8cm 2cm 7cm, clip]{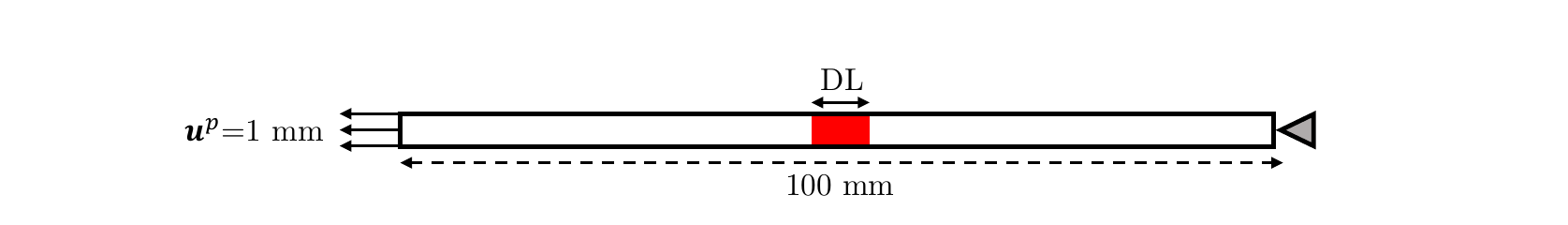}
    \end{minipage}
    \caption{Geometry and boundary conditions for the 1D bar problem. DL refers to the damaged length of the domain.}
    \label{fig:1D_bar_schematic}
\end{figure}

The first numerical experiment is a benchmark 1D bar problem which is studied to demonstrate the efficiency of the UAL method. Several researchers like \cite{nguyen2018smoothing,poh2017localizing} have attempted similar problems. The bar shown in Fig.\ref{fig:1D_bar_schematic} has a length of 100 mm, unit cross-sectional area (A=1 $mm^2$), is fixed at one end, and is subject to a prescribed displacement load of 0.01 mm at the other end. Two mesh sizes with two-node linear shape function elements are presented in this example. In the local damage law problems, 51 elements (Coarse) and 101 elements (Fine) are used while the non-local gradient damage law problems use 51 elements (Coarse) and 151 elements (Fine) respectively. The elastic modulus E = 30 GPa, and the Mazars damage model parameters are damage threshold strain $\epsilon_D$ = $10^{-4}$, $\mathscr{A}$ = 0.7 and $\mathscr{B}$ = $10^4$. 

A damaged length (DL) of 4 mm is defined at the middle of the bar to initiate strain localization. This is achieved by reducing the Young's modulus of elements in the DL region using a parameter $\phi$ such that $E_{DL}=\phi \times E$. The magnitude of $\phi$ is tuned accordingly in order to simulate different response scenarios and control the sharpness of the post-peak curve. Also, the convergence tolerance ($tol$) for the NR and UAL model is kept at $10^{-6}$. The $tol$ for the FAL model for the local damage law is set at $10^{-6}$ but is set at $10^{-4}$ for the gradient damage law since the FAL was not able to converge at smaller $tol$ values in that case. All the 1D examples are modeled using the PC scheme. 

\subsubsection{Non-local gradient damage}
\label{NumExpl:1D_NonLocal_Damage}

\begin{figure}[H]
    \centering
    \begin{minipage}{8.2cm}
    \centering     
    \includegraphics[width=1\textwidth,trim = 1cm 7cm 2cm 7cm, clip]{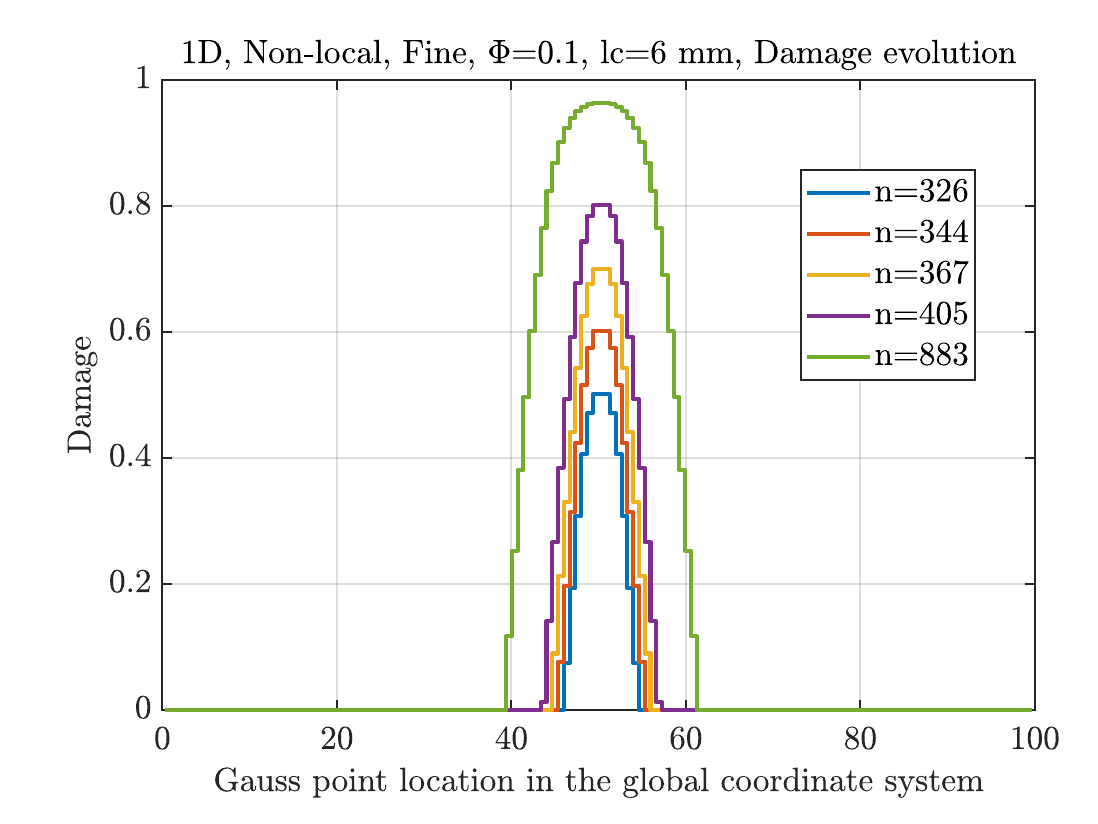}
    \text{(a) Damage evolution} 
    \end{minipage}
    \hfill
    \begin{minipage}{8.2cm}
    \centering     
    \includegraphics[width=1\textwidth,trim = 1cm 7cm 2cm 7cm, clip]{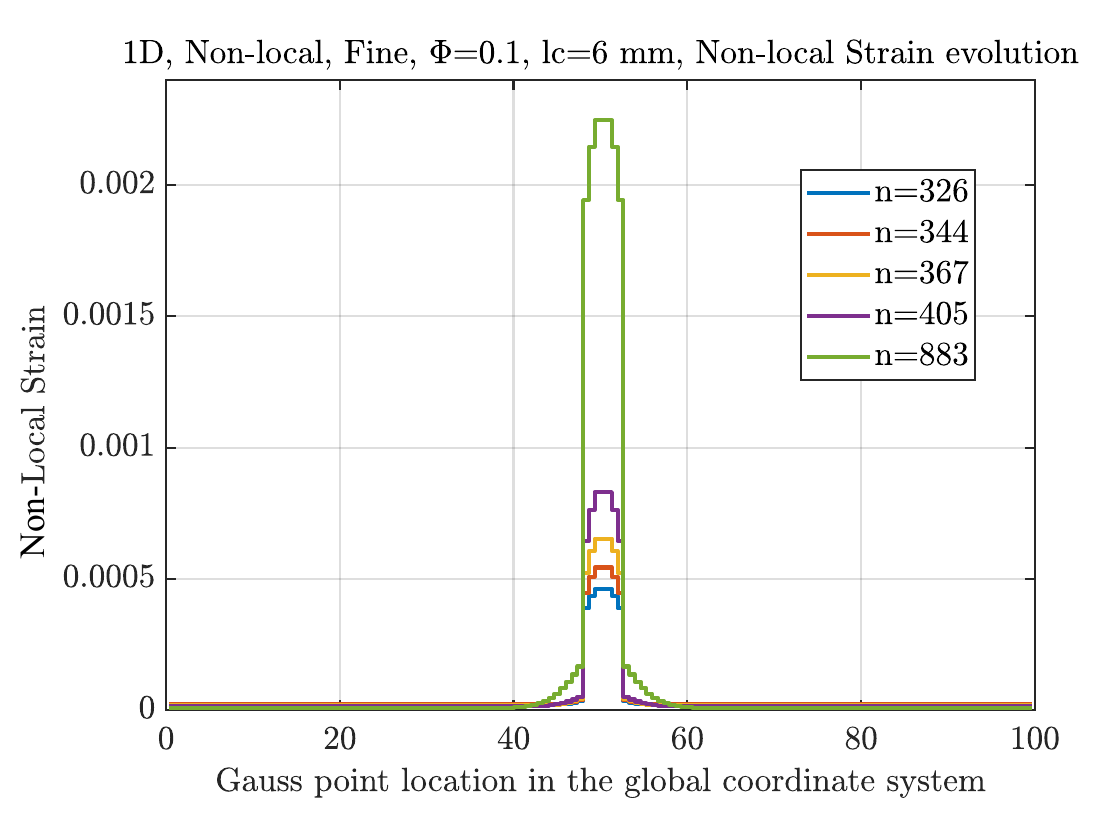}
    \text{(b) Non-local strain evolution}   
    \end{minipage}
    \caption{Evolution of (a) damage and (b) non-local strain in the 1D bar problem with a fine mesh and $\phi = 0.1$, $DL = 4 mm$ and $l_{c} = 6mm$ at different load increments $n$, solved using the UAL method. Both quantities are diffused over a progressively larger region as the applied load increases.}
    \label{fig:1D_gradient_d_and_strain_evolution}
\end{figure}

\begin{figure}
    \centering
    \begin{minipage}{10cm}
    \centering     
    \includegraphics[width=\textwidth,trim =0cm 7.5cm 0cm 7.5cm, clip]{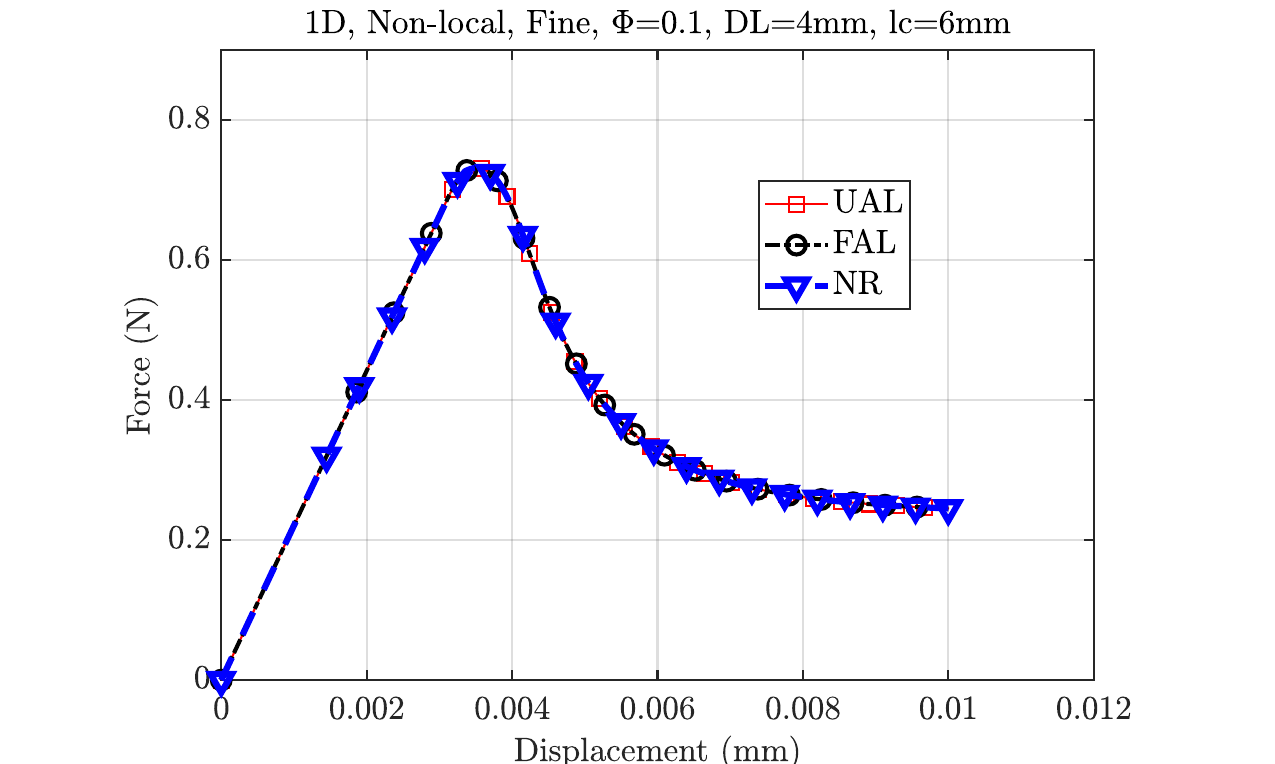}
    \end{minipage}
    \caption{Reaction-displacement curves for the 1D bar shown in Fig.\ref{fig:1D_bar_schematic}, solved using three different non-linear solvers for the non-local gradient damage law with a fine mesh, $\phi=0.1$, $DL = 4mm$ and $l_{c} = 6mm$. This 1D non-local crossplot without a snap-back region is used to verify the performance of the UAL method against existing solvers. For visual clarity, only 1\% of the converged increments are presented in the figure.}
    \label{fig:1D_verification_problem}
\end{figure}

First, the performance of the UAL solver is compared to the NR and FAL solver via a verification problem without a snap-back region. Here, a non-local gradient damage law problem with a fine mesh is used with a $\phi$ = 0.1 and $l_{c} = 6$ mm. Fig. \ref{fig:1D_gradient_d_and_strain_evolution} displays the evolution of damage and non-local strain in the verification problem using the UAL method. With an increase in load, both the damage and non-local strain are distributed over a larger part of the domain, which is consistent with non-local gradient damage model results.    As seen in Fig. \ref{fig:1D_verification_problem}, all three solvers trace an identical path for the entire force-displacement curve. This verifies that the proposed UAL framework captures the global response of the bar with the same level of accuracy as the conventional methods. However, as reported in Table \ref{Table:1D_Gradient_computational_table}, the number of increments and the simulation time taken are significantly different, and they are all in favor of UAL. In particular, UAL takes the least number of increments and takes the least computational time across all methods, which is the first evidence of the proposed method's efficiency. Having therefore showcased that UAL operates at least at the same accuracy level as NR and FAL, we proceed with a more detailed investigation of various test cases to establish the computational superiority of UAL against the other two conventional frameworks. 

\begin{figure}[H]
    \centering
    \begin{minipage}{8.2cm}
    \centering     
    \includegraphics[width=1\textwidth,trim = 1.5cm 7cm 2cm 7cm, clip]{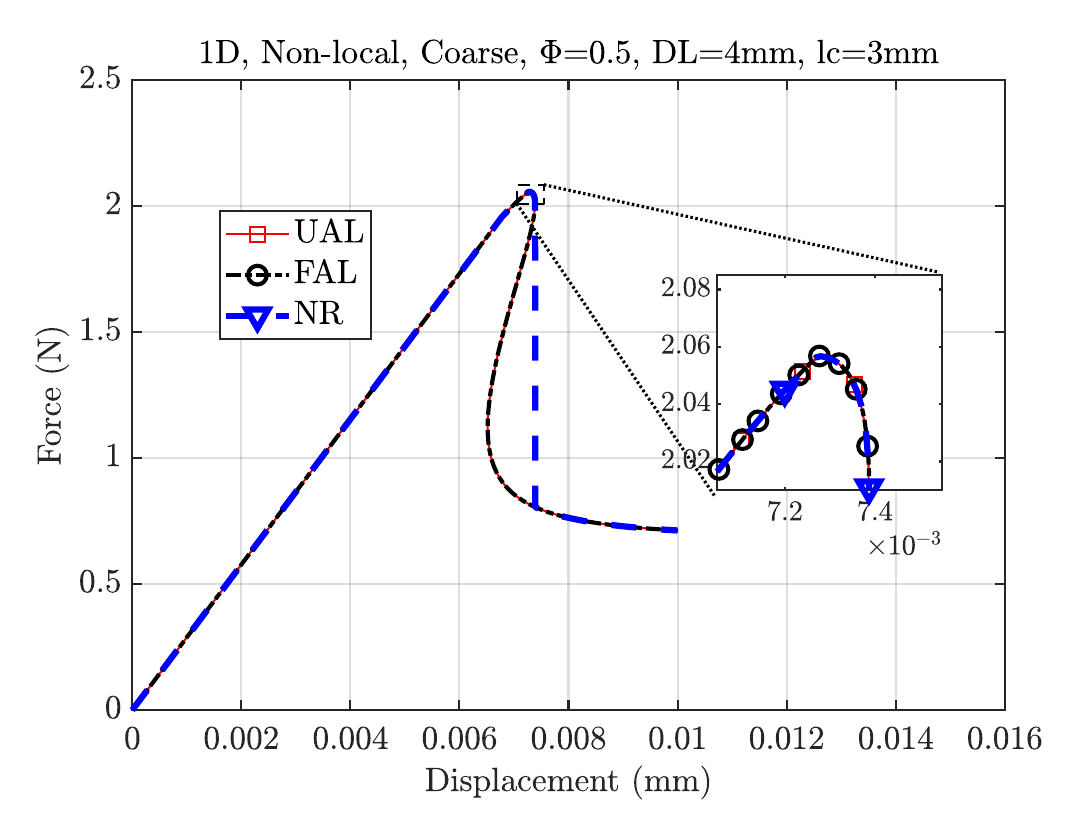}
    \text{(a) Coarse ($l_{c} = 3$)} 
    \end{minipage}
    \hfill
    \begin{minipage}{8.2cm}
    \centering     
    \includegraphics[width=1\textwidth,trim = 1.5cm 7cm 2cm 7cm, clip]{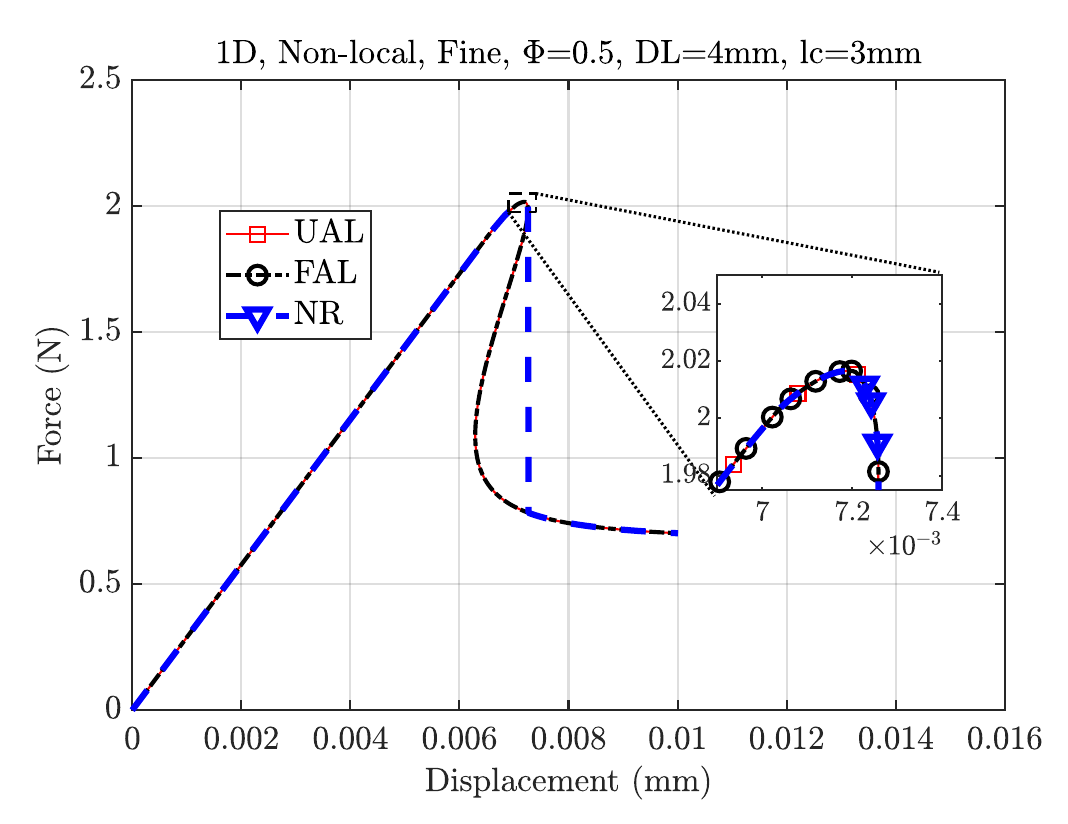}
    \text{(b) Fine ($l_{c} = 3$)}   
    \end{minipage}
    \caption{Reaction-displacement curves for the 1D bar shown in Fig.\ref{fig:1D_bar_schematic}, solved using the three non-linear solvers for the non-local gradient damage law with $\phi=0.5$, $DL = 4mm$ and $l_{c} = 3mm$. For visual clarity, only 1\% of the converged increments are presented in the zoomed-in figures.}
    \label{fig:1D_gradient_lc=3}
\end{figure}

\begin{figure}[H]
    \centering
    \begin{minipage}{8.2cm}
    \centering 
    \includegraphics[width=1\textwidth,trim = 1.5cm 7cm 2cm 7cm, clip]{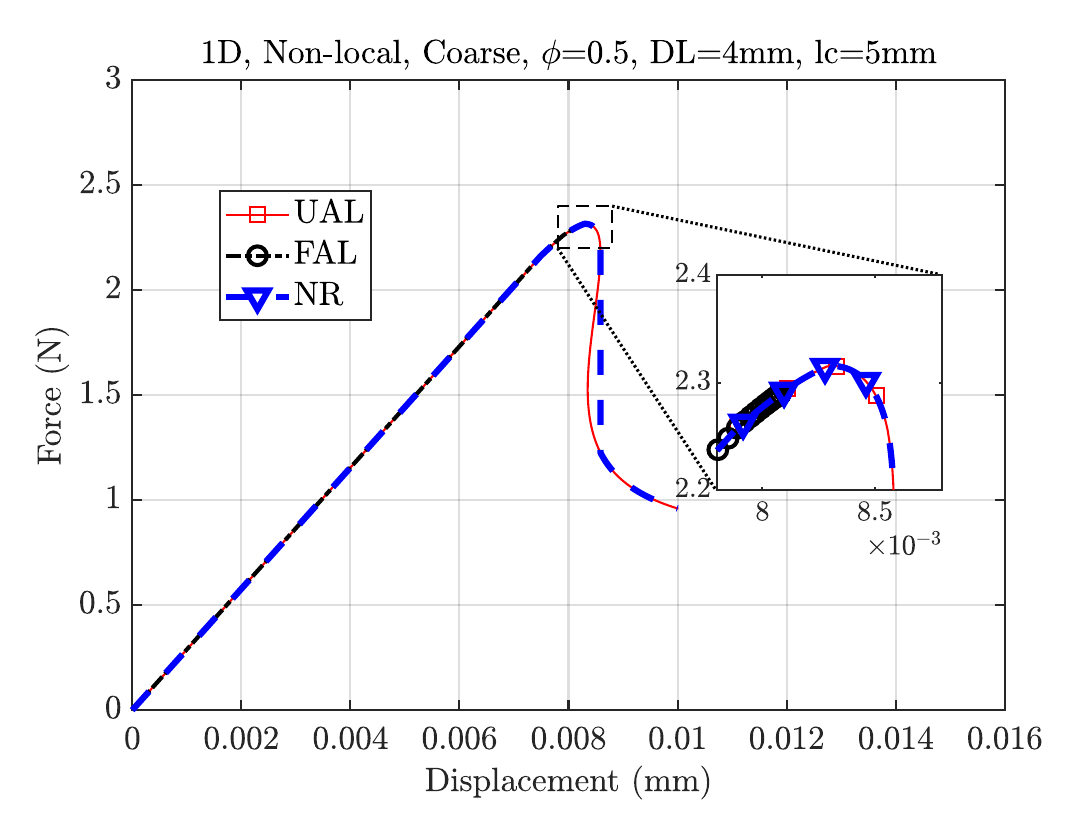}
    \text{(a) Coarse ($l_{c} = 5$)}  
    \end{minipage}
    \hfill
    \begin{minipage}{8.2cm}
    \centering 
    \includegraphics[width=1\textwidth,trim = 1.5cm 7cm 2cm 7cm, clip]{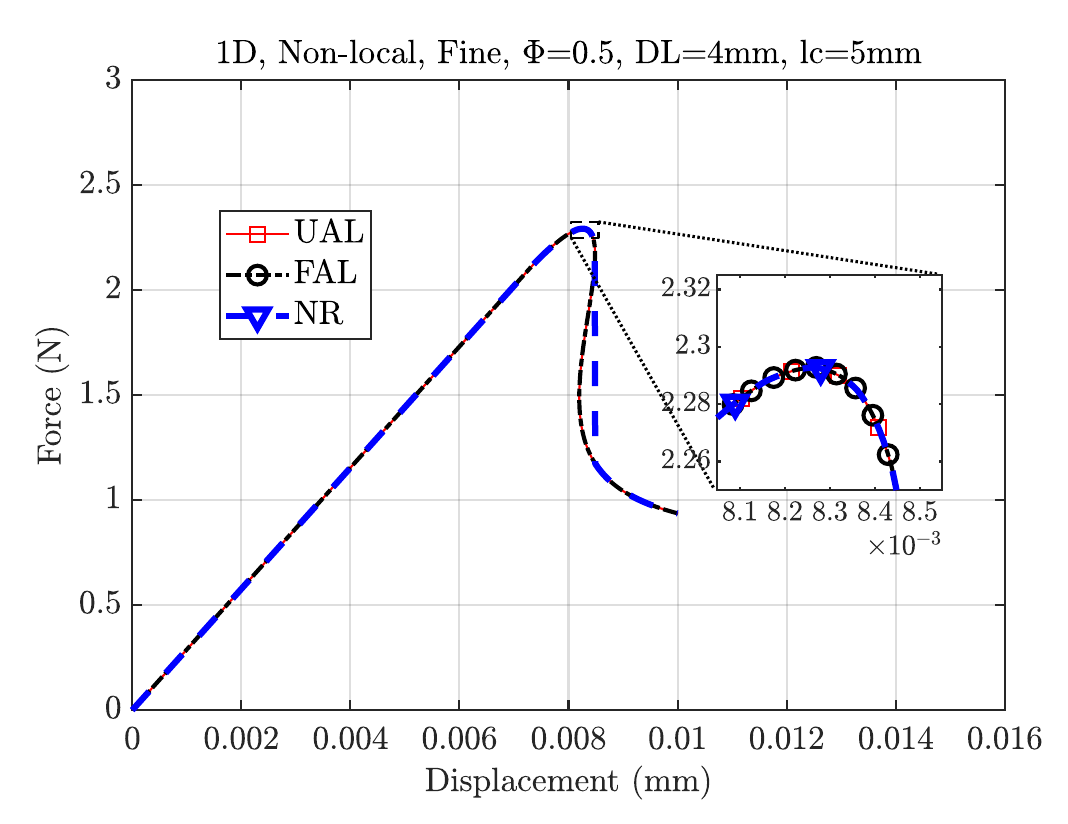}
    \text{(b) Fine ($l_{c} = 5$)}  
    \end{minipage}
    \caption{Reaction-displacement curves for the 1D bar shown in Fig.\ref{fig:1D_bar_schematic}, solved using three different non-linear solvers for the non-local gradient damage law with $\phi=0.5$, $DL = 4mm$ and $l_{c} = 5mm$. For visual clarity, only 1\% of the converged increments are presented in the figures.In the coarse mesh case, the FAL solver struggles to proceed as it approaches the peak of the curve and plot displays the last converged increment before the analysis was terminated.}
    \label{fig:1D_gradient_lc=5}
\end{figure}

For the rest of 1D non-local problems, a value of $\phi$ = 0.5 is applied for both coarse and fine meshes. Table \ref{Table:1D_NonLocal_parameters} displays the parameters used in this study unless specified otherwise in the text. The results of this investigation comparing two characteristic length values $l_c=\{0.3,0.5\}$ mm are shown in Fig. \ref{fig:1D_gradient_lc=3} and \ref{fig:1D_gradient_lc=5}. 

\begin{table}[H]
\centering
\caption{Parameters used in 1D non-local gradient damage examples}
\begin{tabular}{ l l }
    \hline
    \bf Parameter  & \bf Values\\
    \hline
    Implementation scheme & PC \\
    Mesh resolution & Coarse (51 elements), Fine (151 elements) \\
    Young's modulus & E = 30 GPa\\
    Young's modulus remaining percentage & $\phi$ = 0.5 \\
    Characteristic length & $l_{c}$ = \{3 mm , 5 mm\} \\
    Mazars model parameters & $\epsilon_D$ = $10^{-4}$, $\mathscr{A}$ = 0.7, $\mathscr{B}$ = $10^4$ \\
    Damaged length & $DL = 4 mm$ \\
    Intial load factors (UAL) & $\alpha$ = $10^{-1}$ \\
    Intial load factors (FAL, NR) &  $\Delta \lambda_0$ = $10^{-1}$, $\Delta \lambda_0$ = $10^{-1}$ \\
    $tol$ (UAL, FAL, NR) & $10^{-6}$, $10^{-4}$, $10^{-6}$ \\
    Arc-length limits ({$\Delta l_{max}$}) & $10^{-4}$  \\
    \hline
\end{tabular}
\label{Table:1D_NonLocal_parameters}
\end{table} 

It is observed that a smaller $l_{c}$ value leads to a higher strain localization and therefore a sharper snap-back region. Also, as expected and reported in the literature, the non-local gradient damage model results remain mesh-insensitive \cite{peerlings1996gradient}. Across the three solvers, the UAL outperforms the other two by capturing the post-peak response with the least number of increments and the lowest computational time. NR is unable to trace the post peak snap-back region and follows an almost vertical descending path. Table \ref{Table:1D_Gradient_computational_table} shows the computational superiority of UAL over FAL for simulating 1D non-local gradient damage problems. The UAL solver traced the entire non-linear equilibrium path of this numerically challenging 1D problem in less than 5 seconds, while the FAL took 1-2 orders of magnitude longer to trace the same path at a $tol$ value that was 100 times larger than the UAL model. These factors point to the strengths of the UAL method in simulating 1D non-local gradient damage problems over FAL. 

\begin{table}[H]
\caption{Comparison of the performance of three non-linear solvers for various non-local gradient damage problems with DL=4 mm. All the analyses are done on a Dell Precision 5820 workstation with a Intel® Xeon® W-2223 CPU @ 3.60GHz, 64 GB RAM processor with 185.6 GFLOPS computing power \cite{Intelmicroprocessors}}
\centering
\begin{tabular}{ | c | c | c | c | c | c | c | c | c | c | c | c|}
\multicolumn{12}{c}{{\textbf{Non-local Model}}} \\
\hline
\multirow{2}{3em}{Mesh} & \multirow{2}{1em}{$\phi$} & \multirow{2}{1em}{$l_{c}$} & \multicolumn{3}{|c}{No. of increments} & \multicolumn{3}{|c}{Convergence Tolerance} & \multicolumn{3}{|c|}{Simulation Time (secs)} \\
\cline{4-12}
{} & {} & {} & UAL & NR & FAL & UAL & NR & FAL & UAL & NR & FAL \\\hline

\multirow{2}{3em}{Coarse} & \multirow{4}{1.5em}{0.5} & {3} & 309 & 9017 & 816 & \multirow{5}{0.75cm}{$10^{-6}$} & \multirow{5}{0.75cm}{$10^{-6}$} & \multirow{5}{0.75cm}{$10^{-4}$} & 0.72 & 26.04 & 7.35\\\cline{3-6}\cline{10-12}

{} & {} & {5} & 249 & 9014 & 32607 & {} & {} & {} & 0.57 & 20.88 & 705.52 \\\cline{1-1}\cline{3-6}\cline{10-12} 

\multirow{3}{3em}{Fine} & {} & {3} & 568 & 4702 & 1078 & {} & {} & {} & 4.25 & 50.99 & 44.88 \\\cline{3-6}\cline{10-12} 

{} & {} & {5} & 444 & 1902 & 1243 & {} & {} & {} & 2.98 & 14.92 & 43.26 \\\cline{2-6}\cline{10-12} 

{} & {0.1} & {6} & 883 & 9003 & 1247 & {} & {} & {} & 4.49 & 85.32 & 21.33 \\\hline
\end{tabular} 
\label{Table:1D_Gradient_computational_table}
\end{table}


\subsubsection{Local Damage}
\label{NumExpl:1D_Local_Damage}

\begin{figure}[H]
    \centering
    \begin{minipage}{8.2cm}
    \centering     
    \includegraphics[width=1\textwidth,trim = 1.25cm 7cm 2cm 7cm, clip]{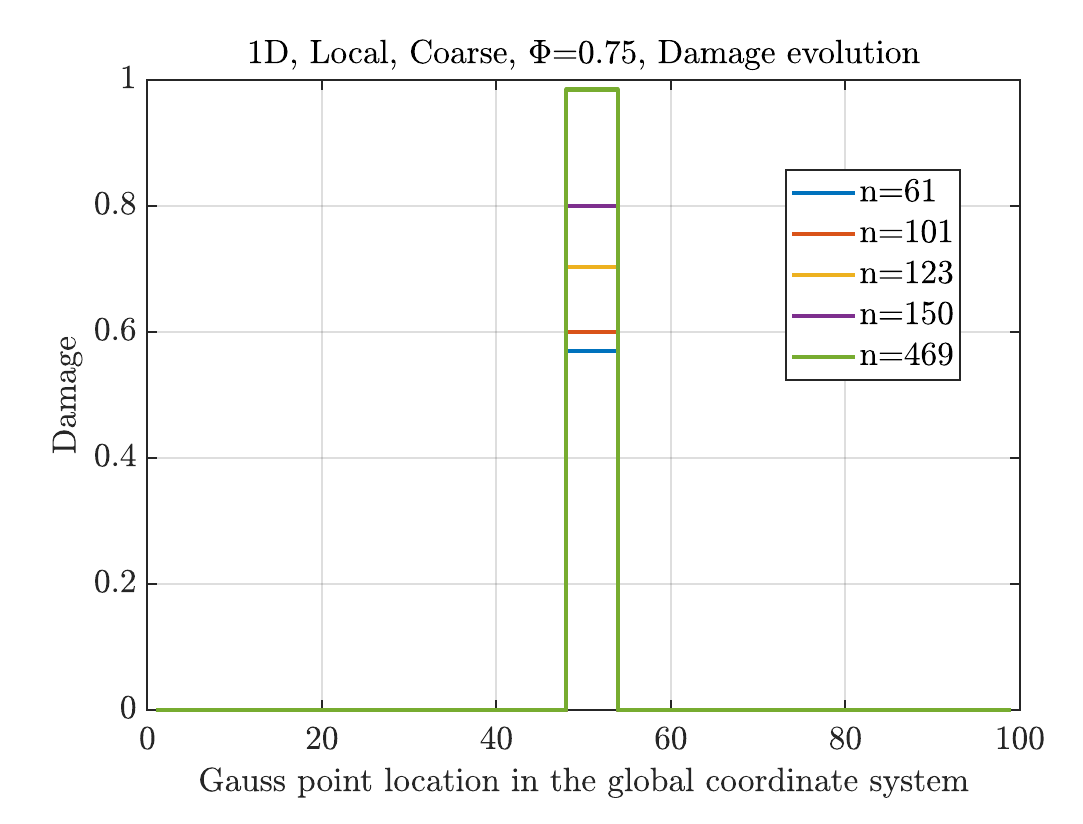}
    \text{(a) Damage evolution} 
    \end{minipage}
    \hfill
    \begin{minipage}{8.2cm}
    \centering     
    \includegraphics[width=1\textwidth,trim = 1.25cm 7cm 2cm 7cm, clip]{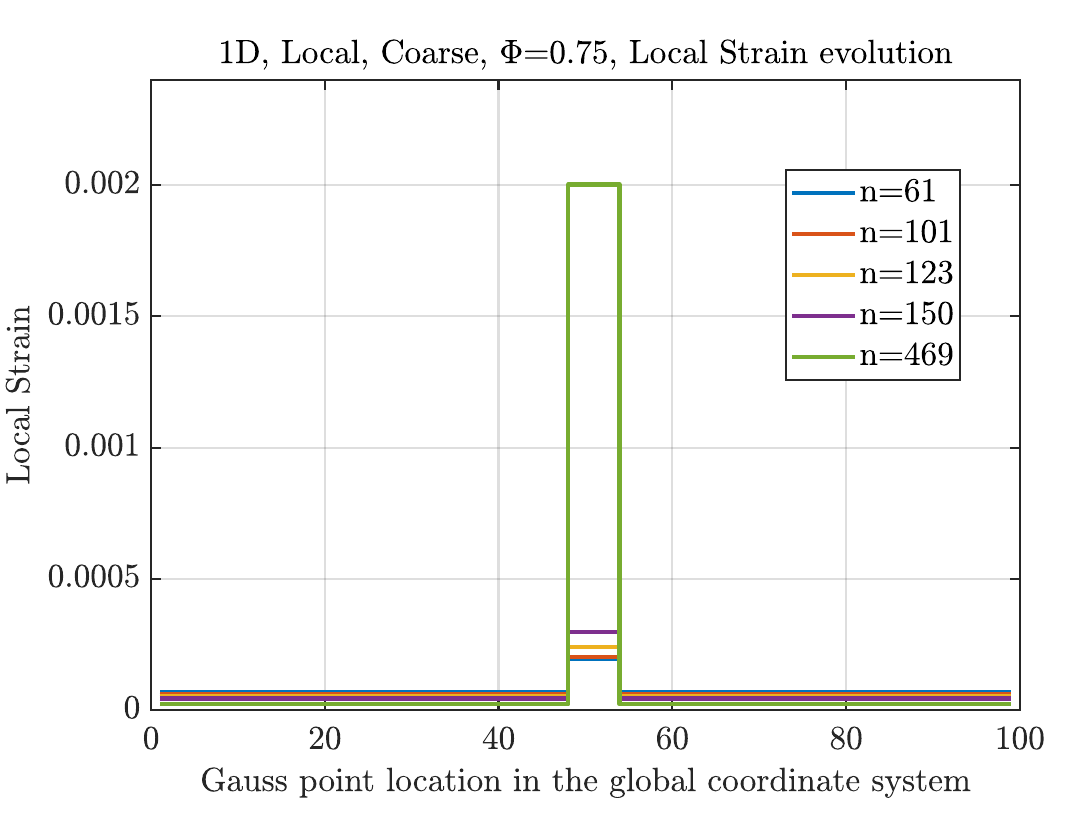}
    \text{(b) Local strain evolution}   
    \end{minipage}
    \caption{Evolution of (a) damage and (b) local strain in the 1D bar problem with a coarse mesh, $\phi = 0.75$ and $DL = 4mm$ at different load increments $n$. The strain remains localized at the damaged length in the middle as the externally applied displacement increases, and it is not diffused as in the non-local case in Fig. \ref{fig:1D_gradient_d_and_strain_evolution}.}
    \label{fig:1D_local_d_and_strain_evolution}
\end{figure}

\begin{figure}[H]
    \centering
    \begin{minipage}{8.3cm}
    \centering     
    \includegraphics[width=1\textwidth,trim = 1.5cm 7cm 2cm 7.5cm, clip]{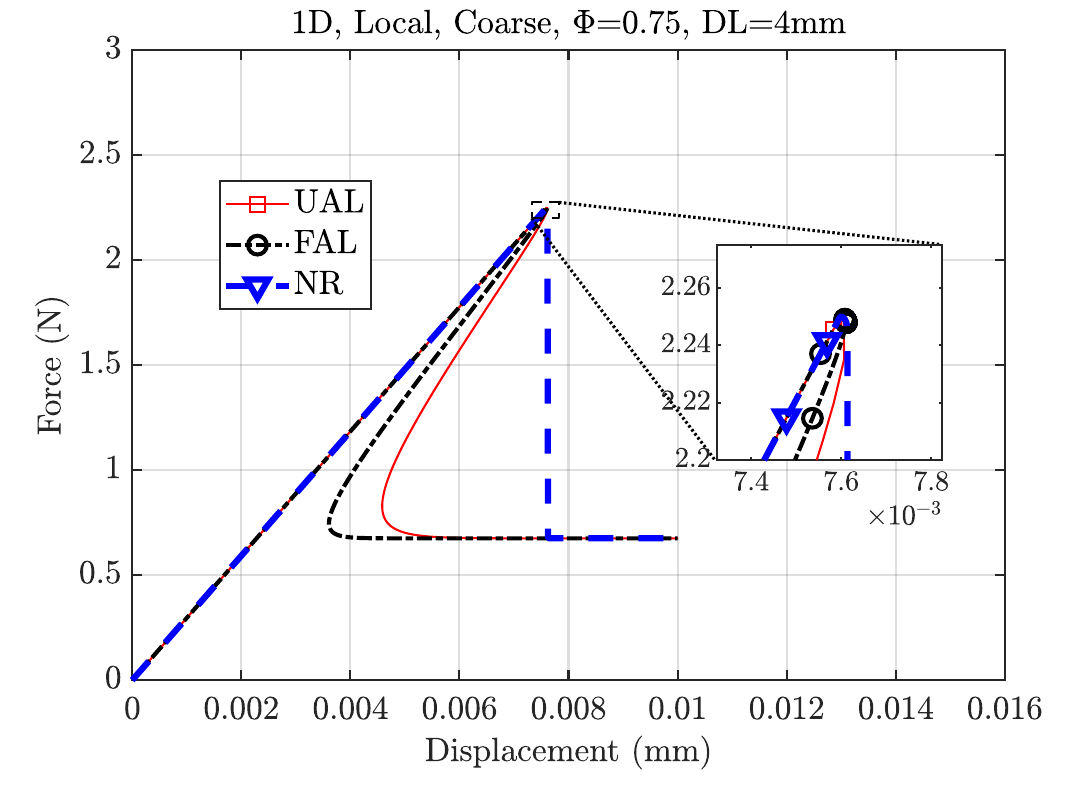}
    \text{(a) Coarse}   
    \end{minipage}
    \hfill
    \begin{minipage}{8.3cm}
    \centering 
    \includegraphics[width=1\textwidth,trim = 1.5cm 7cm 2cm 7.5cm, clip]{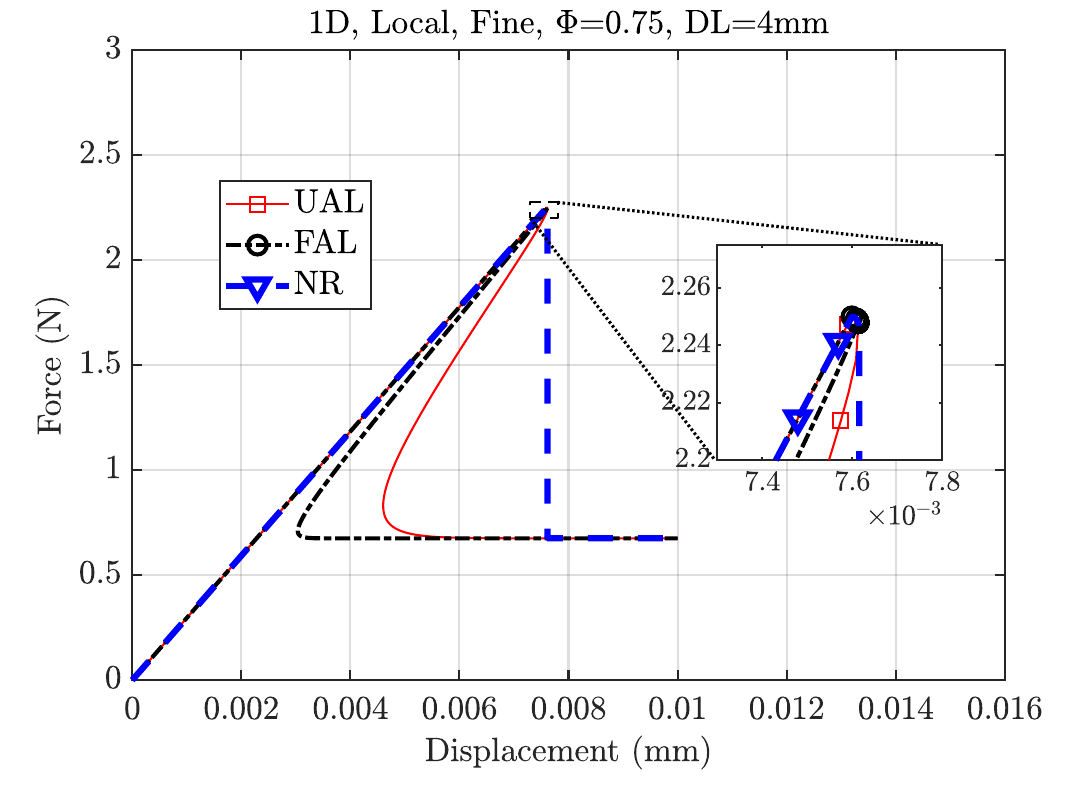}
    \text{(b) Fine}  
    \end{minipage}
    \caption{Reaction-displacement curves for the 1D bar shown in Fig.\ref{fig:1D_bar_schematic}, solved using three different non-linear solvers for the local damage law with $\phi=0.75$ and DL=4 mm. For visual clarity, only 2\% of the converged increments are presented in the figure.}
    \label{fig:1D_local_notch=0.75}
\end{figure}

\begin{figure}[H]
    \centering
    \begin{minipage}{8.2cm}
    \centering     
    \includegraphics[width=1\textwidth,trim = 1.5cm 7cm 2cm 7cm, clip]{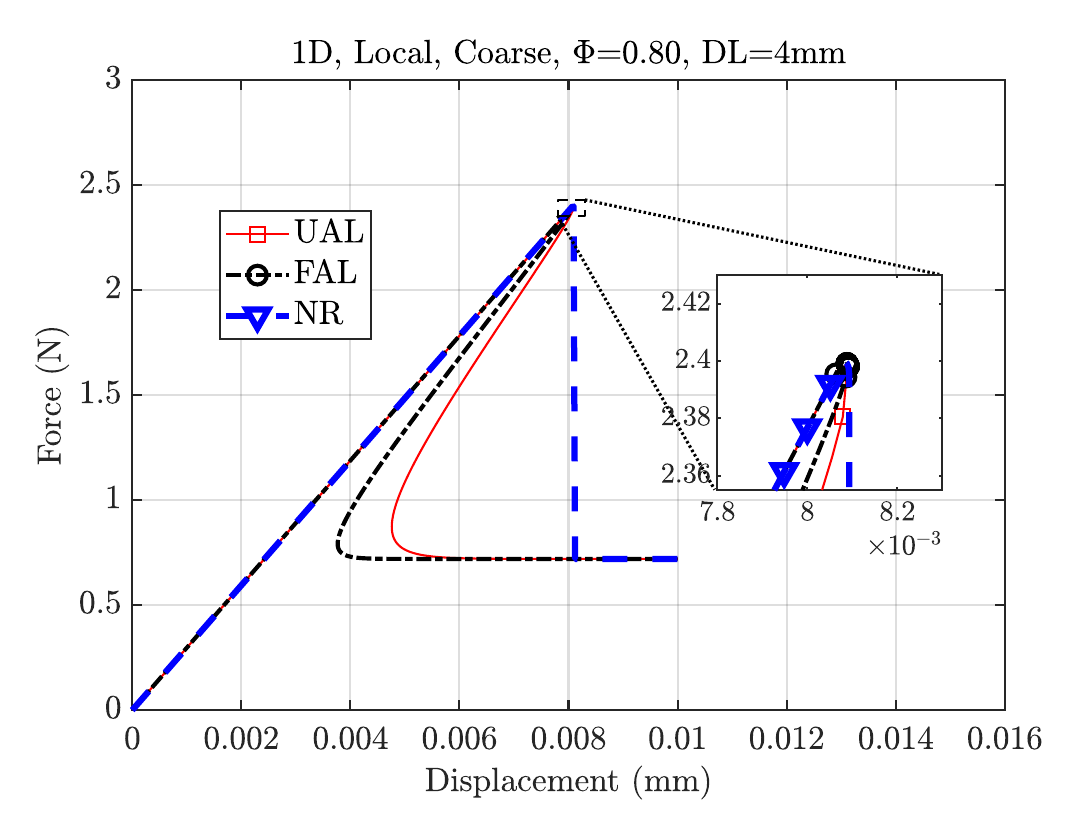}
    \text{(a) Coarse}   
    \end{minipage}
    \hfill
    \begin{minipage}{8.2cm}
    \centering 
    \includegraphics[width=1\textwidth,trim = 1.5cm 7cm 2cm 7cm, clip]{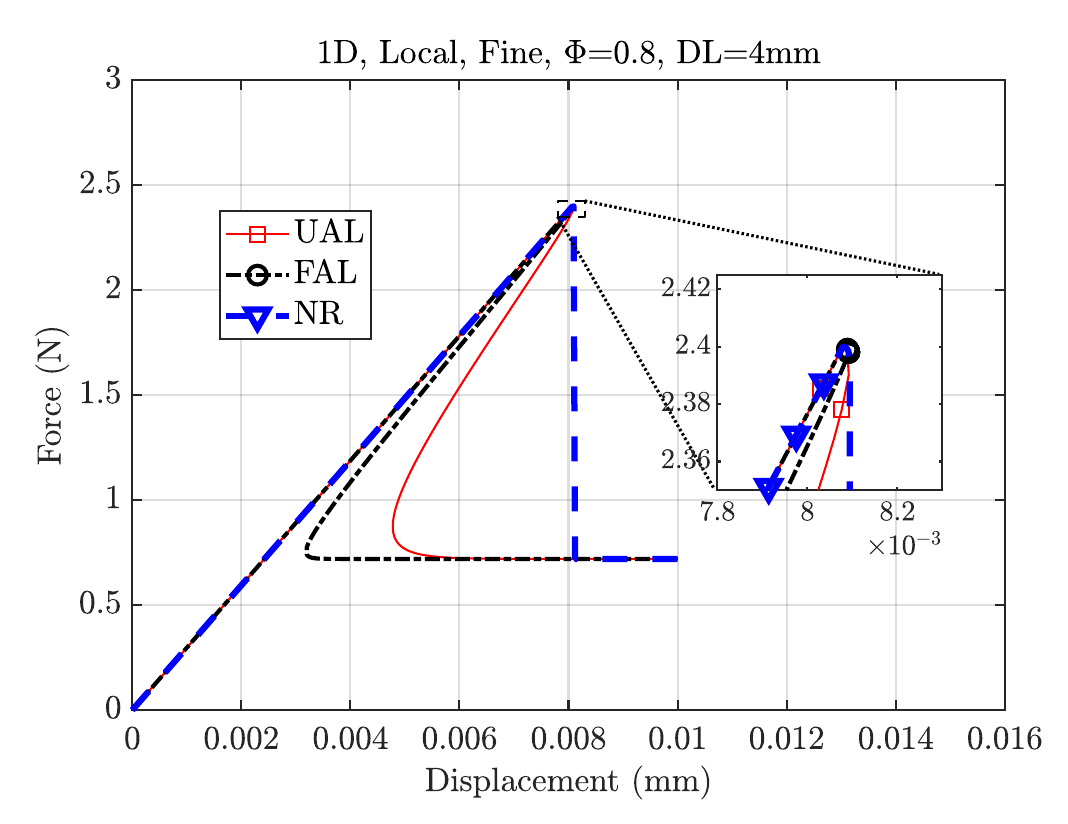}
    \text{(b) Fine}  
    \end{minipage}
    \caption{Reaction-displacement curves for the 1D bar shown in Fig.\ref{fig:1D_bar_schematic}, solved using three different non-linear solvers for the local damage law with $\phi=0.80$ and DL=4 mm. For visual clarity, only 2\% of the converged increments are presented in the figure.}
    \label{fig:1D_local_notch=0.8}
\end{figure}

In this subsection, the 1D bar problem is investigated using the local damage law. This study is performed on the coarse and fine mesh to compare the performance of UAL, FAL, and NR for three different $\phi$ values, which are reported along with the other relevant parameters in Table \ref{Table:1D_Local_parameters}. The three $\phi$ values along with the local damage law are expected to yield a very sharp and highly non-linear response, which will allow the examination of each non-linear solver's capabilities and limits under more challenging conditions than the non-local model. We first plot in Fig. \ref{fig:1D_local_d_and_strain_evolution} the evolution of damage and local strain in a 1D bar problem with $\phi = 0.75$. We observe that the values of both quantities monotonically increase with the applied load, and they are localized in the damaged length region as expected from the local damage model. We then move to the comparison of the three algorithms for the local damage examples, by analyzing the cases reported in Table \ref{Table:1D_Local_computational_table}. The results of this investigation are displayed in Figs. \ref{fig:1D_local_notch=0.75} and \ref{fig:1D_local_notch=0.8}.

\begin{table}[H]
\centering
\caption{Parameters used in 1D local damage examples}
\begin{tabular}{ l l }
    \hline
    \bf Parameter & \bf Values\\
    \hline
    Implementation scheme & PC \\
    Mesh resolution & Coarse (51 elements), Fine (101 elements) \\
    Young's modulus & E = 30 GPa\\
    Damaged length & $DL = 4 mm$ \\
    Young's modulus remaining percentage in DL & $\phi$ = \{0.75, 0.8, 0.85\} \\
    Mazars model parameters & $\epsilon_D$ = $10^{-4}$, $\mathscr{A}$ = 0.7, $\mathscr{B}$ = $10^4$ \\ 
    Initial load factor (UAL) & $\alpha$ = $10^{-1}$ to $10^{-2}$  \\ 
    Initial load factors (FAL, NR) & $\Delta \lambda_0$ = $10^{-1}$ to $10^{-2}$, $\Delta \lambda_0$ = $10^{-1}$ to $10^{-2}$ \\
    $tol$ (UAL, FAL, NR) & $10^{-6}$, $10^{-6}$, $10^{-6}$ \\
    Arc-length limits ({$\Delta l_{max}$}) & $10^{-4}$ \\
    \hline
\end{tabular}
\label{Table:1D_Local_parameters}
\end{table}

Between the three non-linear solvers, NR is clearly the most deficient, vertically dropping past the peak and completely bypassing the snap-back region. The UAL clearly outperforms both the FAL and NR by tracing the equilibrium path with the snap-back region using an order of magnitude fewer steps than FAL in all cases, as reported in Table \ref{Table:1D_Local_computational_table}. It is also important to note that the different post-peak paths between UAL and FAL are attributed to the nature of the investigated problem. Local damage problems are notorious for being ill-posed, and they exhibit major path dependence issues \cite{peerlings1996gradient, bavzant1976instability} Practically, even a minor numerical difference at the peak values of the force-displacement curve will cause the arc-length to follow mildly different paths, hence the variation in the post-peak branches. On the contrary, in a diffused model such as the non-local gradient, this numerical algorithmic artifact is counteracted by the non-locality of the problem, and as such we observe no difference in Figs. \ref{fig:1D_verification_problem} - \ref{fig:1D_gradient_lc=5}. Finally, inspection of Figs. \ref{fig:1D_local_notch=0.75} and \ref{fig:1D_local_notch=0.8} leads to the following mechanics-targeted observations:

\begin{itemize}
    \item The smaller the $\phi$ value, the sharper the post-peak snap-back region becomes for both mesh sizes.
    \item Maximum force at the peak of the equilibrium path increases with an increase in $\phi$ value.
\end{itemize}

These behavioral patterns remain consistent regardless of the selected numerical technique, and therefore this certifies the ability of UAL to capture correctly the structural response of the 1D bar across different scenarios.

\begin{table}[H]
\caption{Comparison of the performance of three non-linear solvers for various Local damage problems with DL=4 mm. All the analyses are on a Dell Precision 5820 workstation with a Intel® Xeon® W-2223 CPU @ 3.60GHz, 64 GB RAM processor with 185.6 GFLOPS computing power \cite{Intelmicroprocessors}}
\centering
\begin{tabular} {|c |c |c |c |c |c |c |c |c |c |c|} 
\multicolumn{11}{c}{{\textbf{Local Model}}} \\
\hline
\multirow{2}{3em}{Mesh} & \multirow{2}{3em}{$\phi$} & \multicolumn{3}{|c}{No. of increments} & \multicolumn{3}{|c}{Convergence Tolerance} & \multicolumn{3}{|c|}{Simulation Time (secs)} \\
\cline{3-11}
{} & {} & UAL & NR & FAL & UAL & NR & FAL & UAL & NR & FAL \\\hline

\multirow{2}{3em}{Coarse} &{0.75} & 469 & 9009 & 5290 & \multirow{5}{0.75cm}{$10^{-6}$} & \multirow{5}{0.75cm}{$10^{-6}$} & \multirow{5}{0.75cm}{$10^{-6}$} & 0.53 & 21.47 & 43.02\\\cline{2-5}\cline{9-11}

{} & {0.80} & 467 & 25647 & 5019 & {} & {} & {} & 0.53 & 89.04 & 39.89 \\\cline{1-5}\cline{9-11}

\multirow{2}{3em}{Fine} & {0.75} & 609 & 9017 & 12948 & {} & {} & {} & 1.43 & 40.53 & 114.63 \\\cline{2-5}\cline{9-11}

{} & {0.80} & 623 & 9021 & 12863 & {} & {} & {} & 1.39 & 35.15 & 164.58 \\\cline{2-5}\cline{9-11}

\hline

\end{tabular} 
\label{Table:1D_Local_computational_table}
\end{table}

\subsubsection{Computational performance}
\label{Sec:Computational_performance}

To better illustrate the efficiency of our proposed scheme, in this subsection we conduct an in-depth investigation of the computational performance of UAL and FAL. Using the local damage model, we analyze a 1D bar problem of $100mm$ length with $51$ elements, $\phi=0.80$, $E = 30GPa$ and $DL=4mm$ using both arc-length approaches. The convergence tolerance is set at $10^{-6}$, with an initial arc-length $\Delta l_0$ = $10^{-4}$, $\alpha$ = $10^{-2}$, $\Delta \lambda_{0}$ = $10^{-2}$, $\Delta l_{max}$ = $10^{-4}$ and $\Delta l_{min}$ = 0. These settings remain strictly identical between UAL and FAL, to ensure a fair one-to-one comparison between the methods. The simulations were performed on a Dell Precision 5820 workstation with an Intel Xeon W-2223 CPU @ 3.60GHz, 64 GB RAM processor with 185.6 GFLOPS computing power. 

\begin{figure}[H]
    \centering  
    \includegraphics[width=0.65\textwidth,trim = 1cm 8cm 2cm 8cm, clip]{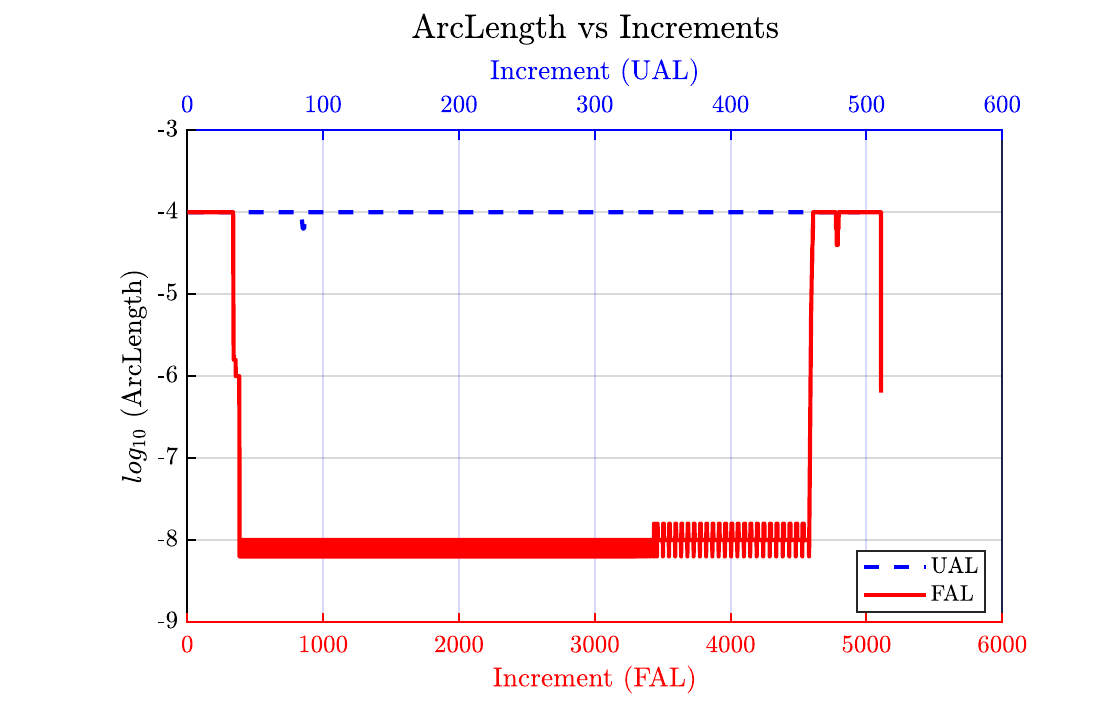}
    \caption{Comparison of the evolution of $\Delta l$ in UAL and FAL model using local damage for 1D bar problem of length=100mm, Coarse mesh, $\phi=0.80$. The input parameters and computational times are reported in Tables \ref{Table:1D_Local_parameters} and \ref{Table:1D_Local_computational_table} respectively. Note that the x axis representing the number of increments for the UAL model is at the top while that of the FAL model is at the bottom.}
    \label{fig:arclength_increment}
\end{figure}

\begin{figure}[H]
    \centering     \includegraphics[width=0.65\textwidth,trim = 1cm 8cm 2cm 8cm, clip]{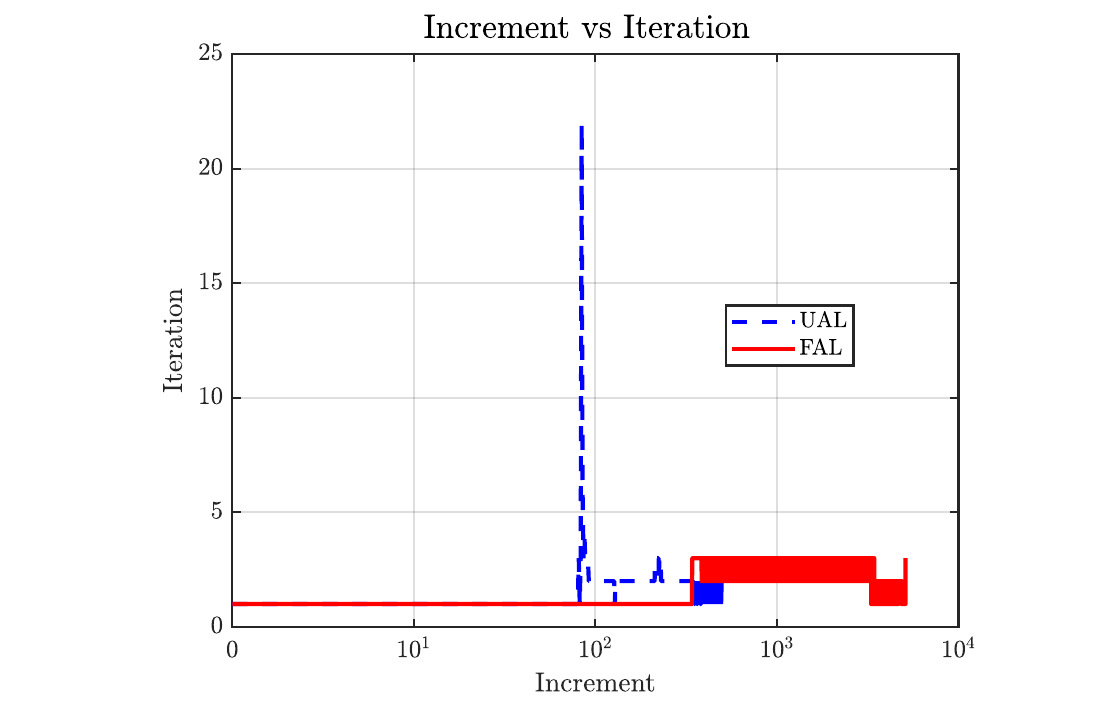}
    \caption{Comparison of the iterations per increment in UAL and FAL model using local damage for 1D bar problem of length=100mm, Coarse mesh, $\phi=0.80$. The input parameters and computational times are reported in Tables \ref{Table:1D_Local_parameters} and \ref{Table:1D_Local_computational_table} respectively. Though UAL takes more iterations in a few critical increments, it takes an order of magnitude fewer increments overall.} \label{fig:increment_iteration}
\end{figure}

Fig. \ref{fig:arclength_increment} shows the evolution of the arc-length for both methods throughout the analysis. Evidently, UAL is capable of maintaining 3-4 orders of magnitude larger arc-length values than FAL, while utilizing 1 order of magnitude fewer increments. As damage propagates, UAL is capable of operating at an arc-length value around $10^{-4}$, whereas FAL requires a much smaller value of $~ 10^{-8}$. This trend, which is present regardless of the problem setup, highlights the flexibility of UAL and provides a clear justification of the computational superiority of UAL compared to FAL. Also, in Fig. \ref{fig:increment_iteration} we plot the number of iterations per increment for the same problem and for both methods. Though at a few critical increments close to the peak of the curve the UAL model needs more iterations, this graph clearly shows that the total number of both the increments and the iterations with UAL is several times less than with FAL. Therefore, even though UAL employs a larger $J$ matrix and the cost of one UAL iteration is expected to be greater than that of one FAL iteration, the total computational cost is significantly lower within our scheme.

\begin{figure}[H]
    \centering  
    \includegraphics[width=0.5\textwidth,trim = 1cm 7cm 2cm 7cm, clip]{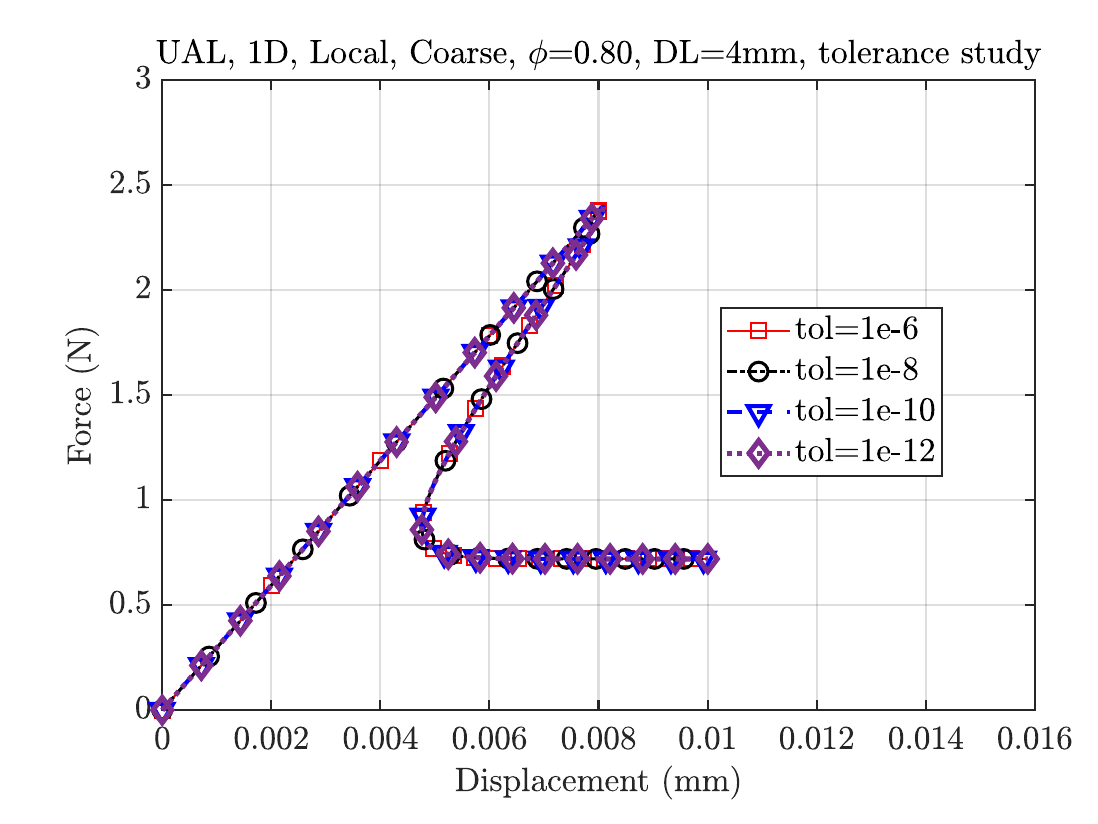}
    \caption{A comparative study on the impact of convergence tolerance ($tol$) on the equilibrium path traced by UAL. Smaller $tol$ values do not affect the equilibrium path.}
    \label{fig:Tolernace_study_crossplot}
\end{figure}

Furthermore, the accuracy and robustness of UAL can be assessed by monitoring the reaction-displacement curves for different tolerances. The same 1D problem described above is modeled with $tol = {1e-6, 1e-8, 1e-10, 1e-12}$, and the results of this study are presented in Fig. \ref{fig:Tolernace_study_crossplot}. The overlap of the curves essentially shows that a) the choice of convergence tolerance has a negligible influence on our algorithm's performance, and b) UAL can converge under strikingly low $tol$ values, which evidently demonstrate the robustness and accuracy of UAL.

Overall, the results and discussion presented in Section \ref{1D_bar_problem} provide sufficient evidence that in the 1D case the proposed UAL method can:

\begin{itemize}
    \item Capture the sharp post-peak snap-back path for a wide range of scenarios encompassing different damage evolution laws, FEM mesh resolutions, initial damage conditions, etc. 
    \item Trace the complete equilibrium path with a computational effort (increments, tolerance and total time) which is orders of magnitude less than the FAL approach.   
\end{itemize}


\subsection{2D problems}
\label{NumEx: 2D problems}

In this subsection, four two-dimensional (2D) models are utilized to illustrate the generalizability of the UAL framework in the 2D case, as well as to examine the features and control parameters of the UAL model that influence its efficiency. The Single Notch Tension (SNT) problem is the primary example used to illustrate the role of the control parameters (Section \ref{Control_parameters}) in the implementation of UAL in 2D problems. The Symmetric Single Notch Tension (SSNT) and the Two Notch Tension (TNT) problems are used to demonstrate the ability of the UAL model to handle problems with different number of notches to initial strain localization. Finally, the Single Notch Shear (SNS) problem, is presented to display the robustness of the UAL method in problems involving shear loads.  In this section we compare the UAL performance only against NR, since we already established the clear superiority of UAL against FAL for the 1D case.

\begin{table}[H]
\centering
\caption{General parameters used in 2D examples; The variation in parameters used in the different 2D studies are mentioned within their respective subsections wherever applicable}
\begin{tabular}{ l l }
    \hline
    \bf Parameter  & \bf Values\\
    \hline
    Material parameters & $\mu$ = 125 MPa, $\nu$=0.2\\
    Mazars model parameters & $\epsilon_D$ = $10^{-4}$, $\mathscr{A}$ = 0.7, $\mathscr{B}$ = $10^4$\\
    Solver-specific parameters (UAL, NR) & $\alpha$ = $10^{-4}$, $\Delta \lambda_0 = 10^{-3}$  \\ 
    $tol$ (UAL, NR) & $10^{-8}$, $10^{-6}$ \\
    Arc-length limits ({$\Delta l_{max}$}) & $10^{-3}$  \\
    \hline
\end{tabular}
\label{Table:2D_parameters}
\end{table} 


\subsubsection{Single Notch Tension (SNT) problem}
\label{NumEx:SNT}

\begin{figure}[H]
    \centering
    \begin{subfigure}{8.2cm}
    \centering     
    \includegraphics[width=0.9\textwidth,trim = 11cm 4cm 11cm 4cm, clip]{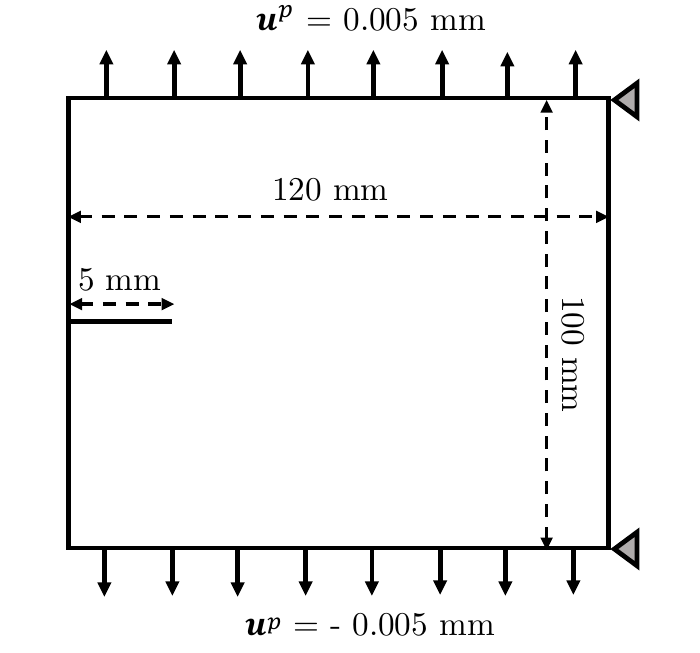}
    \caption{SNT schematic}
    \end{subfigure}
    \hfill
    \centering
    \begin{subfigure}{8.2cm}
    \centering     
    \includegraphics[width=0.9\textwidth,trim = 9cm 2.25cm 9cm 2cm, clip]{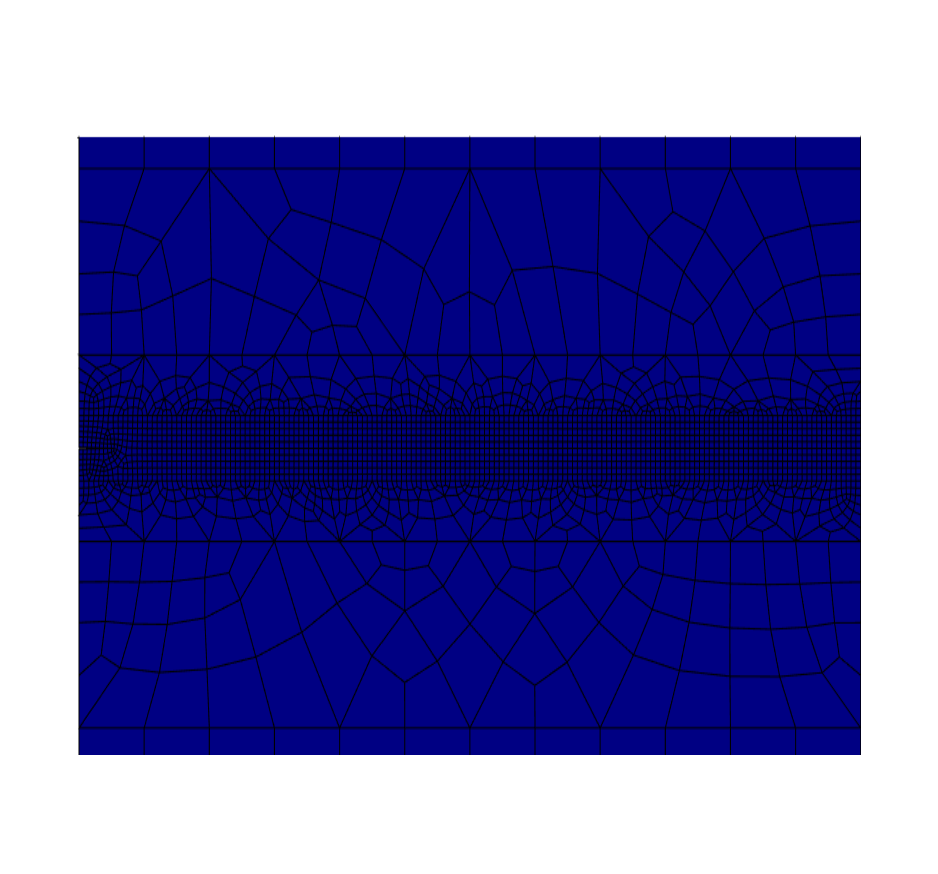}
    \caption{SNT Mesh}
    \end{subfigure}
    \caption{A schematic illustration of the (a) geometry and (b) mesh of the SNT problem. The unstructured coarse mesh displayed here has 2703 quadrilateral elements.}
    \label{fig:SNT_schematic&mesh}
\end{figure}

The Single Notch Tension (SNT) problem is a benchmark example which has been commonly investigated in the literature \cite{miehe2010phase,miehe2010thermodynamically}. In this subsection, the results of the study on the SNT problem shown in Fig. \ref{fig:SNT_schematic&mesh} are presented. The domain has a rectangular shape and its dimensions are 120 mm $\times$ 100 mm, with a notch of 5 mm in the middle of the left boundary protruding into the domain. The FEM idealization of the domain comprises 2703 elements in total, and the size of the elements in the refined zone of the domain is 0.7 $\times$ 0.7 mm. The horizontal displacements of the top right and bottom right nodes are constrained, and a tensile load of $5 \times 10^{-3}$ mm acts upon the top and bottom surfaces. The material parameters are shear modulus $\mu$ = 125 MPa and Poisson's ratio $\nu$ = 0.2. The Mazars law parameters are $\epsilon_D$ = $10^{-4}$, $\mathscr{A}$ = 0.7 and $\mathscr{B}$ = $10^4$, and the equivalent strain is defined based on Eqn. \eqref{ApxEq:Equivalent_strain_tension} of \ref{Appendix: Mazar_Damage_model}. Plane strain conditions are considered, and the convergence tolerance for the UAL and NR solvers are set to $10^{-8}$ and $10^{-6}$ respectively. The initial values of $\Delta l$  and $\lambda$ are set to $10^{-4}$ in the UAL and NR schemes respectively. Regarding the implementation schemes, the PNC scheme is used to implement the UAL model for the local damage case only.  

\begin{figure}[H]
    \centering
    \begin{minipage}{15cm}
    \centering     
    \includegraphics[width=\textwidth,trim =0cm 6cm 0cm 5cm, clip]{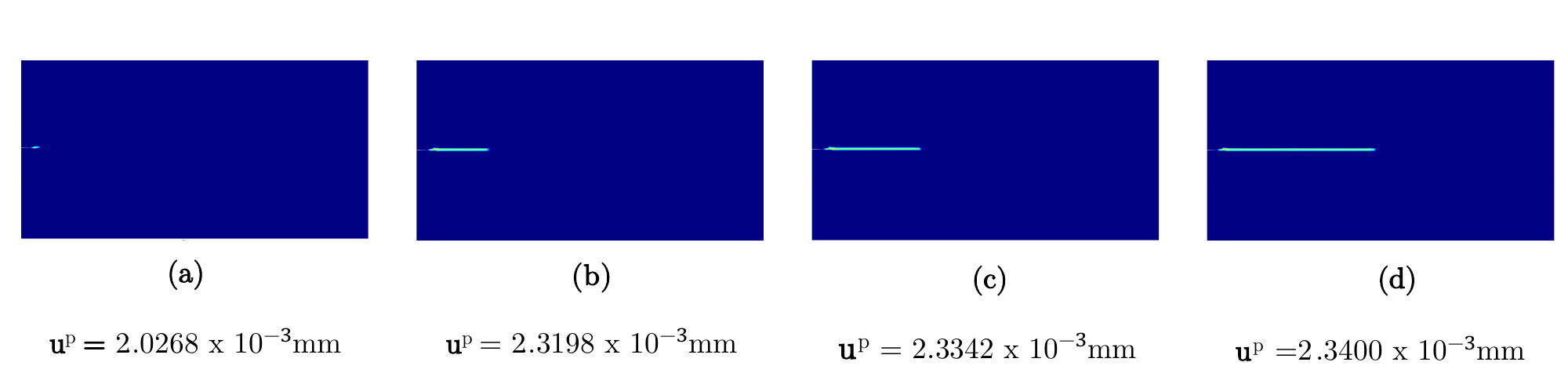}
    \end{minipage}
    \caption{Contour of the damage profile for the SNT problem at several representative load increments.}
    \label{fig:SNT_damage_evolution_contours}
\end{figure}

\begin{figure}[H]
    \centering
    \begin{subfigure}{8.2cm}
    \centering     
    \includegraphics[width=\textwidth,trim = 2cm 7cm 2cm 7cm, clip]{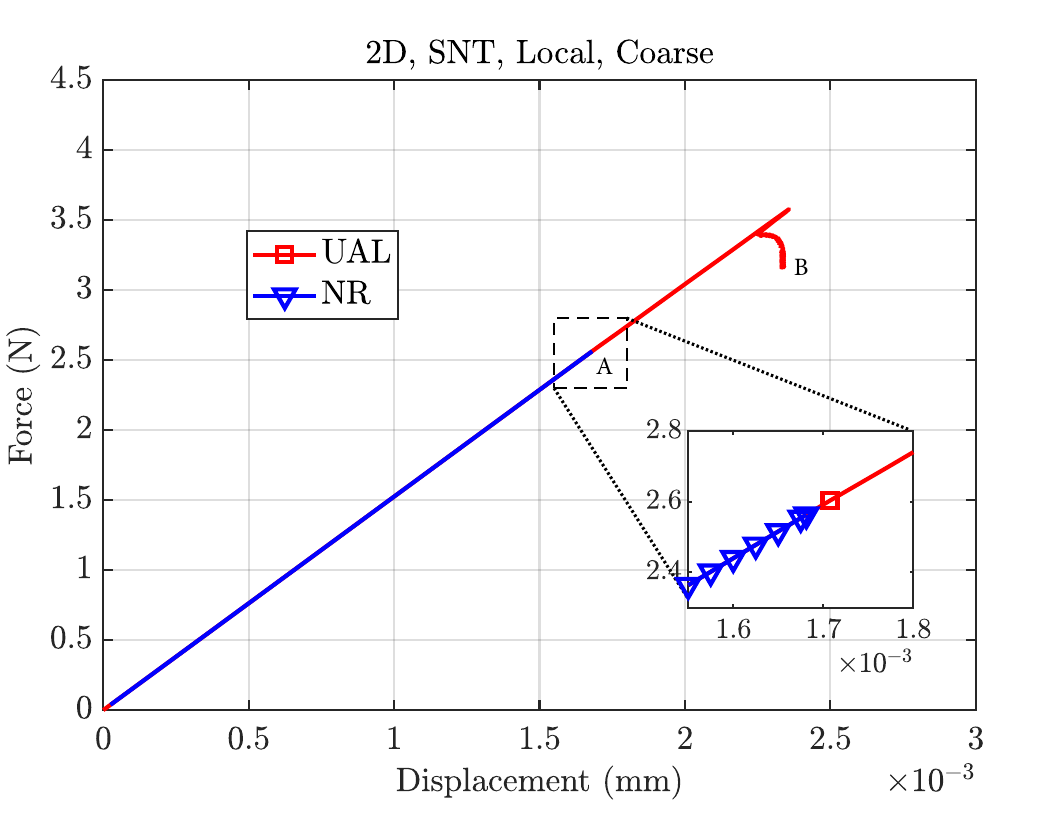}
    \end{subfigure}
    \hfill
    \centering
    \begin{subfigure}{8.2cm}
    \centering     
    \includegraphics[width=\textwidth,trim =14cm 7cm 2cm 5cm, clip]{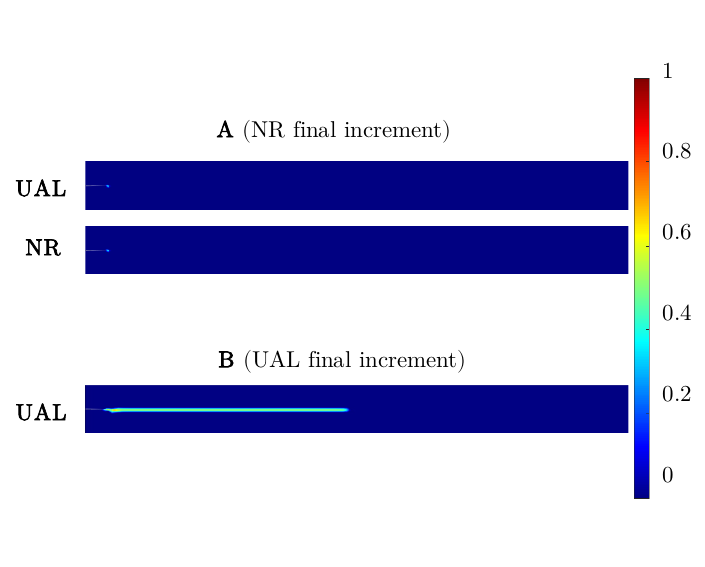}
    \end{subfigure}
    \caption{Comparison of the UAL and NR solvers for the SNT problem. The NR algorithm fails to capture the response of the system beyond the point of damage initiation i.e. at 1.65 mm, while UAL is able to do so well past the peak of the force-displacement response. The graphs on the right indicate the damage profiles at the end NR (top) and UAL (bottom). Here we have set the control parameters as $d_{max}$ = 0.999, {$\Delta l_{max}$} = $10^{-3}$ and ST = $10^{-6}$}.
    \label{fig:SNT_damage_contours}
\end{figure}

Fig. \ref{fig:SNT_damage_evolution_contours} presents the evolution of damage within the SNT problem at different displacement values. An initial investigation compares the local damage propagation contours and equilibrium paths traced by the UAL and NR solvers. NR struggles to proceed once damage initiates at a displacement load of 1.65 $\times$ $10^{-3}$ mm as can be observed in Fig. \ref{fig:SNT_damage_contours}, while UAL is able to go over the peak of the equilibrium path and track damage until it reaches the middle of the domain. This gives confidence in the ability of UAL to simulate 2D problems with strong damage localization significantly better than the NR solver. In this example, the NR solver takes unrealistically small steps once damage initiates and fails to trace the equilibrium path any further. The limitations of the standard NR solver in problems with a narrow damage zone and rapid crack growth is well documented\cite{de2012nonlinear,de1985non,rots1987analysis} and is thus used to benchmark the performance of UAL. Three more studies are conducted to better portray the influence of the control parameters described in Section \ref{Control_parameters} on the performance of the UAL solver.

\begin{figure}[H]
    \centering    
     \includegraphics[width=0.5\textwidth,trim = 0cm 7cm 2cm 7cm, clip]{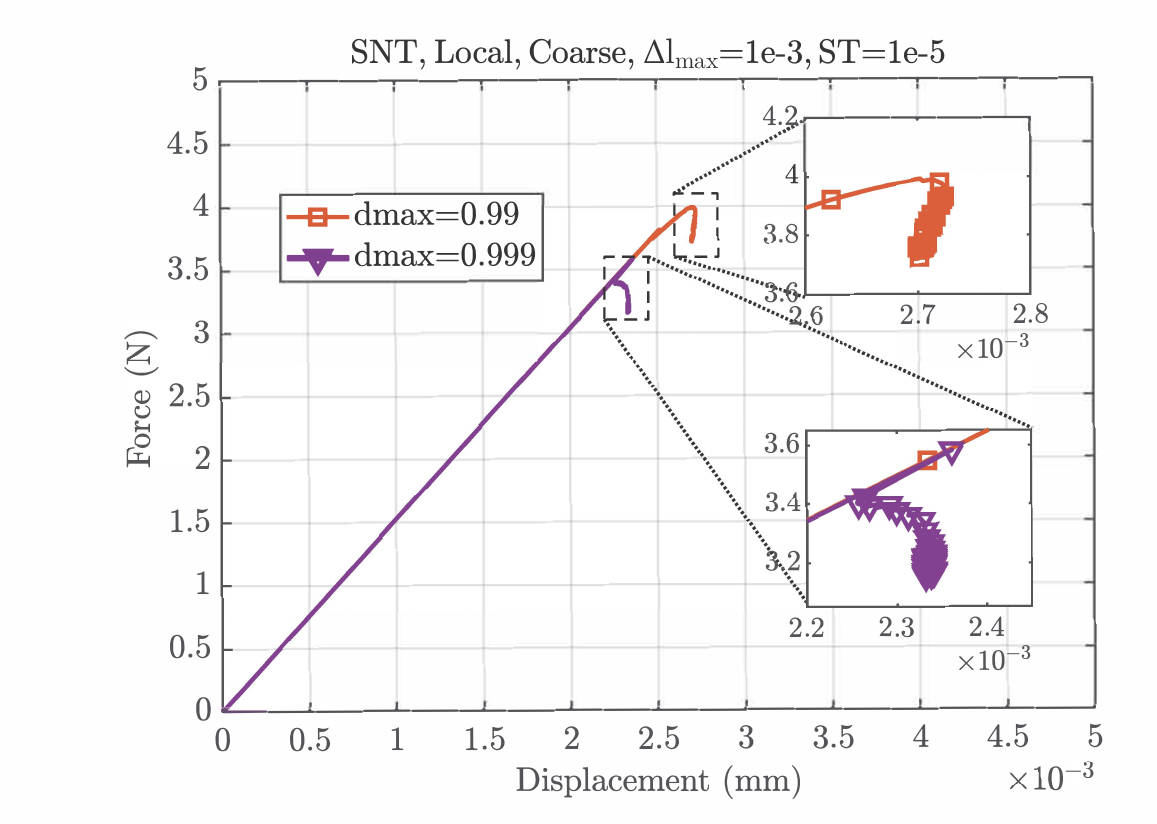}
    \caption{A comparative study on the influence of $d_{max}$ for the SNT problem. Lower $d_{max}$ values result to higher residual strengths and therefore higher total loads that can be borne by the domain.}
    \label{fig:SNT_dmax_crossplot}
\end{figure}

The first parametric study for the SNT case is on the influence of the maximum allowable damage value within an element, $d_{max}$. Fig. \ref{fig:SNT_dmax_crossplot} presents two force-displacement curves for two problems with a $d_{max} = 0.99$ and $d_{max} = 0.999$ respectively. The model with a lower $d_{max}$ value is expected to bear a higher load due to the higher residual strength of the fully damaged elements, and this trend is confirmed in Fig. \ref{fig:SNT_dmax_crossplot} where the case with $d_{max} = 0.99$ peaks at a 12\% higher maximum force than the case with $d_{max} = 0.999$.

\begin{figure}[H]
    \centering  
    \includegraphics[width=0.5\textwidth,trim = 1cm 7cm 2cm 7cm, clip]{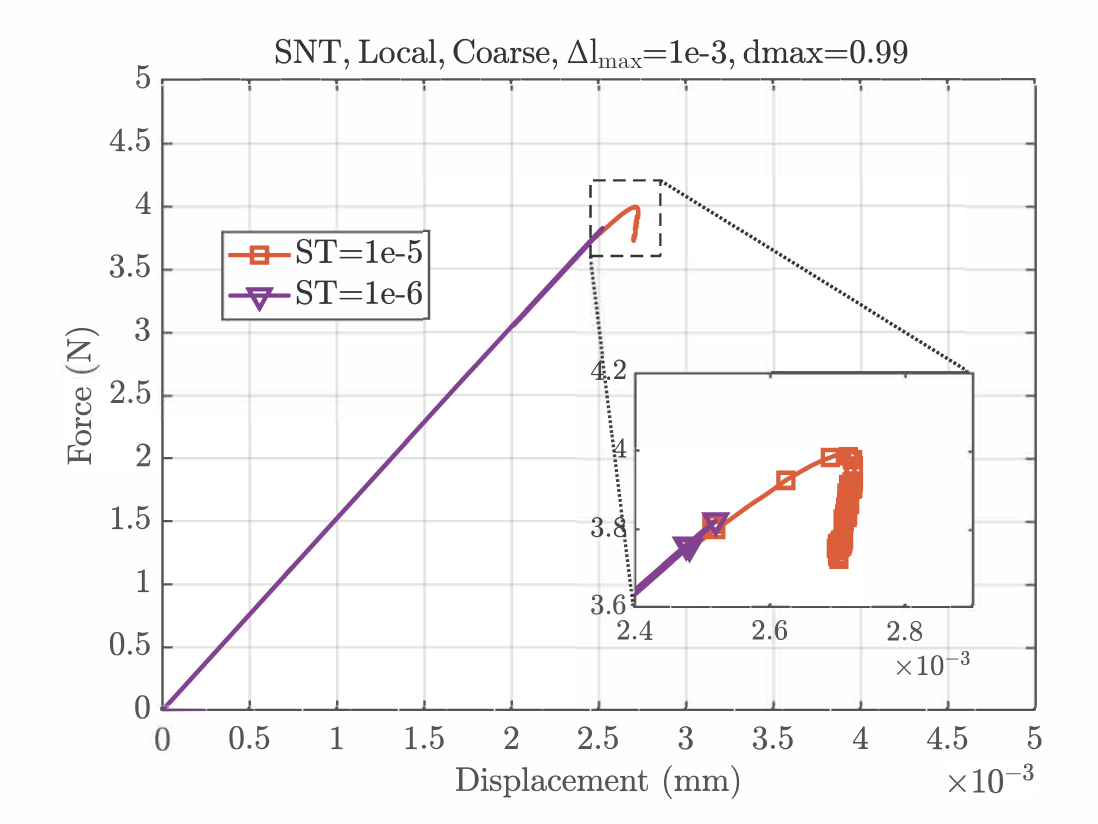}
    \caption{A comparative study on the impact of strain tolerance (ST) on the equilibrium path traced by UAL. Larger ST values tend to allow for further damage propagation inside the domain.}
    \label{fig:SNT_ST_crossplot}
\end{figure}

The second study compares the force-displacement curve computed at two values of ST = $10^{-5}$ and $10^{-6}$. In Fig. \ref{fig:SNT_ST_crossplot} we observe that a higher ST value allows the system to move further along the equilibrium path. This is the because a higher ST accounts for a larger number of instabilities arising from the numerical limit of the device used for running the model. 

\begin{figure}[H]
    \centering
    \includegraphics[width=0.5\textwidth,trim = 1cm 7cm 2cm 7cm, clip]{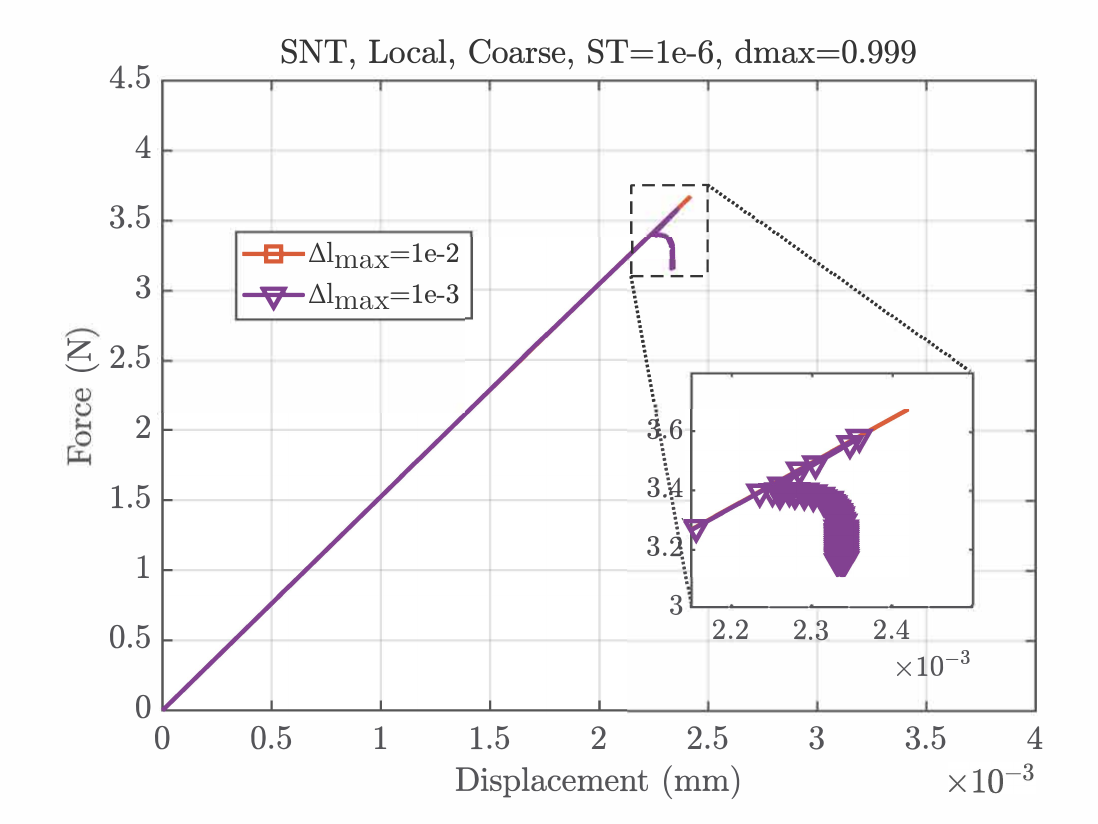}     
    \caption{A comparative study on the influence of the arc-length upper limit ({$\Delta l_{max}$}) values for the SNT problem. A larger {$\Delta l_{max}$} leads the solver to diverge from the equilibrium path.}
    \label{fig:SNT_AL_crossplot}
\end{figure}

Finally, we investigate the influence of the arc-length limit {$\Delta l_{max}$} on the performance of the UAL solver. In Fig. \ref{fig:SNT_AL_crossplot}, the force-displacement crossplots for {$\Delta l_{max}$} = $10^{-2}$ and {$\Delta l_{max}$} = $10^{-3}$ are compared, and we observe that a higher {$\Delta l_{max}$} value leads to the solver overshooting the peak of the equilibrium path and failing to trace the post-peak behavior. This is in contrast with the lower {$\Delta l_{max}$} value, in which case the solver can effectively capture the snap-back path.

Despite performing better than NR, the UAL method also faced challenges in this highly localized damage problem and Figs.\ref{fig:SNT_dmax_crossplot}, \ref{fig:SNT_ST_crossplot} and \ref{fig:SNT_AL_crossplot} show the influence of the control parameters on the performance of the UAL method in this limiting setup. The PNC scheme was chosen as the PC scheme faced considerable challenges in solving the SNT problem. To trace the complete equilibrium path an optimization algorithm such as BFGS \cite{wu2020bfgs} or a staggered scheme \cite{lampron2021efficient} could be employed. However they both share a high computational cost \cite{kristensen2020phase, khalil2022generalised}, and their utilization lies beyond the objectives of this work.

Overall, this subsection has clearly demonstrated the ability of our proposed framework to trace the damage path well-beyond the final converged increment of NR, thus showcasing its vast superiority in terms of predictive accuracy. Also the complementary numerical investigations have shed light on the influence of several control parameters on improving the UAL performance, and we underline that fine-tuning of these hyperparameters is a problem-specific task that needs to be performed in order to maximize the gains from implementing the UAL algorithm.


\subsubsection{Symmetric Single Notch Tension (SSNT) problem}
\label{NumEx:SSNT}

\begin{figure}[H]
    \centering
    \begin{subfigure}{8.2cm}
    \centering    
    \includegraphics[width=1\textwidth,trim = 10.5cm 4cm 10.5cm 4cm, clip]{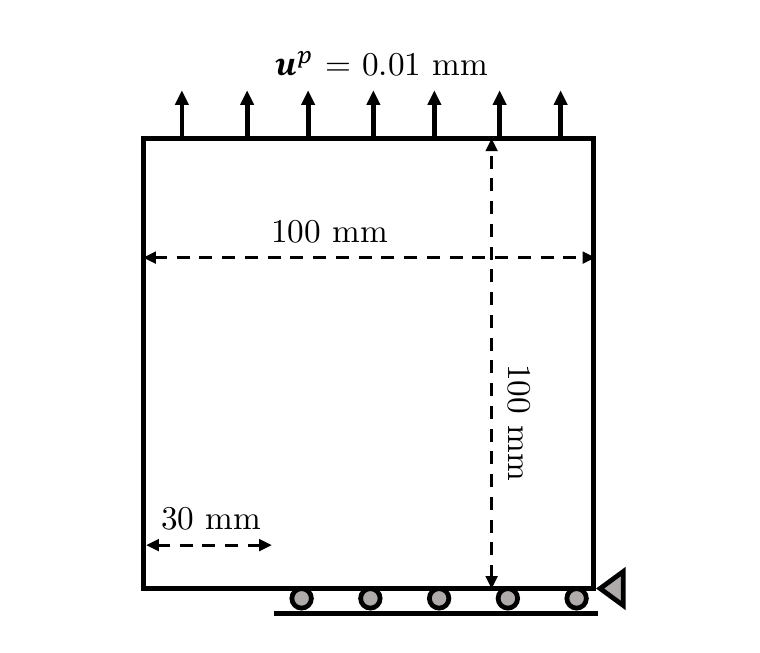}
    \caption{SSNT schematic}
    \label{fig:SSNT_schematic}
    \end{subfigure}
    \hfill
    \begin{subfigure}{8.2cm}
    \centering     
    \includegraphics[width=0.8\textwidth,trim = 13cm 5.75cm 13cm 5cm, clip]{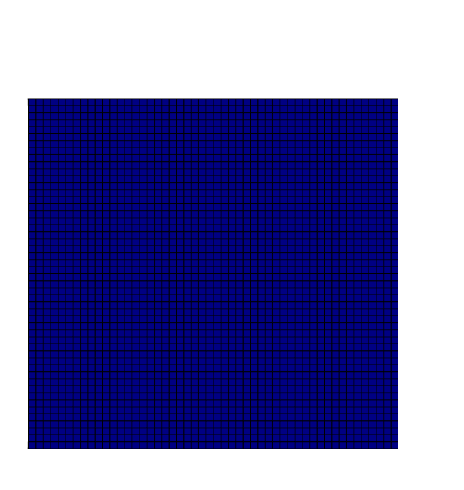}
    \caption{SSNT Mesh}
    \label{fig:SSNT_mesh}
    \end{subfigure}
    \caption{A schematic representation of the (a) geometry and (b) mesh of the SSNT problem. The structured coarse mesh displayed here has 2500 quadrilateral elements.}
    \label{fig:SSNT_schematic&mesh}
\end{figure}

\begin{figure}[H]
    \centering
    \begin{minipage}{15cm}
    \centering     
    \includegraphics[width=\textwidth,trim =0cm 6cm 0cm 5cm, clip]{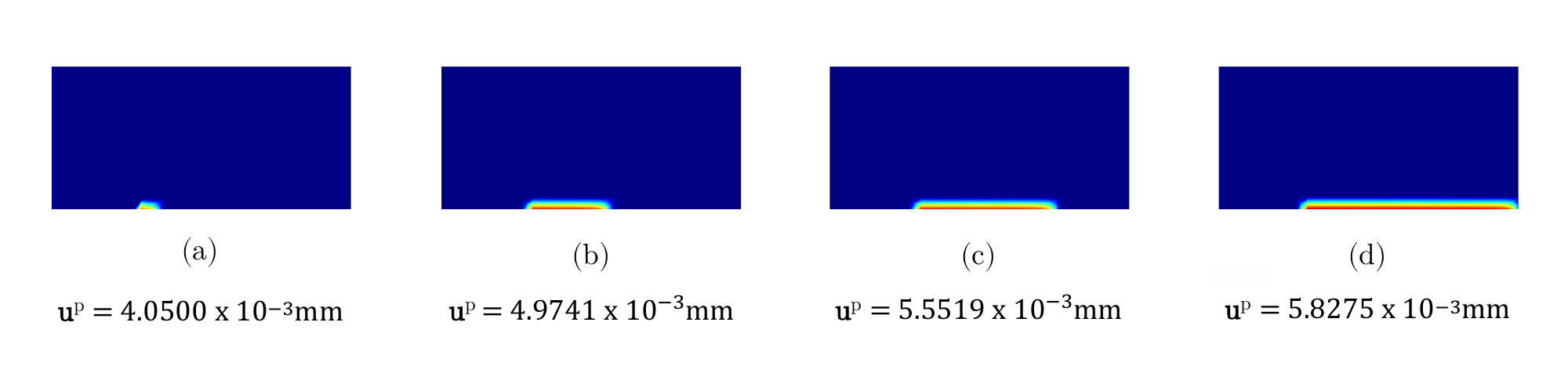}
    \end{minipage}
    \caption{Damage propagation contours for the SSNT problem at several representative load increments.}
    \label{fig:SSNT_damage_evolution_contours}
\end{figure}

The Symmetric Single Notch Tension (SSNT) problem is the second 2D example presented in this paper, following the work of Pantidis et al. \cite{pantidis2023integrated}. Here, a rectangular domain with dimensions  $100mm \times 100mm$ and a structured mesh is chosen. Two mesh sizes of 2500 (Coarse) and 6400 (Fine) elements are studied, with an element size of  $2mm \times 2mm$ and  $1.25mm \times 1.25mm$ respectively. The bottom of the domain is constrained using rollers as shown in Fig. \ref{fig:SSNT_schematic} and a tensile displacement load of 0.01 mm is applied at the top of the domain.  The material parameters chosen are shear modulus $\mu$ = 125 MPa and Poisson's ratio $\nu$ = 0.2. The Mazars law parameters are $\epsilon_D$ = $10^{-4}$, $\mathscr{A}$ = 0.7 and $\mathscr{B}$ = $10^4$. Plain strain conditions are considered, and the convergence tolerance for the UAL and NR solvers are set to $10^{-8}$ and $10^{-6}$ respectively. The initial values of $\Delta l$  and $\lambda$ are set to $10^{-4}$ and the PC scheme is used to implement the UAL model for the local and non-local gradient damage law case respectively. For the non-local gradient damage law case, $l_{c} = 5mm$ is chosen for both mesh sizes. Fig. \ref{fig:SSNT_damage_evolution_contours} displays the evolution of damage in the SSNT problem, from initiation to propagation until the domain boundary, for different displacement values. 

\begin{figure}[H]
    \centering
    \begin{subfigure}{8.2cm}
    \centering     
    \includegraphics[width=\textwidth,trim = 2cm 7cm 2cm 7cm, clip]{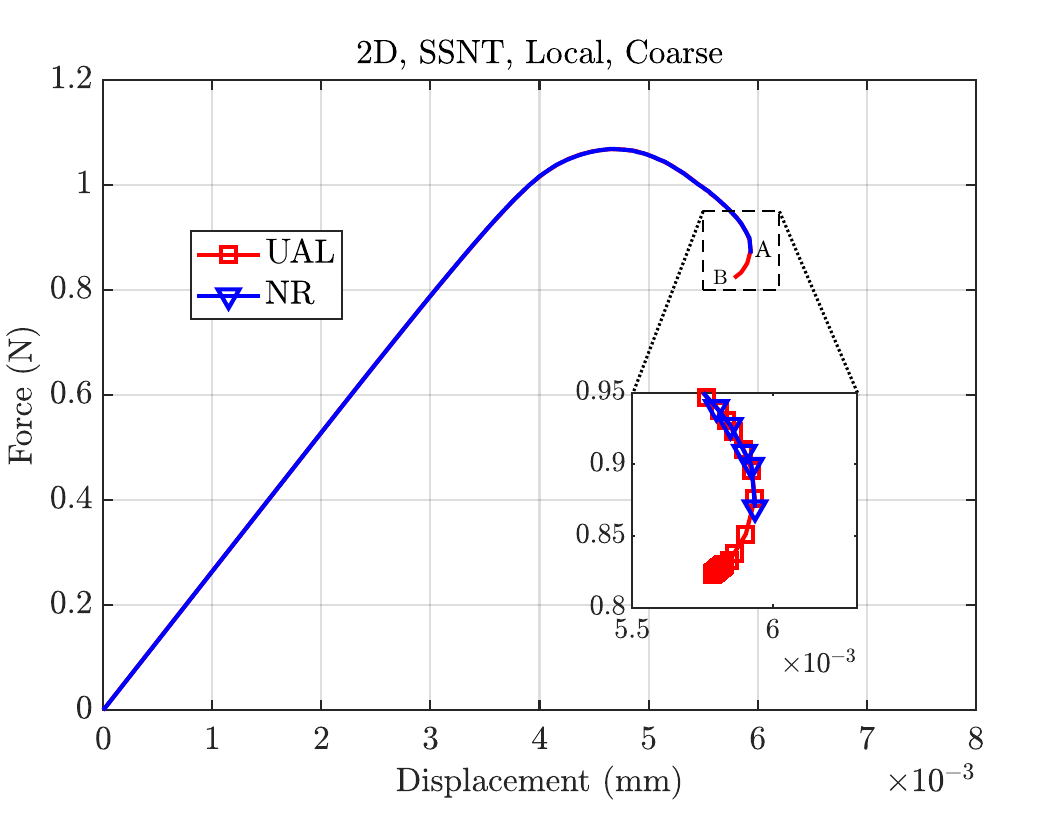}
   \end{subfigure}
   \centering
    \begin{subfigure}{8.2cm}
    \centering     
    \includegraphics[width=\textwidth,trim = 14cm 6cm 0cm 5cm, clip]{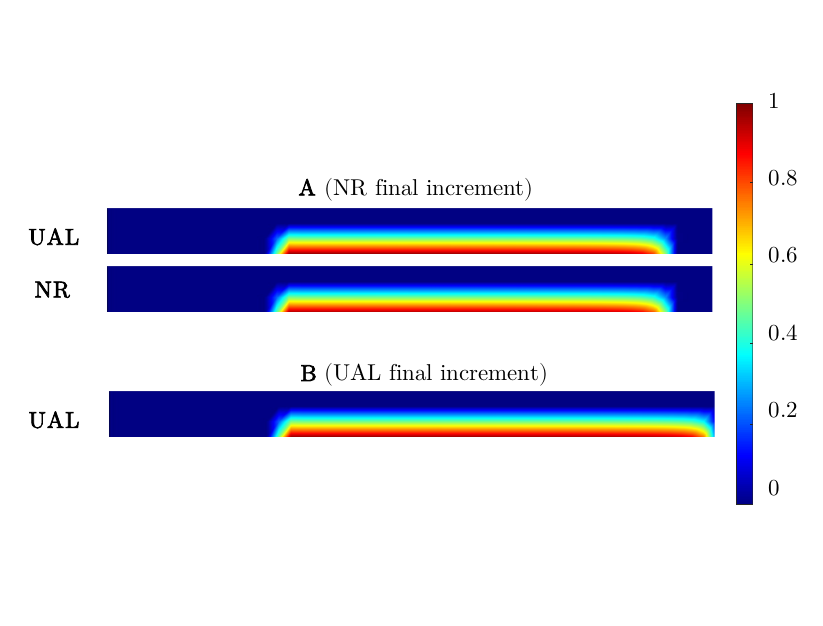}
   \end{subfigure}
    \caption{Comparison of NR and UAL using the local damage law for the SSNT problem. Damage propagation contours for different points on the equilibrium path and the force-displacement crossplots show that UAL outperforms NR.}
    \label{fig:SSNT_validation}
\end{figure}

First, a comparative study of the performance of UAL and NR was conducted and the results are shown in Fig. \ref{fig:SSNT_validation}. The reaction-displacement curves are shown on the left side of Fig. \ref{fig:SSNT_validation} and the damage contours captured at the last converged increment of each solver are displayed on the right side of Fig. \ref{fig:SSNT_validation}. The NR solver advances only until point A, after which the snap-back response of the structure ensues. NR fails to capture the snap-back feature, as was the case in 1D problems and terminates due to convergence issues. The UAL solver successfully advances until point B, traces the snap-back response of the structure, and is able to capture the propagation of damage until the right end of the domain. This is a clear example of the robustness and accuracy of UAL compared to conventional methods.

\begin{figure}[H]
    \centering
    \begin{subfigure}{8.2cm}
    \centering     
    \includegraphics[width=1\textwidth,trim = 1cm 7cm 2cm 7cm, clip]{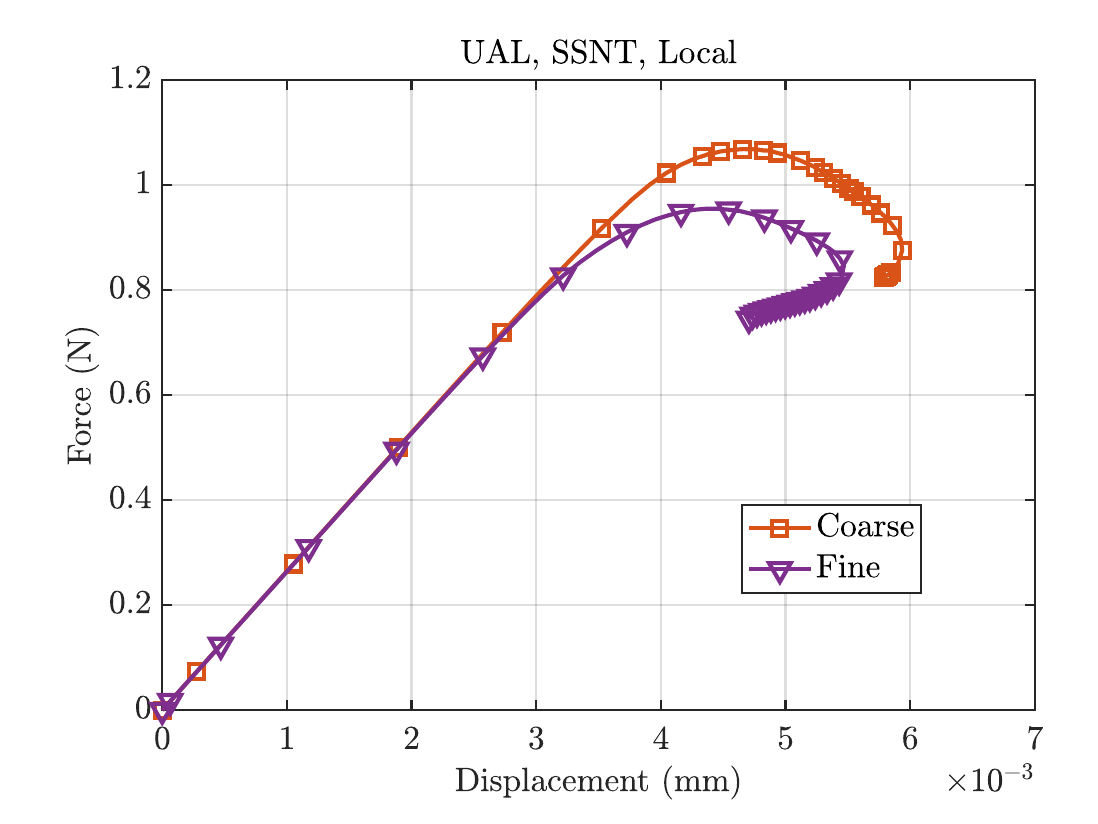}
    \caption{UAL, SSNT, Local}
    \label{fig:SSNT_Local_2mesh_crossplot}
    \end{subfigure}
    \hfill
    \centering
    \begin{subfigure}{8.2cm}
    \centering     
    \includegraphics[width=1\textwidth,trim = 1cm 7cm 2cm 7cm, clip]{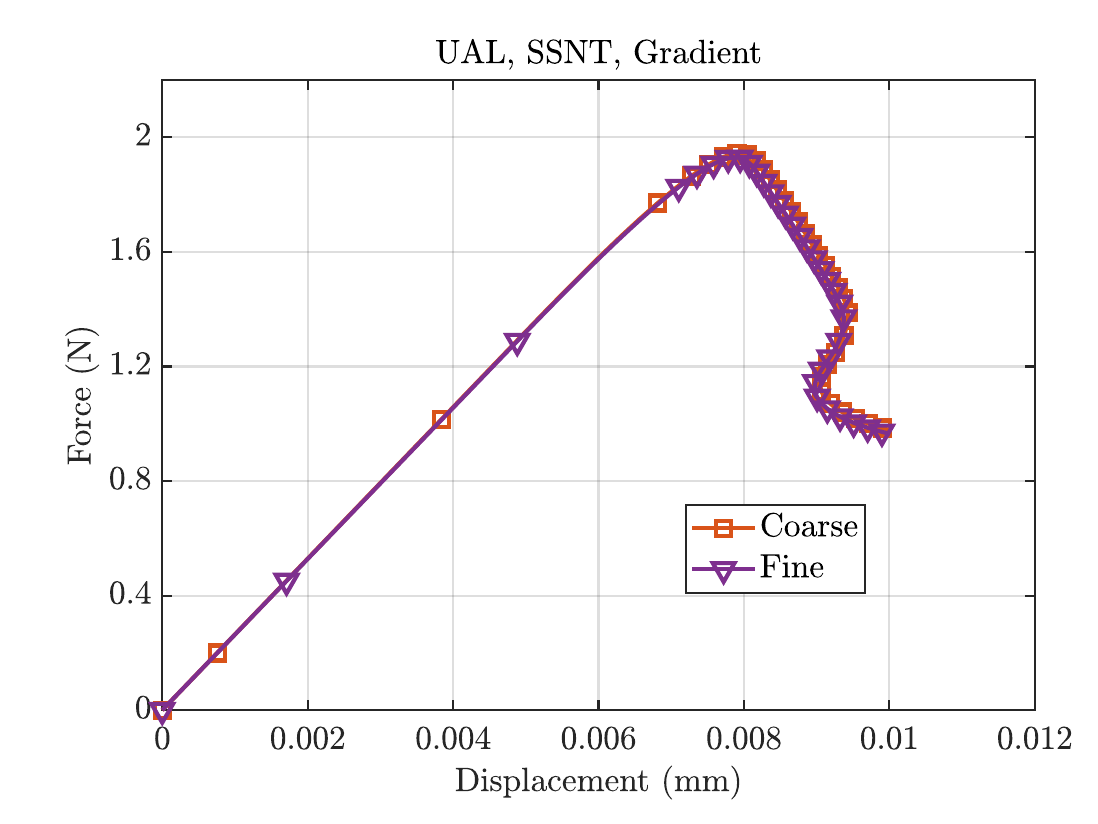}
    \caption{UAL, SSNT, Gradient}
    \label{fig:SSNT_NonLocal_2mesh_crossplot}
    \end{subfigure}
    \caption{Comparison of the mesh resolution impact for the local and non-local gradient damage laws in the SSNT problem. In the local damage case, the equilibrium paths diverge once damage initiates, while the behaviours in the non-local case are clearly mesh independent.}
    \label{fig:SSNT_Local_Gradient_mesh}
\end{figure}

Having established the strengths of UAL in this problem, we proceed with an additional investigation study in order to compare its performance using different mesh resolutions and damage laws. Fig. \ref{fig:SSNT_Local_Gradient_mesh} shows the results of study, where the UAL is applied at the Coarse and Fine discretizations using the local (Fig. \ref{fig:SSNT_Local_2mesh_crossplot}) and non-local gradient (Fig. \ref{fig:SSNT_NonLocal_2mesh_crossplot}) models. In the first case, the smaller element size of the Fine model results in a more pronounced strain localization, and therefore this model reaches a lower capacity than the Coarse model and a different equilibrium path beyond the damage initiation point. In the second case, the mesh-objectivity of the non-local gradient damage law is clearly observed, since the reaction-displacement curves of both models overlap throughout the entire analysis. Both of these trends are consistent with the literature \cite{pijaudier1987nonlocal,peerlings1996gradient}, which further verifies the validity of the UAL solver.

\subsubsection{Two Notch Tension (TNT) problem}
\label{NumEx:TNT}

\begin{figure}[H]
    \centering
    \begin{subfigure}{8.2cm}
    \centering    
    \includegraphics[width=0.8\textwidth,trim = 12cm 4cm 12cm 4cm, clip]{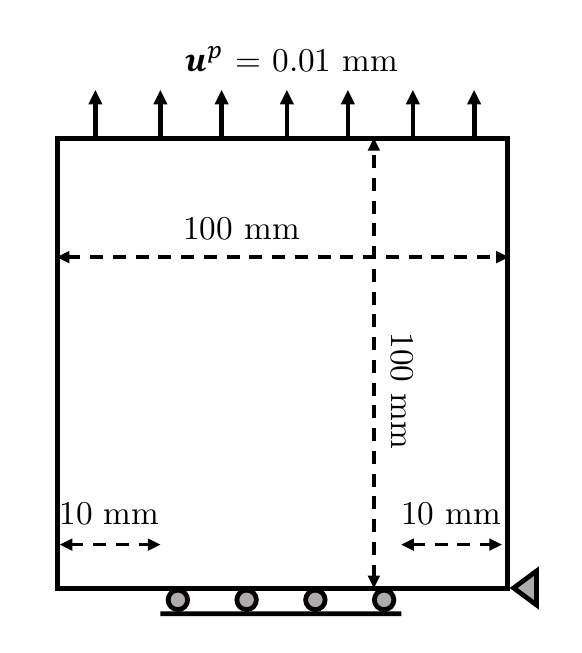}
    \caption{Two Notch Tension schematic}
    \label{fig:TNT_schematic}
    \end{subfigure}
    \hfill
    \begin{subfigure}{8.2cm}
    \centering     
    \includegraphics[width=0.9\textwidth,trim = 12cm 5.5cm 12cm 5cm, clip]{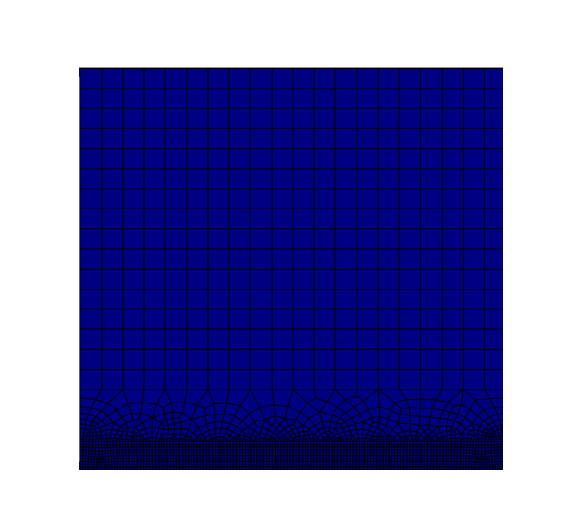}
    \caption{Two Notch Tension Mesh}
    \label{fig:TNT_mesh}
    \end{subfigure}
    \caption{A schematic illustration of the geometry and mesh of the TNT problem; the unstructured mesh displayed here has 2347 elements.}
    \label{fig:TNT_schematic&mesh}
\end{figure}

\begin{figure}[H]
    \centering
    \begin{minipage}{15cm}
    \centering     
    \includegraphics[width=\textwidth,trim =0cm 6cm 0cm 7cm, clip]{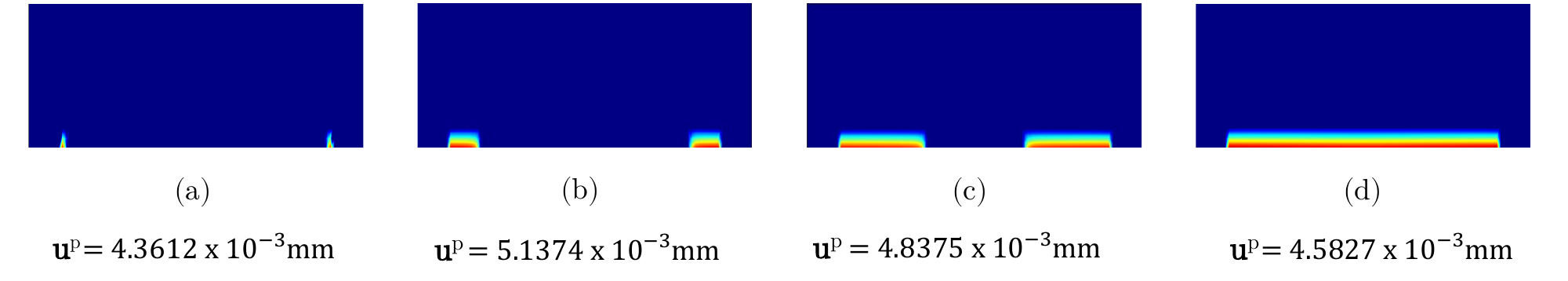}
    \end{minipage}
    \caption{Damage propagation contours for the TNT problem at several representative load increments.}
    \label{fig:TNT_damage_evolution_contours}
\end{figure}

The third 2D example presented in this section is the Two Notch Tension (TNT) problem, a modification of the model explored in the work by Peng et al.\cite{peng2020phase}. The domain has dimensions $100mm \times 100mm$ and an unstructured mesh, as shown in Fig. \ref{fig:TNT_schematic&mesh}. One mesh resolution of 2347 (Coarse) elements is studied, with an element size of $0.7mm \times 0.7mm $. The bottom of the domain is constrained using rollers as shown in Fig. \ref{fig:TNT_schematic}. The material parameters, damage parameters, initial $\Delta l$  and $\lambda$ values and convergence tolerance used here are the same as that in Section \ref{NumEx:SSNT}. The PC scheme is used to implement the UAL model for the TNT problem in the local damage law case only. Fig. \ref{fig:TNT_damage_evolution_contours} displays the contours of damage in the TNT problem at different displacement values. 
 
\begin{figure}[H]
    \centering
    \begin{subfigure}{8.2cm}
    \centering     
    \includegraphics[width=\textwidth,trim = 2cm 7cm 2cm 7cm, clip]{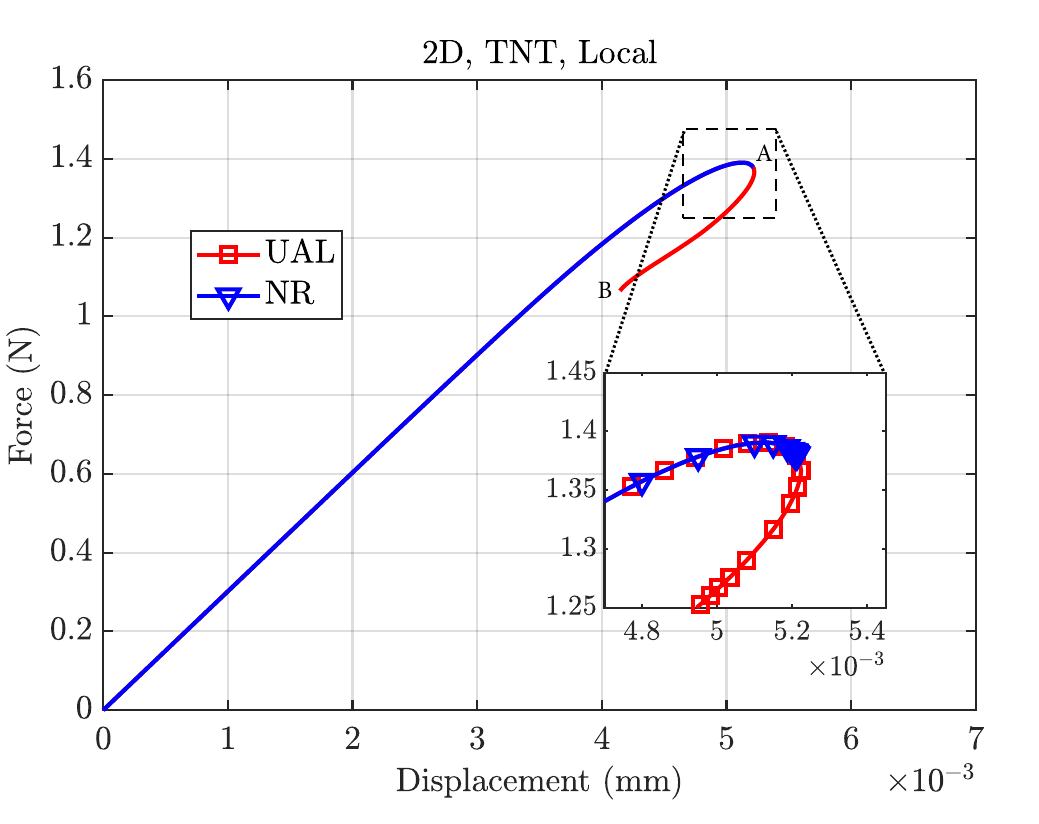}
    \end{subfigure}
    \centering
    \begin{subfigure}{8.2cm}
    \centering     
    \includegraphics[width=\textwidth,trim =14cm 7cm 0cm 5cm, clip]{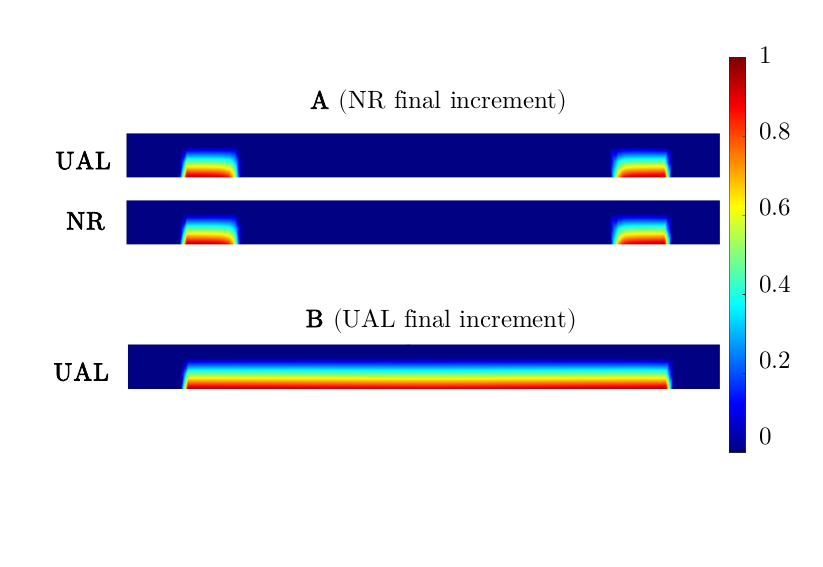}
    \end{subfigure}
    \caption{Comparison of NR and UAL equilibrium paths for the TNT problem using the local damage law. The damage propagation contours and the force-displacement curves demonstrate that the UAL algorithm proceeds well-beyond the final point of NR, being able to capture the entire equilibrium path.}
    \label{fig:DNH_damage}
\end{figure}

A comparison of the response and damage contour propagation between the NR and UAL algorithms are shown in Fig.\ref{fig:DNH_damage}. Similar to the previous examples, NR does not progress beyond the inflection point where the snap-back behavior initiates. In the UAL case however, the damage progresses from both sides of the domain until the two damage paths merge in the middle. Here we emphasize that the UAL analysis continues until the point where damage has fully propagated over the entire bottom side of the domain, and the solver is capable of capturing the entire snap-back response without numerical issues.

\subsection{Single Notch Shear (SNS) problem}
\label{NumEx:SNS}

\begin{figure}[H]
    \centering
    \begin{subfigure}{8.2cm}
    \centering     
    \includegraphics[width=1\textwidth,trim = 11cm 4cm 11cm 4cm, clip]{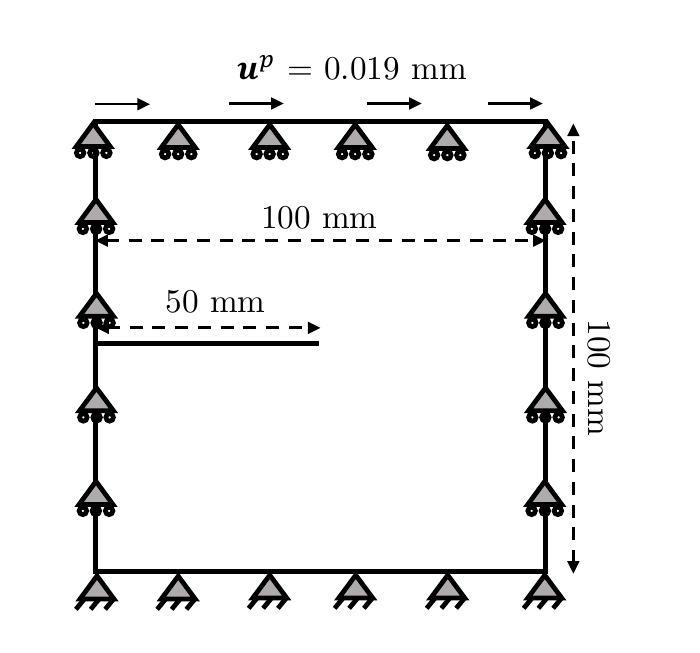}
    \caption{SNS schematic}
    \end{subfigure}
    \hfill
    \centering
    \begin{subfigure}{8.2cm}
    \centering     
    \includegraphics[width=1\textwidth,trim = 11cm 4.5cm 11cm 4.5cm, clip]{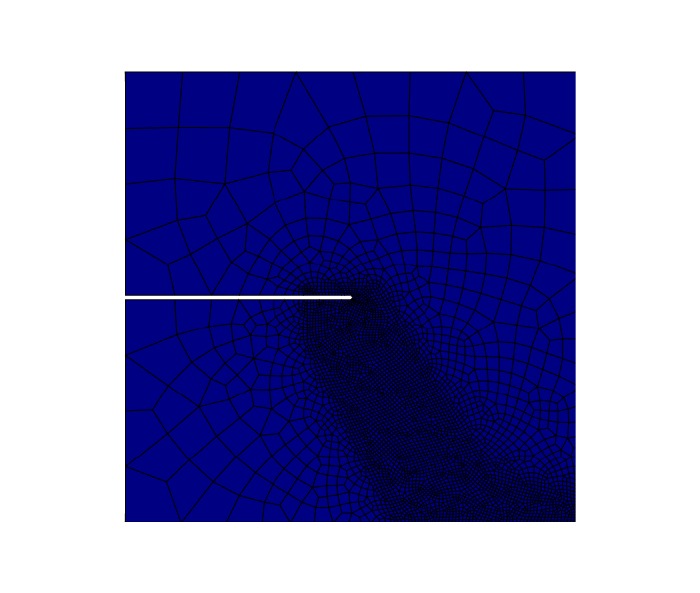}
    \caption{SNS Mesh}
    \end{subfigure}
    \caption{A schematic illustration of the (a) geometry and (b) mesh of the SNS problem. The unstructured mesh displayed here has 4129 quadrilateral elements.}
    \label{fig:SNS Schematic & mesh}
\end{figure}

\begin{figure}[H]
    \centering
    \begin{minipage}{15cm}
    \centering     
    \includegraphics[width=\textwidth,trim =0cm 4cm 0cm 6cm, clip]{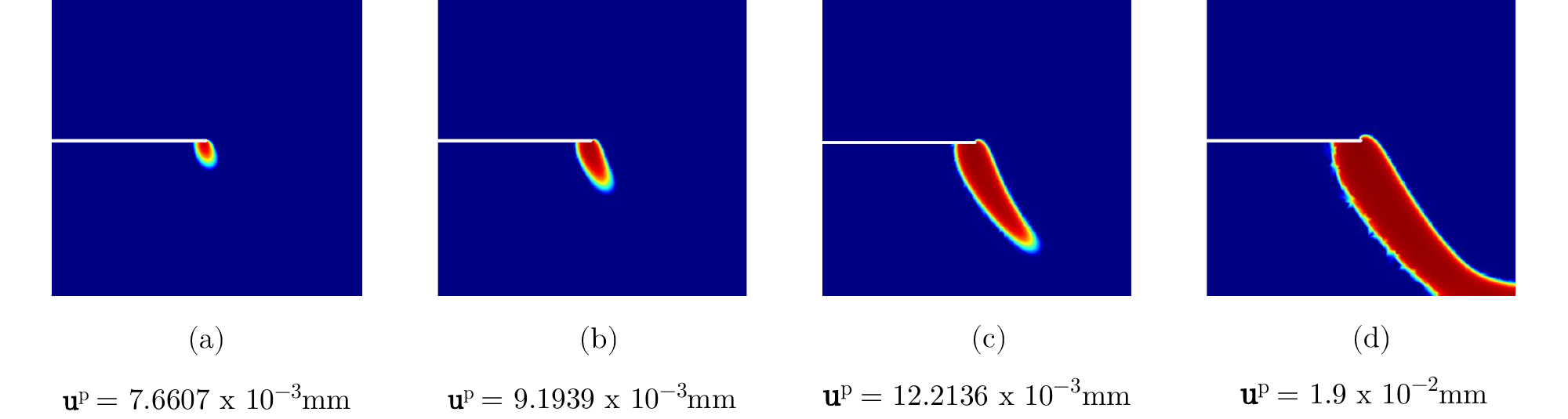}
    \end{minipage}
    \caption{Damage propagation contours for the SNS problem at several representative load increments.}
   \label{fig:SNS_damage_evolution_contours}
\end{figure}

The last 2D example presented is the Single Notch Shear (SNS) problem, and the objective here is to test the UAL solver against a problem involving shear loads. A similar geometry was studied by Treifi et al. \cite{treifi2009computations}, Miehe et al.\cite{miehe2010phase} and Zhou et al.\cite{zhou2018phase}. The domain in the SNS problem has dimensions $100mm \times 100mm$  with an unstructured mesh as shown in Fig. \ref{fig:SNS Schematic & mesh}. A mesh of 4129 elements with a notch extending to half the domain is chosen. The size of the elements in the refined zone are $0.75mm \times 0.75mm$ and the $l_c$ has a value of $2mm$. A displacement shear load of $0.019mm$ is applied on the top surface of the domain. The material parameters, damage parameters, initial $\Delta l$  and $\lambda$ values, and convergence tolerance are the same as that in Section \ref{NumEx:SSNT}. To account for the influence of shear, the equivalent strain definition in this example is based on Eqn. \eqref{ApxEq:Equivalent_strain_shear} presented in \ref{Appendix: Mazar_Damage_model}. The SNS problem is solved using the PC scheme for the non-local gradient damage law. Fig. \ref{fig:SNS_damage_evolution_contours} displays the evolution of damage within the SNS domain at different displacement values.

\begin{figure}[H]
    \centering
    \begin{subfigure}{8.2cm}
    \centering     
    \includegraphics[width=\textwidth,trim = 2cm 7cm 2cm 7cm, clip]
    {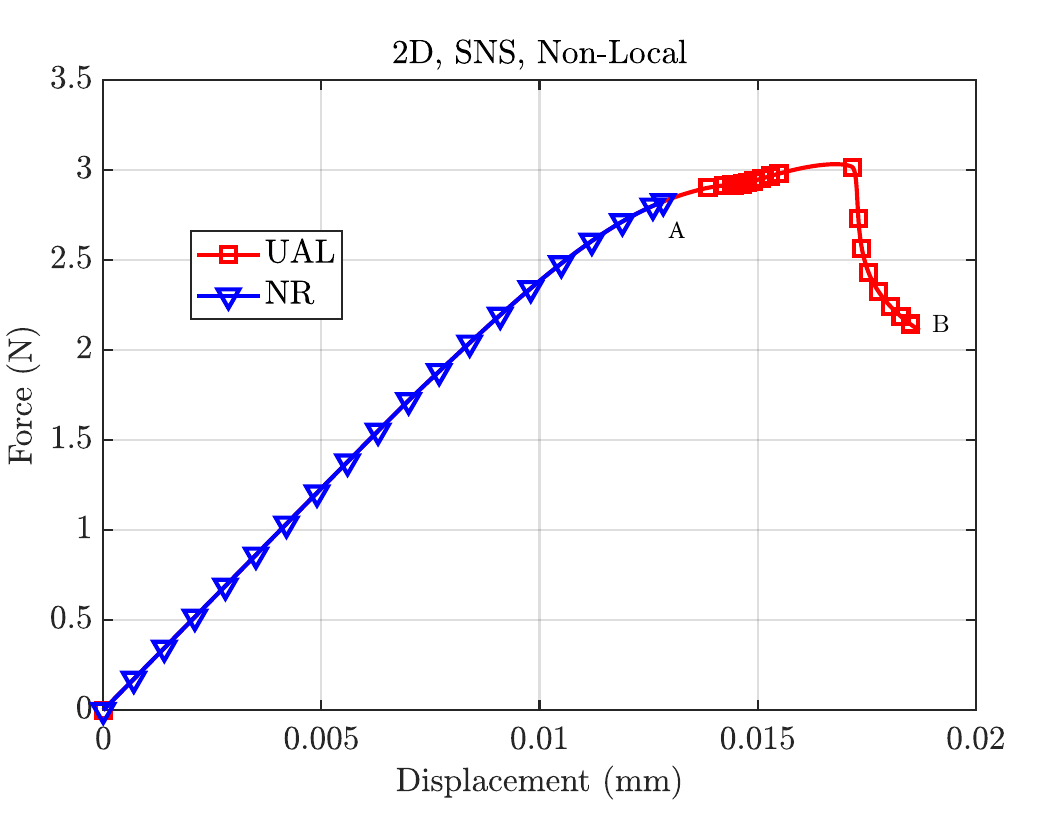}
    \end{subfigure}
    \begin{subfigure}{8.2cm}
    \centering     
    \includegraphics[width=\textwidth,trim = 15cm 2cm 0cm 0cm, clip]
    {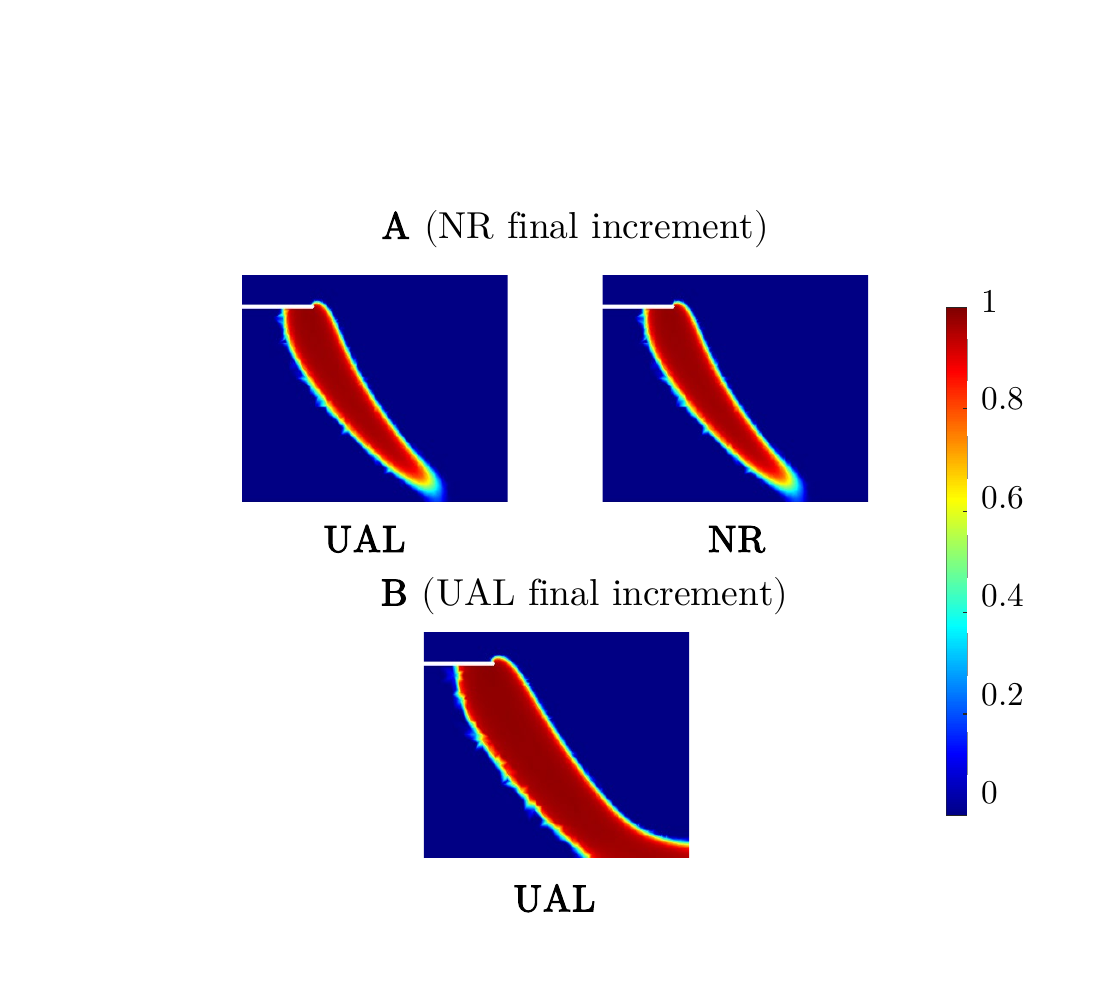}
    \end{subfigure}
    \caption{Comparison of NR and UAL equilibrium paths for the SNS problem using the non-local gradient damage law. Force-displacement curves and damage propagation contours show that the UAL analysis proceeds much further along the equilibrium path than NR.}
    \label{fig:SNS_fd_damage_contours}
\end{figure}

Fig. \ref{fig:SNS_fd_damage_contours} presents the results of the study comparing the performance of UAL and NR for the SNS problem. The response of both solvers is identical until the point where NR fails to converge due to numerical issues. The UAL algorithm once again is more robust, and the analysis continues until damage has fully reached the bottom right boundary of the domain. These results add further confidence to the developed framework, since they demonstrate the ability of UAL to model damage propagation problems involving shear loading as well.

Overall, the results reported in this subsection concur with the already established observations of the 1D bar problem, and the key takeaway points from this investigation are: 

\begin{itemize}

    \item The strengths of the UAL algorithm demonstrated in Section \ref{1D_bar_problem} versus conventional solvers carry forth to 2D problems as well. UAL is substantially more robust than NR, being able to proceed beyond the critical points and providing insight on the entire equilibrium path and damage propagation within the domain.
    
    \item The control parameters discussed in Section \ref{Control_parameters} provide a valuable tool for the end-user to improve the performance of UAL in very challenging problems with highly localized damage.
    
    \item The UAL framework is applicable to both mode  I (pure tension) and mode II (pure shear) type loading scenarios. 

\end{itemize}

\section{Summary and conclusions}
\label{Sec:Conclusion}

This work presents a novel unified arc-length (UAL) method that targets the solution of continuum damage mechanics problems. We have derived the definition of the consistent tangent matrix for both the local and the non-local gradient damage laws, and we have presented the details of appropriate implementation schemes and algorithms for each case. We have implemented the proposed methodology in a series of benchmark 1D and 2D numerical examples with different geometries and loading conditions, and we have compared its performance against the two prevailing monolithic numerical solvers, the Newton-Raphson (NR) and the force-controlled arc-length (FAL). The results of this work clearly highlight the superiority of UAL over the other two conventional approaches, where UAL is 1-2 orders of magnitude faster than FAL and NR. Also, even though we did not perform an explicit comparison, it is expected that the UAL will be vastly faster than staggered schemes, given the well-established additional expense of these schemes against their monolithic counterparts. Whereas NR is incapable of capturing snap-back behaviors and FAL requires tremendous computational time to converge around the critical points, UAL is able to trace the entire equilibrium path, even at the presence of severe snap-back areas. The key element supporting the UAL efficiency is the fact that it treats all the entries of the external force vector as unknown nodal quantities, allowing for different variations in its entries compared to the monolithic changes enforced with FAL. Additionally, numerical issues such as backtracking, which are observed in conventional arc-length based solvers, are overcome with UAL via appropriate implementation schemes and control parameters, as extensively discussed in this work. This work paves the way for the adoption of the proposed UAL solver in real-world highly non-linear engineering problems, in lieu of current methods which suffer from the accuracy-speed trade-off.

\section*{Data availability}
\label{Sec:Data availability}
The code and datasets used in this work will be made publicly available upon publication of the article.  

\section*{Acknowledgements}
\label{Sec:Acknowledgements}
This work was partially supported by the Sand Hazards and Opportunities for Resilience, Energy, and Sustainability (SHORES) Center, funded by Tamkeen under the NYUAD Research Institute.

\bibliography{bibliography}

\begin{thebibliography}{100}
\expandafter\ifx\csname url\endcsname\relax
  \def\url#1{\texttt{#1}}\fi
\expandafter\ifx\csname urlprefix\endcsname\relax\def\urlprefix{URL }\fi
\expandafter\ifx\csname href\endcsname\relax
  \def\href#1#2{#2} \def\path#1{#1}\fi

\bibitem{lemaitre2012course}
J.~Lemaitre, A course on damage mechanics, Springer Science \& Business Media, 2012.

\bibitem{li2001continuum}
F.~Li, Z.~Li, Continuum damage mechanics based modeling of fiber reinforced concrete in tension, International journal of solids and structures 38~(5) (2001) 777--793.

\bibitem{mobasher2017non}
M.~E. Mobasher, L.~Berger-Vergiat, H.~Waisman, Non-local formulation for transport and damage in porous media, Computer Methods in Applied Mechanics and Engineering 324 (2017) 654--688.

\bibitem{mobasher2021dual}
M.~E. Mobasher, H.~Waisman, Dual length scale non-local model to represent damage and transport in porous media, Computer Methods in Applied Mechanics and Engineering 387 (2021) 114154.

\bibitem{bonora1998low}
N.~Bonora, G.~Newaz, Low cycle fatigue life estimation for ductile metals using a nonlinear continuum damage mechanics model, International Journal of Solids and Structures 35~(16) (1998) 1881--1894.

\bibitem{voyiadjis2002modeling}
G.~Z. Voyiadjis, A.~Palazotto, X.~Gao, Modeling of metallic materials at high strain rates with continuum damage mechanics, Appl. Mech. Rev. 55~(5) (2002) 481--493.

\bibitem{talreja2006multi}
R.~Talreja, Multi-scale modeling in damage mechanics of composite materials, Journal of materials science 41 (2006) 6800--6812.

\bibitem{williams2001application}
K.~V. Williams, R.~Vaziri, Application of a damage mechanics model for predicting the impact response of composite materials, Computers \& Structures 79~(10) (2001) 997--1011.

\bibitem{zioupos1998recent}
P.~Zioupos, Recent developments in the study of failure of solid biomaterials and bone:‘fracture’and ‘pre-fracture’toughness, Materials Science and Engineering: C 6~(1) (1998) 33--40.

\bibitem{bjorstad2001domain}
P.~E. Bj{\o}rstad, J.~Koster, P.~Krzy{\.z}anowski, Domain decomposition solvers for large scale industrial finite element problems, in: Applied Parallel Computing. New Paradigms for HPC in Industry and Academia: 5th International Workshop, PARA 2000 Bergen, Norway, June 18--20, 2000 Proceedings 5, Springer, 2001, pp. 373--383.

\bibitem{iankov2003finite}
R.~Iankov, Finite element simulation of profile rolling of wire, Journal of Materials Processing Technology 142~(2) (2003) 355--361.

\bibitem{oden2006finite}
J.~T. Oden, Finite elements of nonlinear continua, Courier Corporation, 2006.

\bibitem{ghrib1995nonlinear}
F.~Ghrib, R.~Tinawi, Nonlinear behavior of concrete dams using damage mechanics, Journal of Engineering Mechanics 121~(4) (1995) 513--527.

\bibitem{giancane2010fatigue}
S.~Giancane, R.~Nobile, F.~Panella, V.~Dattoma, Fatigue life prediction of notched components based on a new nonlinear continuum damage mechanics model, Procedia Engineering 2~(1) (2010) 1317--1325.

\bibitem{cocks1989inelastic}
A.~Cocks, Inelastic deformation of porous materials, Journal of the Mechanics and Physics of Solids 37~(6) (1989) 693--715.

\bibitem{eiger1984bisection}
A.~Eiger, K.~Sikorski, F.~Stenger, A bisection method for systems of nonlinear equations, ACM Transactions on Mathematical Software (TOMS) 10~(4) (1984) 367--377.

\bibitem{demir2008trisection}
A.~Demir, Trisection method by k-lucas numbers, Applied mathematics and computation 198~(1) (2008) 339--345.

\bibitem{brouwer1911abbildung}
L.~E.~J. Brouwer, {\"U}ber abbildung von mannigfaltigkeiten, Mathematische annalen 71~(1) (1911) 97--115.

\bibitem{azure2019comparative}
I.~Azure, G.~Aloliga, L.~Doabil, Comparative study of numerical methods for solving non-linear equations using manual computation, Math. Lett 5~(4) (2019) 41--46.

\bibitem{zeid1985fixed}
I.~Zeid, Fixed-point iteration to nonlinear finite element analysis. part ii: Formulation and implementation, International journal for numerical methods in engineering 21~(11) (1985) 2049--2069.

\bibitem{allgower2003introduction}
E.~L. Allgower, K.~Georg, Introduction to numerical continuation methods, SIAM, 2003.

\bibitem{allgower2012numerical}
E.~L. Allgower, K.~Georg, Numerical continuation methods: an introduction, Vol.~13, Springer Science \& Business Media, 2012.

\bibitem{rheinboldt1983locally}
W.~C. Rheinboldt, J.~V. Burkardt, A locally parameterized continuation process, ACM Transactions on Mathematical Software (TOMS) 9~(2) (1983) 215--235.

\bibitem{bathe1975finite}
K.-J. Bathe, E.~Ramm, E.~L. Wilson, Finite element formulations for large deformation dynamic analysis, International journal for numerical methods in engineering 9~(2) (1975) 353--386.

\bibitem{hughes1978consistent}
T.~J. Hughes, K.~S. Pister, Consistent linearization in mechanics of solids and structures, Computers \& Structures 8~(3-4) (1978) 391--397.

\bibitem{hartmann2005remark}
S.~Hartmann, A remark on the application of the newton-raphson method in non-linear finite element analysis, Computational Mechanics 36 (2005) 100--116.

\bibitem{ehiwario2014comparative}
J.~Ehiwario, S.~Aghamie, Comparative study of bisection, newton-raphson and secant methods of root-finding problems, IOSR Journal of Engineering 4~(4) (2014) 01--07.

\bibitem{jennings1971accelerating}
A.~Jennings, Accelerating the convergence of matrix iterative processes, IMA Journal of Applied Mathematics 8~(1) (1971) 99--110.

\bibitem{nayak1972note}
G.~Nayak, O.~Zienkiewicz, Note on the ‘alpha’-constant stiffness method for the analysis of non-linear problems, International Journal for Numerical Methods in Engineering 4~(4) (1972) 579--582.

\bibitem{wolfe1959secant}
P.~Wolfe, The secant method for simultaneous nonlinear equations, Communications of the ACM 2~(12) (1959) 12--13.

\bibitem{nordmann2019damage}
J.~Nordmann, K.~Naumenko, H.~Altenbach, A damage mechanics based cohesive zone model with damage gradient extension for creep-fatigue-interaction, Key Engineering Materials 794 (2019) 253--259.

\bibitem{riks1972arclen}
E.~Riks, {The Application of Newton’s Method to the Problem of Elastic Stability}, Journal of Applied Mechanics 39~(4) (1972) 1060--1065.

\bibitem{riks1979incremental}
E.~Riks, An incremental approach to the solution of snapping and buckling problems, International journal of solids and structures 15~(7) (1979) 529--551.

\bibitem{wempner1971discrete}
G.~A. Wempner, Discrete approximations related to nonlinear theories of solids, International Journal of Solids and Structures 7~(11) (1971) 1581--1599.

\bibitem{crisfield1981fast}
M.~A. Crisfield, A fast incremental/iterative solution procedure that handles “snap-through”, in: Computational methods in nonlinear structural and solid mechanics, Elsevier, 1981, pp. 55--62.

\bibitem{bathe1983automatic}
K.-J. Bathe, E.~N. Dvorkin, On the automatic solution of nonlinear finite element equations, Computers \& Structures 17~(5-6) (1983) 871--879.

\bibitem{bellini1987improved}
P.~Bellini, A.~Chulya, An improved automatic incremental algorithm for the efficient solution of nonlinear finite element equations, Computers \& Structures 26~(1-2) (1987) 99--110.

\bibitem{de1999determination}
E.~de~Souza~Neto, Y.~Feng, On the determination of the path direction for arc-length methods in the presence of bifurcations andsnap-backs', Computer methods in applied mechanics and engineering 179~(1-2) (1999) 81--89.

\bibitem{park1982family}
K.~Park, A family of solution algorithms for nonlinear structural analysis based on relaxation equations, International Journal for Numerical Methods in Engineering 18~(9) (1982) 1337--1347.

\bibitem{simo1986finite}
J.~Simo, P.~Wriggers, K.~Schweizerhof, R.~Taylor, Finite deformation post-buckling analysis involving inelasticity and contact constraints, International journal for numerical methods in engineering 23~(5) (1986) 779--800.

\bibitem{skeie1990local}
G.~Skeie, C.~Felippa, A local hyperelliptic constraint for nonlinear analysis, in: Proc. 3rd International Conference on Numerical Methods in Engineering: Theory and Apphcations, ed. by GN Pande and J. Middleton, Swansea (UK), 1990, pp. 13--27.

\bibitem{forde1987improved}
B.~W. Forde, S.~F. Stiemer, Improved arc length orthogonality methods for nonlinear finite element analysis, Computers \& structures 27~(5) (1987) 625--630.

\bibitem{de2012nonlinear}
R.~De~Borst, M.~A. Crisfield, J.~J. Remmers, C.~V. Verhoosel, Nonlinear finite element analysis of solids and structures, John Wiley \& Sons, 2012.

\bibitem{may2016new}
S.~May, J.~Vignollet, R.~de~Borst, A new arc-length control method based on the rates of the internal and the dissipated energy, Engineering Computations 33~(1) (2016) 100--115.

\bibitem{singh2016fracture}
N.~Singh, C.~Verhoosel, R.~De~Borst, E.~Van~Brummelen, A fracture-controlled path-following technique for phase-field modeling of brittle fracture, Finite Elements in Analysis and Design 113 (2016) 14--29.

\bibitem{bharali2022robust}
R.~Bharali, S.~Goswami, C.~Anitescu, T.~Rabczuk, A robust monolithic solver for phase-field fracture integrated with fracture energy based arc-length method and under-relaxation, Computer Methods in Applied Mechanics and Engineering 394 (2022) 114927.

\bibitem{memon2004arc}
B.-A. Memon, X.-z. Su, Arc-length technique for nonlinear finite element analysis, Journal of Zhejiang University-Science A 5 (2004) 618--628.

\bibitem{pretti2022displacement}
G.~Pretti, W.~M. Coombs, C.~E. Augarde, A displacement-controlled arc-length solution scheme, Computers \& Structures 258 (2022) 106674.

\bibitem{byrd1987global}
R.~H. Byrd, J.~Nocedal, Y.-X. Yuan, Global convergence of a class of quasi-newton methods on convex problems, SIAM Journal on Numerical Analysis 24~(5) (1987) 1171--1190.

\bibitem{dai2013perfect}
Y.-H. Dai, A perfect example for the bfgs method, Mathematical Programming 138 (2013) 501--530.

\bibitem{levenberg1944method}
K.~Levenberg, Method for the solution of certain problems in least squares siam, J Numer Anal 16 (1944) 588--A604.

\bibitem{marquardt1963algorithm}
D.~W. Marquardt, An algorithm for least-squares estimation of nonlinear parameters, Journal of the society for Industrial and Applied Mathematics 11~(2) (1963) 431--441.

\bibitem{kristensen2020phase}
P.~K. Kristensen, E.~Mart{\'\i}nez-Pa{\~n}eda, Phase field fracture modelling using quasi-newton methods and a new adaptive step scheme, Theoretical and Applied Fracture Mechanics 107 (2020) 102446.

\bibitem{farhat1991method}
C.~Farhat, F.-X. Roux, A method of finite element tearing and interconnecting and its parallel solution algorithm, International journal for numerical methods in engineering 32~(6) (1991) 1205--1227.

\bibitem{mobasher2016adaptive}
M.~E. Mobasher, H.~Waisman, Adaptive modeling of damage growth using a coupled fem/bem approach, International Journal for Numerical Methods in Engineering 105~(8) (2016) 599--619.

\bibitem{pebrel2008nonlinear}
J.~Pebrel, C.~Rey, P.~Gosselet, A nonlinear dual-domain decomposition method: Application to structural problems with damage, International Journal for Multiscale Computational Engineering 6~(3) (2008).

\bibitem{lloberas2011domain}
O.~Lloberas-Valls, D.~Rixen, A.~Simone, L.~Sluys, Domain decomposition techniques for the efficient modeling of brittle heterogeneous materials, Computer Methods in Applied Mechanics and Engineering 200~(13-16) (2011) 1577--1590.

\bibitem{shaidurov2013multigrid}
V.~V. Shaidurov, Multigrid methods for finite elements, Vol. 318, Springer Science \& Business Media, 2013.

\bibitem{rosam2008adaptive}
J.~Rosam, P.~K. Jimack, A.~Mullis, An adaptive, fully implicit multigrid phase-field model for the quantitative simulation of non-isothermal binary alloy solidification, Acta Materialia 56~(17) (2008) 4559--4569.

\bibitem{miehe2010thermodynamically}
C.~Miehe, F.~Welschinger, M.~Hofacker, Thermodynamically consistent phase-field models of fracture: Variational principles and multi-field fe implementations, International journal for numerical methods in engineering 83~(10) (2010) 1273--1311.

\bibitem{hofacker2012continuum}
M.~Hofacker, C.~Miehe, Continuum phase field modeling of dynamic fracture: variational principles and staggered fe implementation, International journal of fracture 178 (2012) 113--129.

\bibitem{crisfield1983arc}
M.~Crisfield, An arc-length method including line searches and accelerations, International journal for numerical methods in engineering 19~(9) (1983) 1269--1289.

\bibitem{hellweg1998new}
H.-B. Hellweg, M.~Crisfield, A new arc-length method for handling sharp snap-backs, Computers \& Structures 66~(5) (1998) 704--709.

\bibitem{crisfield1982local}
M.~Crisfield, TRRL, Local instabilities in the non-linear analysis of reinforced concrete beams and slabs., Proceedings of the Institution of Civil Engineers 73~(1) (1982) 135--145.

\bibitem{peerlings1996gradient}
R.~H. Peerlings, R.~de~Borst, W.~M. Brekelmans, J.~de~Vree, Gradient enhanced damage for quasi-brittle materials, International Journal for numerical methods in engineering 39~(19) (1996) 3391--3403.

\bibitem{murakami2012continuum}
S.~Murakami, Continuum damage mechanics: a continuum mechanics approach to the analysis of damage and fracture, Vol. 185, Springer Science \& Business Media, 2012.

\bibitem{kachanov1986introduction}
L.~Kachanov, Introduction to continuum damage mechanics, Vol.~10, Springer Science \& Business Media, 1986.

\bibitem{pijaudier1987nonlocal}
G.~Pijaudier-Cabot, Z.~P. Ba{\v{z}}ant, Nonlocal damage theory, Journal of engineering mechanics 113~(10) (1987) 1512--1533.

\bibitem{ahmed2021local}
B.~Ahmed, G.~Z. Voyiadjis, T.~Park, Local and non-local damage model with extended stress decomposition for concrete, International Journal of Damage Mechanics 30~(8) (2021) 1149--1191.

\bibitem{pijaudier2004non}
G.~Pijaudier-Cabot, K.~Haidar, J.-F. Dub{\'e}, Non-local damage model with evolving internal length, International journal for numerical and analytical methods in geomechanics 28~(7-8) (2004) 633--652.

\bibitem{jirasek2005non}
M.~Jir{\'a}sek, S.~Marfia, Non-local damage model based on displacement averaging, International Journal for Numerical Methods in Engineering 63~(1) (2005) 77--102.

\bibitem{chaves2013notes}
E.~W. Chaves, Notes on continuum mechanics, Springer Science \& Business Media, 2013.

\bibitem{bathe2006finite}
K.-J. Bathe, Finite element procedures, Klaus-Jurgen Bathe, 2006.

\bibitem{hughes2012finite}
T.~J. Hughes, The finite element method: linear static and dynamic finite element analysis, Courier Corporation, 2012.

\bibitem{padovan1980self}
J.~Padovan, Self adaptive incremental newton-raphson algorithms, NASA CP-2147 (1980) 115--121.

\bibitem{diez2003note}
P.~D{\'\i}ez, A note on the convergence of the secant method for simple and multiple roots, Applied mathematics letters 16~(8) (2003) 1211--1215.

\bibitem{nocedal2006line}
J.~Nocedal, S.~J. Wright, Line search methods, Numerical optimization (2006) 30--65.

\bibitem{findeisen2017characteristics}
C.~Findeisen, J.~Hohe, M.~Kadic, P.~Gumbsch, Characteristics of mechanical metamaterials based on buckling elements, Journal of the Mechanics and Physics of Solids 102 (2017) 151--164.

\bibitem{yang1990solution}
Y.-B. Yang, M.-S. Shieh, Solution method for nonlinear problems with multiple critical points, AIAA journal 28~(12) (1990) 2110--2116.

\bibitem{carrera1994study}
E.~Carrera, A study on arc-length-type methods and their operation failures illustrated by a simple model, Computers \& structures 50~(2) (1994) 217--229.

\bibitem{crisfield1979faster}
M.~Crisfield, A faster modified newton-raphson iteration, Computer methods in applied mechanics and engineering 20~(3) (1979) 267--278.

\bibitem{mldivide}
MathWorks, \href{https://www.mathworks.com/help/matlab/ref/mldivide.html}{mldivide} (2023).
\newline\urlprefix\url{https://www.mathworks.com/help/matlab/ref/mldivide.html}

\bibitem{schweizerhof1986consistent}
K.~Schweizerhof, P.~Wriggers, Consistent linearization for path following methods in nonlinear fe analysis, Computer Methods in Applied Mechanics and Engineering 59~(3) (1986) 261--279.

\bibitem{watson1983quadratic}
L.~T. Watson, S.~M. Holzer, Quadratic convergence of crisfield's method, Computers \& Structures 17~(1) (1983) 69--72.

\bibitem{krishnamoorthy1996post}
C.~Krishnamoorthy, G.~Ramesh, K.~Dinesh, Post-buckling analysis of structures by three-parameter constrained solution techniques, Finite elements in analysis and design 22~(2) (1996) 109--142.

\bibitem{zienkiewicz2005finite}
O.~C. Zienkiewicz, R.~L. Taylor, The finite element method for solid and structural mechanics, Elsevier, 2005.

\bibitem{taylor2014feap}
R.~L. Taylor, Feap-a finite element analysis program (2014).

\bibitem{londono2016prony}
J.~G. Londono, L.~Berger-Vergiat, H.~Waisman, A prony-series type viscoelastic solid coupled with a continuum damage law for polar ice modeling, Mechanics of Materials 98 (2016) 81--97.

\bibitem{chen2021phase}
L.~Chen, R.~de~Borst, Phase-field modelling of cohesive fracture, European Journal of Mechanics-A/Solids 90 (2021) 104343.

\bibitem{nguyen2018smoothing}
T.~H. Nguyen, T.~Q. Bui, S.~Hirose, Smoothing gradient damage model with evolving anisotropic nonlocal interactions tailored to low-order finite elements, Computer Methods in Applied Mechanics and Engineering 328 (2018) 498--541.

\bibitem{poh2017localizing}
L.~H. Poh, G.~Sun, Localizing gradient damage model with decreasing interactions, International Journal for Numerical Methods in Engineering 110~(6) (2017) 503--522.

\bibitem{Intelmicroprocessors}
I.~Corporation, \href{https://www.intel.com/content/www/us/en/support/articles/000005755/processors.html}{Export compliance metrics for intel® microprocessors} (2022).
\newline\urlprefix\url{https://www.intel.com/content/www/us/en/support/articles/000005755/processors.html}

\bibitem{bavzant1976instability}
Z.~P. Ba{\v{z}}ant, Instability, ductility, and size effect in strain-softening concrete, Journal of the engineering mechanics division 102~(2) (1976) 331--344.

\bibitem{miehe2010phase}
C.~Miehe, M.~Hofacker, F.~Welschinger, A phase field model for rate-independent crack propagation: Robust algorithmic implementation based on operator splits, Computer Methods in Applied Mechanics and Engineering 199~(45-48) (2010) 2765--2778.

\bibitem{de1985non}
R.~de~Borst, P.~Nauta, Non-orthogonal cracks in a smeared finite element model, Engineering computations 2~(1) (1985) 35--46.

\bibitem{rots1987analysis}
J.~G. Rots, R.~De~Borst, Analysis of mixed-mode fracture in concrete, Journal of engineering mechanics 113~(11) (1987) 1739--1758.

\bibitem{wu2020bfgs}
J.-Y. Wu, Y.~Huang, V.~P. Nguyen, On the bfgs monolithic algorithm for the unified phase field damage theory, Computer Methods in Applied Mechanics and Engineering 360 (2020) 112704.

\bibitem{lampron2021efficient}
O.~Lampron, D.~Therriault, M.~Levesque, An efficient and robust monolithic approach to phase-field quasi-static brittle fracture using a modified newton method, Computer Methods in Applied Mechanics and Engineering 386 (2021) 114091.

\bibitem{khalil2022generalised}
Z.~Khalil, A.~Y. Elghazouli, E.~Martinez-Paneda, A generalised phase field model for fatigue crack growth in elastic--plastic solids with an efficient monolithic solver, Computer Methods in Applied Mechanics and Engineering 388 (2022) 114286.

\bibitem{pantidis2023integrated}
P.~Pantidis, M.~E. Mobasher, Integrated finite element neural network (i-fenn) for non-local continuum damage mechanics, Computer Methods in Applied Mechanics and Engineering 404 (2023) 115766.

\bibitem{peng2020phase}
F.~Peng, W.~Huang, Z.-Q. Zhang, T.~F. Guo, Y.~E. Ma, Phase field simulation for fracture behavior of hyperelastic material at large deformation based on edge-based smoothed finite element method, Engineering Fracture Mechanics 238 (2020) 107233.

\bibitem{treifi2009computations}
M.~Treifi, S.~O. Oyadiji, D.~K. Tsang, Computations of the stress intensity factors of double-edge and centre v-notched plates under tension and anti-plane shear by the fractal-like finite element method, Engineering Fracture Mechanics 76~(13) (2009) 2091--2108.

\bibitem{zhou2018phase}
S.~Zhou, T.~Rabczuk, X.~Zhuang, Phase field modeling of quasi-static and dynamic crack propagation: Comsol implementation and case studies, Advances in Engineering Software 122 (2018) 31--49.

\bibitem{mazars1986description}
J.~Mazars, A description of micro-and macroscale damage of concrete structures, Engineering Fracture Mechanics 25~(5-6) (1986) 729--737.

\bibitem{mazars1984application}
J.~Mazars, Application de la m{\'e}canique de l'endommagement au comportement non lin{\'e}aire et {\`a} la rupture du b{\'e}ton de structure, THESE DE DOCTEUR ES SCIENCES PRESENTEE A L'UNIVERSITE PIERRE ET MARIE CURIE-PARIS 6 (1984).

\bibitem{de1995comparison}
J.~De~Vree, W.~Brekelmans, M.~van Gils, Comparison of nonlocal approaches in continuum damage mechanics, Computers \& Structures 55~(4) (1995) 581--588.

\end{thebibliography}
\newpage
\appendix

\section{Coefficients used in PC derivation of non-local damage model}
\label{Appendix:Coeff_PC_gradient}
The following are the coefficients used in the definition of predictor and correcter values of Section \ref{PC_Nonlocal}. The coefficients $\boldsymbol A$ to $\boldsymbol H$ depends on the consistent tangent $\boldsymbol{J}$ and the residuals $\boldsymbol{r}$.  
\begin{subequations}

\begin{align}
\boldsymbol{A} = & \displaystyle \left[1 - \left[\prescript{n}{i}{[{{\boldsymbol{J}^{\bar \epsilon \bar \epsilon}}}]^{-1}} \ \prescript{n}{i}{\boldsymbol{J}^{\bar \epsilon u , f}} \ \prescript{n}{i}{[{{\boldsymbol{J}^{ff}}}]^{-1}} \ \prescript{n}{i}{\boldsymbol{J}^{\bar u \epsilon, f}} \right] \right]^{-1} \\ \nonumber &
\left[ - \prescript{n}{i}{[{{\boldsymbol{J}^{\bar \epsilon \bar \epsilon}}}]^{-1}} \prescript{n}{i}{\boldsymbol{r}^{\bar \epsilon}} + \left[\prescript{n}{i}{[{{\boldsymbol{J}^{\bar \epsilon \bar \epsilon}}}]^{-1}} \  \prescript{n}{i}{\boldsymbol{J}^{\bar \epsilon u , f}}  \ \prescript{n}{i}{[{{\boldsymbol{J}^{ff}}}]^{-1}}  \  \prescript{n}{i}{\boldsymbol{r}^f}  \right] \right]
\end{align}

\begin{align}
\boldsymbol{B} = & \displaystyle 
{\left[{1- \prescript{n}{i}{[{{\boldsymbol{J}^{\bar \epsilon \bar \epsilon}}}]^{-1}} \ \prescript{n}{i}{\boldsymbol{J}^{\bar \epsilon u , f}} \ \prescript{n}{i}{[{{\boldsymbol{J}^{ff}}}]^{-1}} \ \prescript{n}{i}{\boldsymbol{J}^{u \bar \epsilon, f}} }\right]}^{-1} \\ \nonumber &
\left[{\prescript{n}{i}{[{{\boldsymbol{J}^{\bar \epsilon \bar \epsilon}}}]^{-1}} \ \prescript{n}{i}{\boldsymbol{J}^{\bar \epsilon u , f}} \ \prescript{n}{i}{[{{\boldsymbol{J}^{ff}}}]^{-1}} \ \prescript{n}{i}{\boldsymbol{J}^{fp}} \ \boldsymbol{u}^p - \prescript{n}{i}{[{{\boldsymbol{J}^{\bar \epsilon \bar \epsilon}}}]^{-1}} \ \prescript{n}{i}{\boldsymbol{J}^{\bar \epsilon u , p}} \boldsymbol{u}^p}\right]  
\end{align}

\begin{equation}
    \boldsymbol{C} = \displaystyle - \prescript{n}{i}{[{{\boldsymbol{J}^{ff}}}]^{-1}} \ \prescript{n}{i}{\boldsymbol{r}^f} - \left[\prescript{n}{i}{[{{\boldsymbol{J}^{ff}}}]^{-1}} \ \prescript{n}{i}{\boldsymbol{J}^{u \bar \epsilon , f}} \ \boldsymbol{A} \right]
\end{equation}

\begin{equation}
\boldsymbol{D} = \displaystyle - \prescript{n}{i}{[{{\boldsymbol{J}^{ff}}}]^{-1}} \ \prescript{n}{i}{\boldsymbol{J}^{u \bar \epsilon , f}} \ \boldsymbol{B}- \left[\prescript{n}{i}{[{{\boldsymbol{J}^{ff}}}]^{-1}} \ \prescript{n}{i}{\boldsymbol{J}^{fp}} \ \boldsymbol{u}^p \right] 
\end{equation}

\begin{equation}
\boldsymbol{E} = \displaystyle - \prescript{n}{i}{\boldsymbol{r}^p} - \prescript{n}{i}{\boldsymbol{J}^{pf}} \ \boldsymbol{C} - \prescript{n}{i}{\boldsymbol{J}^{u \bar \epsilon , p}} \ \boldsymbol{A}
\end{equation}

\begin{equation}
\boldsymbol{F} = \displaystyle - \prescript{n}{i}{\boldsymbol{J}^{pf}} \ \boldsymbol{D} - \prescript{n}{i}{\boldsymbol{J}^{pp}} \ \boldsymbol{u}^p - \prescript{n}{i}{\boldsymbol{J}^{u \bar \epsilon , p}} \ \boldsymbol{B}
\end{equation} 

\begin{equation}
 \boldsymbol{G} = \displaystyle 1+ \frac{2 \ \prescript{n}{i}{[{\Delta \boldsymbol{u}^f}]^T} \ \boldsymbol{D} + 2 \ \beta^2 \ \prescript{n}{i}{[{\Delta \boldsymbol{f}^{ext}}]^T} \ \boldsymbol{F} + 2 \ \prescript{n}{i}{[{\Delta \boldsymbol{\bar \epsilon}}]^T} \ \boldsymbol{B}}{ 2 \ \prescript{n}{i}{[{\Delta \boldsymbol{u}^p}]^T}\ \boldsymbol{u}^p} 
 \end{equation}

\begin{equation}
 \boldsymbol{H} = \displaystyle \frac{-g - 2 \ \prescript{n}{i}{[{\Delta \boldsymbol{u}^f}]^T} \boldsymbol{C} - 2 \ \beta^2 \ \prescript{n}{i}{[{\Delta \boldsymbol{f}^{ext}}]^T} \ \boldsymbol{E} -  2 \ \prescript{n}{i}{[{\Delta \boldsymbol{\bar \epsilon}}]^T} \ \boldsymbol{A}}{ 2 \ \prescript{n}{i}{[{\Delta \boldsymbol{u}^p}]^T}\ \boldsymbol{u}^p} 
 \end{equation}

Note that proposed methods are implemented in MATLAB and the matrix inversion operations in the derivations are performed by using mldivide \cite{mldivide} to solve the system of equations.

\label{eq:AtHcoeff}
\end{subequations}

\section{Coefficients used in PNC derivation of local damage model}
\label{Appendix:Coeff_PNC_local}
The following are the coefficients used in the definition of correcter values in Section \ref{PNC_Local_Damage}. 

\begin{subequations}
\begin{equation}
\displaystyle a = \left[{\boldsymbol{u}^p}\right]^T \left[{\boldsymbol{u}^p}\right] 
+ 
\prescript{n}{i}{\left[{{\delta \boldsymbol{u}^{f,B}}}\right]^T} \ \prescript{n}{i}{\left[{{\delta \boldsymbol{u}^{f,B}}}\right]} 
+
\beta^2  \prescript{n}{i}{\left[{\delta \boldsymbol{f}^{ext,B}}\right]^T} \ \prescript{n}{i}{\left[{\delta \boldsymbol{f}^{ext,B}}\right]}
\label{PNC_Quadratic_Eq_a}
\end{equation} 
 
\begin{align}
\displaystyle b = & \left[\prescript{n}{i}{\Delta \boldsymbol{u}^p }\right]^T \ \left[\boldsymbol{u}^p \right] 
+ 
\left[{\boldsymbol{u}^p}\right]^T \ \prescript{n}{i}{\left[{\Delta \boldsymbol{u}^p}\right]} 
+ 
\prescript{n}{i}{\left[{\Delta \boldsymbol{u}^f}\right]^T} \ \prescript{n}{i}{\left[{\delta {\boldsymbol{u}^{f,B}}}\right]} + 
\prescript{n}{i}{\left[{\delta {\boldsymbol{u}^{f,B}}}\right]^T} \ \prescript{n}{i}{\left[{\Delta \boldsymbol{u}^f}\right]} \nonumber \\ & 
+
\prescript{n}{i}{\left[{\delta {\boldsymbol{u}^{f,A}}}\right]^T} \  \prescript{n}{i}{\left[{\delta {\boldsymbol{u}^{f,B}}}\right]} 
+ 
\prescript{n}{i}{\left[{\delta {\boldsymbol{u}^{f,B}}}\right]^T} \  \prescript{n}{i}{\left[{\delta {\boldsymbol{u}^{f,A}}}\right]} 
+ 
\beta^2 \prescript{n}{i}{\left[{\Delta \boldsymbol{f}^{ext}}\right]^T}\ \prescript{n}{i}{\left[{\delta \boldsymbol{f}^{ext,B}}\right]}\nonumber \\ & 
+ 
\beta^2  \prescript{n}{i}{\left[{\delta \boldsymbol{f}^{ext,B}}\right]^T} \ \prescript{n}{i}{\left[{\Delta \boldsymbol{f}^{ext}}\right]} 
+
\beta^2 \prescript{n}{i}{\left[{\delta \boldsymbol{f}^{ext,A}}\right]^T} \  \prescript{n}{i}{\left[{\delta \boldsymbol{f}^{ext,B}}\right]} 
+ 
\beta^2 \prescript{n}{i}{\left[{\delta \boldsymbol{f}^{ext,B}}\right] ^T} \ \prescript{n}{i}{\left[{\delta \boldsymbol{f}^{ext,A}}\right]}
\label{PNC_Quadratic_Eq_b}
\end{align} 

\begin{align}
\displaystyle c = & \prescript{n}{i}{[\Delta \boldsymbol{u}^p]^T} \ \prescript{n}{i}{[\Delta \boldsymbol{u}^p]} + 
\prescript{n}{i}{[\Delta \boldsymbol{u}^f]^T} \ \prescript{n}{i}{[\Delta \boldsymbol{u}^f]} + 
\prescript{n}{i}{[\Delta \boldsymbol{u}^f]^T} \ \prescript{n}{i}{\delta {\boldsymbol{u}^{f,A}}} + 
\prescript{n}{i}{\delta {\boldsymbol{u}^{f,A}}}^T \ \prescript{n}{i}{[\Delta \boldsymbol{u}^f]} \nonumber \\ & +
\prescript{n}{i}{\delta {\boldsymbol{u}^{f,A}}}^T \ \prescript{n}{i}{\delta {\boldsymbol{u}^{f,A}}} +
\beta^2 \prescript{n}{i}{[\Delta \boldsymbol{f}^{ext}]^T} \ \prescript{n}{i}{[\Delta \boldsymbol{f}^{ext}]} + 
\beta^2 \prescript{n}{i}{[\Delta \boldsymbol{f}^{ext}]^T} \ \prescript{n}{i}{[\delta \boldsymbol{f}^{ext,A}]} \nonumber \\ & + 
\beta^2 \prescript{n}{i}{[\delta \boldsymbol{f}^{ext,A}]}^T \ \prescript{n}{i}{[\Delta \boldsymbol{f}^{ext}]}  + 
\beta^2 \prescript{n}{i}{[\delta \boldsymbol{f}^{ext,A}]}^T \ \prescript{n}{i}{[\delta \boldsymbol{f}^{ext,A}]}
\label{PNC_Quadratic_Eq_c}
\end{align} 
\label{PNC_correctors}
\end{subequations}

\section{Unified arc-length Jacobian Matrix (Local Damage)}
\label{Appendix: UAL Jacobian (Local Damage)}

We begin the derivation of consistent tangent stiffness matrix ($\boldsymbol{J}$) for the UAL Local damage law from the following residual equations,
\begin{subequations}
\begin{equation}
\prescript{n}{i}{\boldsymbol{r}^p} = \prescript{n}{i}{\left[{\underbrace{\int_\Omega [\boldsymbol{B}^u]^T \boldsymbol{\sigma} d\Omega}_{\boldsymbol{f}^{int,p}} - \underbrace{\int_\Gamma [\boldsymbol{N}^u]^T t d\Gamma}_{\boldsymbol{f}^{ext,p}}}\right]}
\label{ApxC_Eq:Local_Damage_Res1}
\end{equation}

\begin{equation}
\prescript{n}{i}{\boldsymbol{r}^f} = \prescript{n}{i}{\left[{\underbrace{\int_\Omega [\boldsymbol{B}^u]^T \boldsymbol{\sigma} d\Omega}_{\boldsymbol{f}^{int,f}}}\right]}
\label{ApxC_Eq:Local_Damage_Res2}
\end{equation}

\begin{equation}
\prescript{n}{i}{\ \boldsymbol{g}} = \prescript{n}{i}{[\Delta \boldsymbol{x}]^T} \ \prescript{n}{i}{[\Delta \boldsymbol{x}]} + {\beta}^{2} \prescript{n}{i}{[{\Delta \boldsymbol{f}^{ext}}]^T} \ \prescript{n}{i}{[{\Delta \boldsymbol{f}^{ext}}]} - \Delta l^2 
\label{ApxC_Eq:Local_Damage_Res3}
\end{equation} 
\label{ApxC_Eq:UAL_Local_damage_Res_Eqns}
\end{subequations}

The linearized form of Eqns. \eqref{ApxC_Eq:UAL_Local_damage_Res_Eqns} expressed as $\boldsymbol{J}\delta \boldsymbol{x}=-\boldsymbol{r}$ is presented below:

\begin{equation}
 \begin{bmatrix} {\dfrac{\partial {\boldsymbol{r}^p}}{\partial f^{ext,e}}} & {\dfrac{\partial {\boldsymbol{r}^p}}{\partial u^{f,e}}} & {\dfrac{\partial {\boldsymbol{r}^p}}{\partial \bar m}} \\  & \\ {\dfrac{\partial {\boldsymbol{r}^f}}{\partial f^{ext,e}}} & {\dfrac{\partial {\boldsymbol{r}^f}}{\partial u^{f,e}}} & {\dfrac{\partial {\boldsymbol{r}^f}}{\partial \bar m}} \\  & \\ {\dfrac{\partial g}{\partial f^{ext,e}}} & {\dfrac{\partial g}{\partial u^{f,e}}} & {\dfrac{\partial g}{\partial \bar m}} \end{bmatrix}
 \begin{bmatrix} \prescript{}{}{\delta \boldsymbol{f}^{ext}} \\ \\ \prescript{}{}{\delta \boldsymbol{u}^f} \\ \\ \prescript{}{}{\delta \bar{m}} \end{bmatrix}
 =
 - {\begin{bmatrix} \prescript{n}{i}{\boldsymbol{r}^p} \\ \\ \prescript{n}{i}{\boldsymbol{r}^f} \\ \\ \prescript{n}{i}{g} \end{bmatrix}}
 \label{ApxC_Eq:UAL_Local_Jacobian}
 \end{equation}

 where: 
\begin{align}
a) \dfrac{\partial {\boldsymbol{r}^p}}{\partial f^{ext,e}} = \dfrac{\partial {\left[{\int_\Omega [\boldsymbol{B}^u]^T \boldsymbol{\sigma} d\Omega - \int_\Gamma [\boldsymbol{N}^u]^T t d\Gamma}\right]}}{\partial f^{ext,e}} = \dfrac{\partial \left[\cancel{\boldsymbol{f}^{int,p}} + \boldsymbol{f}^{ext,p}\right]}{\partial f^{ext,e}} = \dfrac{\partial \boldsymbol{f}^{ext,p}}{\partial f^{ext,e}} = \boldsymbol{I} 
\end{align}

\begin{align}
b) \dfrac{\partial {\boldsymbol{r}^p}}{\partial u^{f,e}} = \dfrac{\partial {\left[{\int_\Omega [\boldsymbol{B}^u]^T \boldsymbol{\sigma} d\Omega - \int_\Gamma [\boldsymbol{N}^u]^T t d\Gamma}\right]}}{\partial  u^{f,e}} = \dfrac{\partial \left[\boldsymbol{f}^{int,p} + \cancel{\boldsymbol{f}^{ext,p}}\right]}{\partial u^{f,e}} = \dfrac{\partial \boldsymbol{f}^{int,p} }{\partial u^{f,e}} = \prescript{n}{i}{\boldsymbol{J}^{pf}}
\end{align}

\begin{align}
c) \dfrac{\partial {\boldsymbol{r}^p}}{\partial \bar m} = \dfrac{\partial {\left[{\int_\Omega [\boldsymbol{B}^u]^T \boldsymbol{\sigma} d\Omega - \int_\Gamma [\boldsymbol{N}^u]^T t d\Gamma}\right]}}{\partial  \bar m} =\dfrac{\partial \left[\boldsymbol{f}^{int,p} + \cancel{\boldsymbol{f}^{ext,p}}\right]}{\partial \bar m} = \dfrac{\partial \boldsymbol{f}^{int,p} }{\partial {u}^{p,e}} \boldsymbol{u}^p = \prescript{n}{i}{\boldsymbol{J}^{pp}} \ \boldsymbol{u}^p
\end{align}

\begin{align}
d) \dfrac{\partial {\boldsymbol{r}^f}}{\partial {f}^{ext,e}}  & = \dfrac{\partial \ {{\left[{\int_\Omega [\boldsymbol{B}^u]^T \boldsymbol{\sigma} d\Omega}\right]}}}{\partial {f}^{ext,e}} =\dfrac{\partial \boldsymbol{f}^{int,f}}{\partial {f}^{ext,e}} = \boldsymbol{0}
\end{align}

\begin{align}
e) \dfrac{\partial {\boldsymbol{r}^f}}{\partial u^{f,e}}  & =  \dfrac{\partial \ {{\left[{\int_\Omega [\boldsymbol{B}^u]^T \boldsymbol{\sigma} d\Omega}\right]}}}{\partial u^{f,e}} =\dfrac{\partial \boldsymbol{f}^{int,f}}{\partial u^{f,e}} = {\prescript{n}{i}{\boldsymbol{J}^{ff}}}
\end{align}

\begin{align}
f) \dfrac{\partial {\boldsymbol{r}^f}}{\partial \bar m}  & = \dfrac{\partial \ {{\left[{\int_\Omega [\boldsymbol{B}^u]^T \boldsymbol{\sigma} d\Omega}\right]}}}{\partial {\bar m}} =\dfrac{\partial \boldsymbol{f}^{int,f}}{\partial \bar m} =\dfrac{\partial \boldsymbol{f}^{int,f}}{\partial u^{p,e}} {\boldsymbol{u}^p} =\prescript{n}{i}{\boldsymbol{J}^{fp}} \ \boldsymbol{u}^p
\end{align}

\begin{align}
g) \dfrac{\partial {g}}{\partial f^{ext,e}}
& = 2 {\beta}^{2} {\left[\prescript{n}{i}{\Delta \boldsymbol{f}^{ext}}\right]}^T
\end{align}

\begin{align}
h) \dfrac{\partial {g}}{\partial u^{f,e}}
& = 2 {\prescript{n}{i}{[\Delta \boldsymbol{u}^f}]}^T
\end{align}

\begin{align}
i) \dfrac{\partial {g}}{\partial \bar m} 
& = \dfrac{\partial \left[{{\prescript{n}{i}{[\Delta \boldsymbol{u}^p]^T} \ \prescript{n}{i}{[\Delta \boldsymbol{u}^p]}}}\right]}{\partial \bar m} = \dfrac{\partial \left[{{\prescript{n}{i}{[\Delta \boldsymbol{u}^p]^T} \ \prescript{n}{i}{[\Delta \boldsymbol{u}^p]}}} \boldsymbol{u}^p \right]}{\partial {u}^p} = 2 {\left[\prescript{n}{i}{\Delta \boldsymbol{u}^p}\right]}^T \boldsymbol{u}^p
\end{align}

\begin{gather}
 \begin{bmatrix} \boldsymbol{I} & \prescript{n}{i}{\left[\boldsymbol{J}^{pf}\right]} & \prescript{n}{i}{\left[{{\boldsymbol{J}^{pp}} \ \boldsymbol{u}^p}\right]} \\  & \\ 0 & \prescript{n}{i}{\left[\boldsymbol{J}^{ff}\right]} & \prescript{n}{i}{\left[{\boldsymbol{J}^{fp}} \ \boldsymbol{u}^p\right]} \\  & \\  2 {\beta}^{2} {\left(\prescript{n}{i}{\Delta \boldsymbol{f}^{ext}}\right)}^T & 2 {\left(\prescript{n}{i}{\Delta \boldsymbol{u}^f}\right)}^T &  2 {\left(\prescript{n}{i}{\Delta \boldsymbol{u}^p}\right)}^T \ \boldsymbol{u}^p \end{bmatrix}
\begin{bmatrix} \prescript{n}{i+1}{\delta \boldsymbol{f}^{ext}} \\ \\ \prescript{n}{i+1}{\delta \boldsymbol{u}^f} \\ \\ \prescript{n}{i+1}{\delta \bar{m}} \end{bmatrix}
 =
 - \begin{bmatrix} \prescript{n}{i}{\boldsymbol{r}^p} \\ \\ \prescript{n}{i}{\boldsymbol{r}^f} \\ \\ \prescript{n}{i}{g} \end{bmatrix}
 \end{gather}

\section{Unified arc-length Jacobian Matrix (non-local gradient damage)}
\label{Appendix: UAL Jacobian (Non-Local Damage)}

\begin{subequations}
\begin{equation}
\prescript{n}{i}{\boldsymbol{r}^p} = \prescript{n}{i}{\left[{\underbrace{\int_\Omega [\boldsymbol{B}^u]^T \boldsymbol{\sigma} d\Omega}_{\boldsymbol{f}^{int,p}} - \underbrace{\int_\Gamma [\boldsymbol{N}^u]^T t d\Gamma}_{\boldsymbol{f}^{ext,p}}}\right]}
\label{ApxD_Eq:NonLocal_Damage_Res1}
\end{equation}

\begin{equation}
\prescript{n}{i}{\boldsymbol{r}^f} = \prescript{n}{i}{\left[{\underbrace{\int_\Omega [\boldsymbol{B}^u]^T \boldsymbol{\sigma} d\Omega}_{\boldsymbol{f}^{int,f}}}\right]}
\label{ApxD_Eq:NonLocal_Damage_Res2}
\end{equation}

\begin{equation}
\prescript{n}{i}{\boldsymbol{r}^{\bar\epsilon}} = \underbrace{{\prescript{n}{i}{\left[{{\int_\Omega [\boldsymbol{N}^\epsilon]^T \boldsymbol{\bar{\epsilon}} \ d\Omega + \int_\Omega [\boldsymbol{B}^\epsilon]^T \ c \ \nabla \boldsymbol{\bar{\epsilon}} \ d\Omega} -{ \int_\Omega [\boldsymbol{N}^\epsilon]^T \boldsymbol{\epsilon} \ d\Omega }}\right]}}}_{\boldsymbol{f}^{int, \bar\epsilon}}
\label{ApxD_Eq:NonLocal_Damage_Res3}
\end{equation}

\begin{equation}
\prescript{n}{i}{\ g} = \prescript{n}{i}{[\Delta \boldsymbol{x}]^T} \ \prescript{n}{i}{[\Delta \boldsymbol{x}]} + {\beta}^{2} \prescript{n}{i}{[\Delta \boldsymbol{f}^{ext}]^T} \ \prescript{n}{i}{[\Delta \boldsymbol{f}^{ext}]} - \Delta l^2 
\label{ApxD_Eq:NonLocal_Damage_Res4}
\end{equation} 
\label{ApxD_Eq:NonLocal_damage_Res_Eqns}
\end{subequations}

The linearized form of Eqns. \eqref{ApxD_Eq:NonLocal_damage_Res_Eqns} expressed as $\boldsymbol{J}\delta \boldsymbol{x}=-\boldsymbol{r}$ is presented below:

\begin{gather}
 \begin{bmatrix} {\dfrac{\partial {\boldsymbol{r}^p}}{\partial f^{ext,e}}} & {\dfrac{\partial {\boldsymbol{r}^p}}{\partial u^{f,e}}} & {\dfrac{\partial {\boldsymbol{r}^p}}{\partial \bar m}} & {\dfrac{\partial {\boldsymbol{r}^p}}{\partial {\bar \epsilon}^e}} \\  & \\ {\dfrac{\partial {\boldsymbol{r}^f}}{\partial f^{ext,e}}} & {\dfrac{\partial {\boldsymbol{r}^{f}}}{\partial u^{f,e}}} & {\dfrac{\partial {\boldsymbol{r}^f}}{\partial \bar m}} & {\dfrac{\partial {\boldsymbol{r}^f}}{\partial {\bar \epsilon}^e}}\\  & \\ {\dfrac{\partial g}{\partial f^{ext,e}}} & {\dfrac{\partial g}{\partial u^{f,e}}} & {\dfrac{\partial g}{\partial \bar m}} & {\dfrac{\partial g}{\partial {\bar \epsilon}^e}} \\& \\ {\dfrac{\partial \boldsymbol{r}^{\bar\epsilon}}{\partial f^{ext,e}}} & {\dfrac{\partial \boldsymbol{r}^{\bar\epsilon}}{\partial u^{f,e}}} & {\dfrac{\partial \boldsymbol{r}^{\bar\epsilon}}{\partial \bar m}} & {\dfrac{\partial \boldsymbol{r}^{\bar\epsilon}}{\partial {\bar \epsilon}^e}}
 \end{bmatrix}
\begin{bmatrix} {\delta \boldsymbol{f}^{ext}} \\ \\ {\delta \boldsymbol{u}^f} \\ \\ \delta \bar{m} \\ \\ \delta {\bar {\boldsymbol{\epsilon}}} \end{bmatrix}
 =
 - \begin{bmatrix} {\boldsymbol{r}^p} \\ \\  {\boldsymbol{r}^f} \\ \\ {g} \\ \\ \boldsymbol{r}^{\bar\epsilon} \end{bmatrix} 
 \end{gather}

\begin{align}
a) \dfrac{\partial {\boldsymbol{r}^p}}{\partial f^{ext,e}} & = \dfrac{\partial {\left[{\int_\Omega [\boldsymbol{B}^u]^T \boldsymbol{\sigma} d\Omega - \int_\Gamma [\boldsymbol{N}^u]^T t d\Gamma}\right]}}{\partial f^{ext,e}} =\dfrac{\partial \left[\cancel{\boldsymbol{f}^{int,p}} + \boldsymbol{f}^{ext,p}\right]}{\partial f^{ext,e}} = \dfrac{\partial \boldsymbol{f}^{ext,p}}{\partial f^{ext,e}} = \boldsymbol{I} 
\end{align}

\begin{align}
b) \dfrac{\partial {\boldsymbol{r}^p}}{\partial u^{f,e}}  & = \dfrac{\partial {\left[{\int_\Omega [\boldsymbol{B}^u]^T \boldsymbol{\sigma} d\Omega - \int_\Gamma [\boldsymbol{N}^u]^T t d\Gamma}\right]}}{\partial  u^{f,e}} =\dfrac{\partial \left[\boldsymbol{f}^{int,p} + \cancel{\boldsymbol{f}^{ext,p}}\right]}{\partial u^{f,e}}=\dfrac{\partial \boldsymbol{f}^{int,p} }{\partial u^{f,e}} = \prescript{n}{i}{\boldsymbol{J}^{pf}}
\end{align}

\begin{align}
c) \dfrac{\partial {\boldsymbol{r}^p}}{\partial \bar m}  
& = \dfrac{\partial {\left[{\int_\Omega [\boldsymbol{B}^u]^T \boldsymbol{\sigma} d\Omega - \int_\Gamma [\boldsymbol{N}^u]^T t d\Gamma}\right]}}{\partial  \bar m} =\dfrac{\partial \left[\boldsymbol{f}^{int,p} + \cancel{\boldsymbol{f}^{ext,p}}\right]}{\partial \bar m} =\dfrac{\partial \boldsymbol{f}^{int,p} }{\partial \bar m} 
=\dfrac{\partial \boldsymbol{f}^{int,p} }{\partial u^{p,e}} \boldsymbol{u}^p \\
& = \prescript{n}{i}{\boldsymbol{J}^{pp}} \ \boldsymbol{u}^p
\end{align}

\begin{align}
d) {\dfrac{\partial {\boldsymbol{r}^p}}{\partial {\bar \epsilon}^e}}  & = \dfrac{\partial \left[{\int_\Omega [\boldsymbol{B}^u]^T \boldsymbol{\sigma} d\Omega - \int_\Gamma [\boldsymbol{N}^u]^T t d\Gamma}\right]}{{\partial {\bar\epsilon}^e}} = \dfrac{\partial \left[{\int_\Omega [\boldsymbol{B}^u]^T \boldsymbol{\sigma} d\Omega }\right]}{{\partial {\bar\epsilon}^e}} - \cancel{\dfrac{\partial \left[{\int_\Gamma [\boldsymbol{N}^u]^T t d\Gamma}\right]}{{\partial {\bar\epsilon}^e}}} \\
& = \dfrac{\partial{\left[\int_\Omega [\boldsymbol{B}^u]^T (1-d){\sigma}_{ij} d\Omega \right]}}{\partial {\bar\epsilon}^e} = \dfrac{\partial{\left[\int_\Omega [\boldsymbol{B}^u]^T (1-d) C_{ijkl} {\epsilon}_{kl} d\Omega \right]}}{\partial{\bar\epsilon}^e} \\
& =- {\int_\Omega [\boldsymbol{B}^u]^T \frac{\partial{d}} {{\partial{\bar\epsilon}^e}} C_{ijkl} {\epsilon}_{kl} d\Omega } =- {\int_\Omega [\boldsymbol{B}^u]^T \frac{\partial{d}}{{\partial{\bar\epsilon}}}\frac{\partial{\bar\epsilon}}{{\partial{\bar\epsilon}^e}} C_{ijkl} {\epsilon}_{kl} d\Omega }\\ 
& =- {\int_\Omega [\boldsymbol{B}^u]^T \frac{\partial{d}}{{\partial{\bar\epsilon}}} \boldsymbol{N}^{{\bar \epsilon}} C_{ijkl} {\epsilon}_{kl} d\Omega } = \prescript{n}{i}{\left[{\boldsymbol{J}^{u \bar\epsilon}}\right]}^p
\end{align}

\begin{align}
e) \dfrac{\partial {\boldsymbol{r}^f}}{\partial f^{ext,e}}  & = \dfrac{\partial \ {{\left[{\int_\Omega [\boldsymbol{B}^u]^T \boldsymbol{\sigma} d\Omega}\right]}}}{\partial f^{ext,e}}  =\dfrac{\partial \boldsymbol{f}^{int,f}}{\partial f^{ext,e}} = \boldsymbol{0}
\end{align}

\begin{align}
f) \dfrac{\partial {\boldsymbol{r}^f}}{\partial u^{f,e}}  & =  \dfrac{\partial \ {{\left[{\int_\Omega [\boldsymbol{B}^u]^T \boldsymbol{\sigma} d\Omega}\right]}}}{\partial u^{f,e}} =\dfrac{\partial \boldsymbol{f}^{int,f}}{\partial u^{f,e}} ={\prescript{n}{i}{\boldsymbol{J}^{ff}}}
\end{align}

\begin{align}
g) \dfrac{\partial {\boldsymbol{r}^f}}{\partial \bar m}  & = \dfrac{\partial \ {{\left[{\int_\Omega [\boldsymbol{B}^u]^T \boldsymbol{\sigma} d\Omega}\right]}}}{\partial {\bar m}} =\dfrac{\partial \boldsymbol{f}^{int,f}}{\partial \bar m} =\dfrac{\partial \boldsymbol{f}^{int,f}}{\partial u^{p,e}} {\boldsymbol{u}_p}  =\prescript{n}{i}{\left[{\boldsymbol{J}^{fp}}\right]} \ \boldsymbol{u}^p
\end{align}

\begin{align}
h) \dfrac{\partial {\boldsymbol{r}^f}}{\partial {\bar \epsilon}^e} & = \dfrac{\partial \ {{\left[{\int_\Omega [\boldsymbol{B}^u]^T \boldsymbol{\sigma} d\Omega}\right]}}}{\partial {\bar \epsilon}^e} = \dfrac{\partial{\left[\int_\Omega [\boldsymbol{B}^u]^T (1-d){\sigma}_{ij} d\Omega \right]}}{\partial {\bar\epsilon}^e} =- {\int_\Omega [\boldsymbol{B}^u]^T \frac{\partial{d}}{{\partial{\bar\epsilon}}}\frac{\partial{\bar\epsilon}}{{\partial{\bar\epsilon}^e}} C_{ijkl} {\epsilon}_{kl} d\Omega }\\
& =- {\int_\Omega [\boldsymbol{B}^u]^T \frac{\partial{d}}{{\partial{\bar\epsilon}}} \boldsymbol{N}^{\epsilon} C_{ijkl} {\epsilon}_{kl} d\Omega} = \prescript{n}{i}{\left[{\boldsymbol{J}^{u \bar\epsilon}}\right]}^f
\end{align}

\begin{align}
i) \dfrac{\partial {g}}{\partial f^{ext,e}} 
& = 2 {\beta}^{2} {\left[\prescript{n}{i}{\Delta \boldsymbol{f}^{ext}}\right]}^T
\end{align}

\begin{align}
j) \dfrac{\partial {g}}{\partial {u}^{f,e}} 
& = 2 {\left[\prescript{n}{i}{\Delta \boldsymbol{u}^f}\right]}^T
\end{align}

\begin{align}
k) \dfrac{\partial {g}}{\partial \bar m} 
& = \dfrac{\partial \left[{{\prescript{n}{i}{[\Delta \boldsymbol{u}^p]^T} \ \prescript{n}{i}{[\Delta \boldsymbol{u}^p]}}}\right]}{\partial \bar m} = \dfrac{\partial \left[{{\prescript{n}{i}{[\Delta \boldsymbol{u}^p]^T} \ \prescript{n}{i}{[\Delta \boldsymbol{u}^p]}}}\right] \boldsymbol{u}^p}{\partial \boldsymbol{u}^p} = 2 {\left[\prescript{n}{i}{\Delta \boldsymbol{u}^p}\right]}^T \boldsymbol{u}^p
\end{align}

\begin{align}
l) \dfrac{\partial {g}}{\partial {\bar{{\epsilon}}}^e} 
& = 2 {\prescript{n}{i}{\Delta \bar{\boldsymbol{\epsilon}}}}^T
\end{align}

\begin{align}
m) {\dfrac{\partial {\boldsymbol{r}^{\bar\epsilon}}}{\partial f^{ext,e}} }  & = 
\dfrac{\partial \ {{\left[{{{\int_\Omega [\boldsymbol{N}^{\bar \epsilon}]^T \bar{\epsilon} \ d\Omega + \int_\Omega [\boldsymbol{B}^{\bar \epsilon}]^T \ c \ \nabla \bar{\epsilon} \ d\Omega} -{ \int_\Omega [\boldsymbol{N}^{\bar \epsilon}]^T \epsilon \ d\Omega }}}\right]}}}{\partial f^{ext,e}} = \boldsymbol{0} 
\end{align}

\begin{align}
n) {\dfrac{\partial {\boldsymbol{r}^{\bar\epsilon}}}{\partial u^{f,e}}}  & = 
\dfrac{\partial \ {{\left[{{{\int_\Omega [\boldsymbol{N}^{\bar \epsilon}]^T \bar{\epsilon} \ d\Omega + \int_\Omega [\boldsymbol{B}^{\bar \epsilon}]^T \ c \ \nabla \bar{\epsilon} \ d\Omega} -{ \int_\Omega [\boldsymbol{N}^{\bar \epsilon}]^T \epsilon \ d\Omega }}}\right]}}}{\partial u^{f,e}} \\
& = {{\int_\Omega [\boldsymbol{N}^{\bar \epsilon}]^T \cancel{\frac{\partial \bar{\epsilon}}{\partial u^{f,e}}} \ d\Omega + \int_\Omega [\boldsymbol{B}^{\bar \epsilon}]^T \ c \ \frac{\partial [\nabla\bar{\epsilon}]}{\partial u^{f,e}} \ d\Omega } -{ \int_\Omega [\boldsymbol{N}^{\bar \epsilon}]^T \frac{\partial {\epsilon}}{\partial u^{f,e}} \ d\Omega } } \\
& = {{\int_\Omega [\boldsymbol{B}^{\bar \epsilon}]^T \ c \ \frac{\partial [\nabla\bar{\epsilon}]}{\partial u^{f,e}} \ d\Omega } -{ \int_\Omega [\boldsymbol{N}^{\bar \epsilon}]^T \frac{\partial {\epsilon}}{\partial u^{f,e}} \ d\Omega } } \\
& = {{\int_\Omega [\boldsymbol{B}^{\bar \epsilon}]^T \ c \ \boldsymbol{B}^{\bar \epsilon} \ \cancel{\frac{\partial \bar{\epsilon}}{\partial u^{f,e}}} \ d\Omega } -{ \int_\Omega [\boldsymbol{N}^{\bar \epsilon}]^T \frac{\partial {\epsilon}}{\partial {\epsilon}_{ij}} \frac{\partial {\epsilon}_{ij}}{\partial u^{f,e}} \ d\Omega } } \\
& = -{ \int_\Omega [\boldsymbol{N}^{\bar \epsilon}]^T \frac{\partial {\boldsymbol{B}^u u^e}}{\partial u^{f,e}} \ d\Omega }  = -{ \int_\Omega [\boldsymbol{N}^{\bar \epsilon}]^T \boldsymbol{B}^u \ d\Omega } = \prescript{n}{i}{\left[{\boldsymbol{J}^{\bar \epsilon u}}\right]}^f
\end{align}

\begin{align}
o) {\dfrac{\partial {\boldsymbol{r}^{\bar\epsilon}}}{\partial \bar m}}  & = 
\left[{\dfrac{\partial \ {{\left[{{{\int_\Omega [\boldsymbol{N}^{\bar \epsilon}]^T \bar{\epsilon} \ d\Omega + \int_\Omega [\boldsymbol{B}^{\bar \epsilon}]^T \ c \ \nabla \bar{\epsilon} \ d\Omega} -{ \int_\Omega [\boldsymbol{N}^{\bar \epsilon}]^T \epsilon \ d\Omega }}}\right]}}}{\partial u^{p,e}}}\right] \boldsymbol{u}_p\\
& = \left[{{\int_\Omega [\boldsymbol{N}^{\bar \epsilon}]^T \cancel{\frac{\partial \bar{\epsilon}}{\partial u^{p,e}}} \ d\Omega + \int_\Omega [\boldsymbol{B}^{\bar \epsilon}]^T \ c \ \frac{\partial [\nabla\bar{\epsilon}]}{\partial u^{p,e}} \ d\Omega } -{ \int_\Omega [\boldsymbol{N}^{\bar \epsilon}]^T \frac{\partial {\epsilon}}{\partial u^{p,e}} \ d\Omega } }\right] \boldsymbol{u}_p \\
& = \left[{{\int_\Omega [\boldsymbol{B}^{\bar \epsilon}]^T \ c \ \frac{\partial [\nabla\bar{\epsilon}]}{\partial u^{p,e}} \ d\Omega } -{ \int_\Omega [\boldsymbol{N}^{\bar \epsilon}]^T \frac{\partial {\epsilon}}{\partial u^{p,e}} \ d\Omega }}\right] \boldsymbol{u}_p \\
& = \left[{{\int_\Omega [\boldsymbol{B}^{\bar \epsilon}]^T \ c \ \boldsymbol{B}^{\bar \epsilon} \ \cancel{\frac{\partial \bar{\epsilon}}{\partial u^{p,e}}} \ d\Omega } -{ \int_\Omega [\boldsymbol{N}^{\bar \epsilon}]^T \frac{\partial {\epsilon}}{\partial {\epsilon}_{ij}} \frac{\partial {\epsilon}_{ij}}{\partial u^{p,e}} \ d\Omega }}\right] \boldsymbol{u}_p \\
& = \left[{-{ \int_\Omega [\boldsymbol{N}^\epsilon]^T \frac{\partial {\boldsymbol{B}^u u^e}}{\partial u^{p,e}} \ d\Omega }}\right] \boldsymbol{u}_p = \left[{-{ \int_\Omega [\boldsymbol{N}^\epsilon]^T \boldsymbol{B}^u \ d\Omega }}\right] \boldsymbol{u}_p = \prescript{n}{i}{\left[{\boldsymbol{J}^{\bar \epsilon u}}\right]}^p \boldsymbol{u}^p
\end{align}

\begin{align}
p) {\dfrac{\partial {\boldsymbol{r}^{\bar\epsilon}}}{\partial {\bar \epsilon}^e}}  & = \dfrac{\partial \ {{\left[{{{\int_\Omega [\boldsymbol{N}^{\bar \epsilon}]^T \bar{\epsilon} \ d\Omega + \int_\Omega [\boldsymbol{B}^{\bar \epsilon}]^T \ c \ \nabla \bar{\epsilon} \ d\Omega} -{ \int_\Omega [\boldsymbol{N}^{\bar \epsilon}]^T \epsilon \ d\Omega }}}\right]}}}{\partial {\bar \epsilon}^e} \\
& = {{\int_\Omega [\boldsymbol{N}^{\bar \epsilon}]^T \frac{\partial \bar{\epsilon}}{\partial {\bar \epsilon}^e} \ d\Omega + \int_\Omega [\boldsymbol{B}^{\bar \epsilon}]^T \ c \ \frac{\partial [\nabla\bar{\epsilon}]}{\partial {\bar \epsilon}^e} \ d\Omega } -{\int_\Omega [\boldsymbol{N}^{\bar \epsilon}]^T \cancel{\frac{\partial {\epsilon}}{\partial {\bar \epsilon}^e}} \ d\Omega }} \\
& = {{\int_\Omega [\boldsymbol{N}^{\bar \epsilon}]^T [\boldsymbol{N}^{\bar \epsilon}] d\Omega + \int_\Omega [\boldsymbol{B}^{\bar \epsilon}]^T \ c \ \boldsymbol{B}^{\bar \epsilon} \ \frac{\partial \bar{\epsilon}}{\partial \bar \epsilon^e} \ d\Omega}}\\
& = {{\int_\Omega [\boldsymbol{N}^ {\bar \epsilon}]^T [\boldsymbol{N}^{\bar \epsilon}]  \ d\Omega + \int_\Omega [\boldsymbol{B}^{\bar \epsilon}]^T \ c \ \boldsymbol{B}^{\bar \epsilon}\ \ d\Omega}}= \prescript{n}{i}{\boldsymbol{J}^{{\bar \epsilon} {\bar \epsilon}}}
\end{align}

\noindent Thus, the final system of equations is:

\begin{gather}
 \begin{bmatrix} I & {\prescript{n}{i}{\boldsymbol{J}^{pf}}} & {\prescript{n}{i}{\boldsymbol{J}^{pp}} \ \boldsymbol{u}^p} & {\prescript{n}{i}{\left[{\boldsymbol{J}^{u \bar\epsilon}}\right]}^p} \\  & \\ \boldsymbol{0} & {\prescript{n}{i}{\boldsymbol{J}^{ff}}} & {\prescript{n}{i}{\left[{\boldsymbol{J}^{fp}}\right]} \ \boldsymbol{u}^p} & {\prescript{n}{i}{\left[{\boldsymbol{J}^{u \bar\epsilon}}\right]}^f}\\  & \\ {2 {\beta}^{2} {\left[\prescript{n}{i}{\Delta \boldsymbol{f}^{ext}}\right]}^T} & {2 {\left[\prescript{n}{i}{\Delta \boldsymbol{u}^f}\right]}^T} & {2 {\left[\prescript{n}{i}{\Delta \boldsymbol{u}^p}\right]}^T \boldsymbol{u}^p} & {2 {\prescript{n}{i}{\Delta \bar{\boldsymbol{\epsilon}}}}^T} \\& \\ \boldsymbol{0} & {\prescript{n}{i}{\left[{\boldsymbol{J}^{\bar \epsilon u}}\right]}^f} & {\prescript{n}{i}{\left[{\boldsymbol{J}^{\bar \epsilon u}}\right]}^p \boldsymbol{u}^p} & {\prescript{n}{i}{\boldsymbol{J}^{{\bar \epsilon} {\bar \epsilon}}}}
 \end{bmatrix}
\begin{bmatrix} \prescript{n+1}{i}{\ \delta \boldsymbol{f}^{ext}} \\ \\ \prescript{n+1}{i}{\ \delta \boldsymbol{u}^f} \\ \\ \prescript{n+1}{i}{\ \delta \bar{m}} \\ \\ \prescript{n+1}{i}{\delta {\ \bar {\boldsymbol{\epsilon}}}} \end{bmatrix}
 =
 - \begin{bmatrix} \prescript{n}{i}{\boldsymbol{r}^p} \\ \\  \prescript{n}{i}{\boldsymbol{r}^f} \\ \\ \prescript{n}{i}{g} \\ \\ \prescript{n}{i}{\boldsymbol{r}^{\bar\epsilon}} \end{bmatrix} 
 \end{gather}

\section{Mazars damage model}
\label{Appendix: Mazar_Damage_model}

In this work, a widely cited damage model first proposed by Mazars \cite{mazars1986description} is used. The following is the condition based on which damage is triggered in the model:

\begin{equation}
    d(\varepsilon^{*}_{eq}) =
    \left\{
    	\begin{array}{ll}
    	0 & \text { if } \varepsilon^{*}_{eq} < \varepsilon_{D} \\
    	1 - \frac{\varepsilon_{D}(1-\mathscr{A})}{\varepsilon^{*}_{eq}} - 
    	\frac{\mathscr{A}}{\exp(\mathscr{B}(\varepsilon^{*}_{eq} - \varepsilon_{D}))} & \text { if } \varepsilon^{*}_{eq} \geq \varepsilon_{D}
    	\end{array}
    \right.
\label{DamageA}
\end{equation}

In the expression above, \ $\varepsilon^{*}_{eq}$ is the local or non-local equivalent strain,\ $\varepsilon_{D}$ is the damage threshold strain at which damage initiates, while $\mathscr{A}$ and $\mathscr{B}$ are material properties. Two definitions of $\varepsilon^{*}_{eq}$ are adopted in this work. In the problems with tensile loads, $\varepsilon^{*}_{eq}$ is calculated following Mazars approach \cite{mazars1984application}: 

\begin{equation}
  \varepsilon^{*}_{eq} = \sqrt{\sum_{I=1}^{3} {\langle \epsilon_I \rangle}^2} 
  \label{ApxEq:Equivalent_strain_tension}
\end{equation}

where, $\epsilon_I$, I = 1,2,3, are the principal strains, and the Macauley brackets denote the positive part  $\displaystyle {\langle \boldsymbol{\cdot} \rangle} = \frac{\lvert \boldsymbol{\cdot} \rvert + \boldsymbol{\cdot} }{2}$. Eqn. \eqref{ApxEq:Equivalent_strain_tension} is used to calculate the equivalent strain in all the problems presented in Section \ref{Sec:NumericalExamples} except SNS. In the Single Notch Shear (SNS) 2D problem, the  $\varepsilon^{*}_{eq}$ is based on the work of \cite{de1995comparison}; $I_{1}$ and $J_{2}$ are the strain invariants and $\bf{\varepsilon}$ is the strain tensor.

\begin{equation}
   \varepsilon^{*}_{eq} = \frac{k-1}{2k(1-2\nu)} + \frac{1}{2k} \sqrt{\frac{(k-1)^2}{(1-2\nu)^{2}}I_{1}^{2} + \frac{2k}{(1+\nu)^2}J_{2}}
\label{ApxEq:Equivalent_strain_shear}
\end{equation}

\noindent where :

\begin{equation}
    I_{1} = tr(\boldsymbol{\varepsilon})
\label{ApxEq:Invariants1}
\end{equation}

\begin{equation}
    J_{2} = 3tr(\boldsymbol{\varepsilon} \cdot \boldsymbol{\varepsilon}) - tr^{2}(\boldsymbol{\varepsilon})
\label{ApxEq:Invariants2}
\end{equation}

\appendixpageoff
\appendixtitleoff

\end{document}